\documentclass[11pt]{article}
\pdfoutput=1
\usepackage[utf8]{inputenc}
\usepackage{amsmath,amsfonts,amssymb}
\usepackage{amsthm}
\usepackage{latexsym}
\usepackage[noadjust]{cite}
\usepackage{graphicx}
\usepackage{xcolor}
\usepackage{dirtytalk}
\usepackage{mathtools}
\usepackage[margin=1in]{geometry}
\usepackage{thm-restate}
\usepackage{enumitem}
\usepackage[linesnumbered,ruled]{algorithm2e}
\usepackage[colorlinks=true, allcolors=blue]{hyperref}
\usepackage[nameinlink,capitalise]{cleveref}
\hypersetup{
    citecolor={violet}
}

\newtheorem{theorem}{Theorem}[section]
\newtheorem{corollary}[theorem]{Corollary}
\newtheorem{lemma}[theorem]{Lemma}

\newtheorem{claim}[theorem]{Claim}
\theoremstyle{definition}
\newtheorem{definition}[theorem]{Definition}
\newtheorem{remark}{Remark}[section]
\newtheorem{fact}[theorem]{Fact}

\newcommand{\F}{\mathbb{F}}
\newcommand{\N}{\mathbb{N}}

\newcommand{\R}{\mathbb{R}}
\newcommand{\Z}{\mathbb{Z}}

\newcommand{\calS}{\mathcal{S}}

\newcommand{\zo}{\{0, 1\}}

\newcommand{\eps}{\epsilon}

\newcommand{\ra}{\rightarrow}

\newcommand{\minkcut}{\mathbf{\Phi}}

\newcommand{\UR}{\mathbf{UR}}

\newcommand{\Supp}{\mathrm{Supp}}

\newcommand{\polylog}{\mathrm{polylog}}

\newcommand{\poly}{\mathrm{poly}}

\newif\ifdraft
\newif\ifanon

\drafttrue
\anonfalse

\title{Near-optimal Size Linear Sketches for Hypergraph Cut Sparsifiers}

\ifanon
\else
\author{Sanjeev Khanna\thanks{School of Engineering and Applied Sciences, University of Pennsylvania, Philadelphia, PA. Email: {\tt sanjeev@cis.upenn.edu}. Supported in part by NSF awards CCF-1934876, CCF-2008305, and CCF-2402284.} \and Aaron (Louie) Putterman\thanks{School of Engineering and Applied Sciences, Harvard University, Cambridge, Massachusetts, USA. Supported in part by the Simons Investigator Awards of Madhu Sudan and Salil Vadhan, NSF Award CCF 2152413 and a Hudson River Trading PhD Research Scholarship. Email: \texttt{aputterman@g.harvard.edu}.} \and Madhu Sudan\thanks{School of Engineering and Applied Sciences, Harvard University, Cambridge, Massachusetts, USA. Supported in part by a Simons Investigator Award and NSF Award CCF 2152413. Email: \texttt{madhu@cs.harvard.edu}.}}
\fi

\date{\today}

\begin{document}

\maketitle

\begin{abstract}

A $(1 \pm \epsilon)$-sparsifier of a hypergraph $G(V,E)$ is a (weighted) subgraph that preserves the value of every cut to within a $(1 \pm \epsilon)$-factor. It is known that every hypergraph with $n$ vertices admits a $(1 \pm \epsilon)$-sparsifier with $\tilde{O}(n/\epsilon^2)$ hyperedges. 
In this work, we explore the task of building such a sparsifier by using only linear measurements (a \emph{linear sketch}) over the hyperedges of $G$, and provide nearly-matching upper and lower bounds for this task.

Specifically, we show that there is a randomized linear sketch of size $\widetilde{O}(n r \log(m) / \epsilon^2)$ bits which with high probability contains sufficient information to recover a $(1 \pm \epsilon)$ cut-sparsifier 
with $\tilde{O}(n/\epsilon^2)$ hyperedges for 
any hypergraph with at most $m$ edges each of which has arity bounded by $r$. This immediately gives a dynamic streaming algorithm for hypergraph cut sparsification with an identical space complexity, improving on the previous best known bound of $\widetilde{O}(n r^2 \log^4(m) / \epsilon^2)$ bits of space (Guha, McGregor, and Tench, PODS 2015).     
We complement our algorithmic result above with a nearly-matching lower bound. We show that for every $\epsilon \in (0,1)$, one needs $\Omega(nr \log(m/n) / \log(n))$ bits to construct a $(1 \pm \epsilon)$-sparsifier via linear sketching, thus showing that our linear sketch achieves an optimal dependence on both $r$ and $\log(m)$.

The starting point for our improved algorithm is importance sampling of hyperedges based on the new notion of $k$-cut strength introduced in the recent work of Quanrud (SODA 2024). The natural algorithm based on this concept leads to $\log m$ levels of sampling where errors can potentially accumulate, and this accounts for the $\text{polylog}(m)$ losses in the sketch size of the natural algorithm. We develop a more intricate analysis of the accumulation in error to show
most levels do not contribute to the error and actual loss is only $\text{polylog}(n)$. Combining with careful preprocessing (and analysis) this enables us to get rid of all extraneous $\log m$ factors in the sketch size, but the quadratic dependence on $r$ remains. This dependence originates from use of correlated $\ell_0$-samplers to recover a large number of low-strength edges in a hypergraph simultaneously by looking at neighborhoods of individual vertices. In graphs, this leads to discovery of $\Omega(n)$ edges in a single shot, whereas in hypergraphs, this may potentially only reveal $O(n/r)$ new edges, thus requiring $\Omega(r)$ rounds of recovery. To remedy this we introduce a new technique of \emph{random fingerprinting} of hyperedges which effectively eliminates the correlations created by large arity hyperedges, and leads to a scheme for recovering hyperedges of low strength with an optimal dependence on $r$. Putting all these ingredients together yields our linear sketching algorithm. Our lower bound is established by a reduction from the universal relation problem in the one-way communication setting.
\end{abstract}

\newpage

\tableofcontents

\newpage
\pagenumbering{arabic}

\section{Introduction}

In this paper we study the task of building cut sparsifiers for hypergraphs in the {\em linear sketching model} and derive nearly matching bounds on the size of the sketch as a function of key hypergraph parameters. 

For any $\epsilon \in (0,1)$, given a hypergraph $H=(V,E)$ where every hyperedge (sometimes simply referred to as an edge) $e \in E$ is a subset of $V$, a $(1 \pm \epsilon)$-sparsifier of $H$ is a re-weighted subhypergraph which preserves the weight of every cut to a $(1 \pm \eps)$ multiplicative factor. The goal in hypergraph sparsification is to construct, or to prove the existence of, a small sparsifier (where the size of the sparsifier is measured by the number of hyperedges) for a given hypergraph. In this work we study the space required to build such a sparsifier in the {\em linear sketching} model, where the sparsifier has to be reconstructed from a linear ``measurement'' of the input.\footnote{Here, a hypergraph on $n$ vertices is viewed as a vector in $\{0,1\}^{2^n}$ and a linear measurement of size $s$ is obtained by mutliplying a (possibly random) $s \times 2^n$ matrix with this vector.} We study the space required as a function of three key parameters: $n$, the number of vertices in $H$; $m$, the number of edges in $H$ and $r$ the arity (size) of largest hyperedge in $H$. It is known that the space required by the smallest linear sketch depends only polynomially on the parameters $n$, $r$ and $\log m$, and in this work we get the exact polynomial that governs the space required, up to polylogarithmic factors in $n$. We review some of the past work before stating our results in greater detail. 

We start with an abbreviated history of the notion of sparsification.
Cut-preserving sparsification of graphs has been a fundamental tool in algorithm design ever since its conception in the seminal works of Karger \cite{Kar93} and Bencz\'ur and Karger \cite{BK96}. 
Subsequent work generalized this in many different directions, for instance, to spectral sparsification \cite{BSS09, ST11}, to cut and spectral sparsification in hypergraphs
\cite{KK15, BST19, SY19, CKN20, KKTY21b, KKTY21a, JLS22, Lee23, JLLS23}, to sparsification of linear codes (which capture graph cuts as a special case) \cite{KPS24}, and to sparsifiying quotients of submodular functions \cite{Qua23}, in each case achieving sparsifiers of essentially the optimal size of $\widetilde{O}(n)$, with $n$ being the number of vertices in the graph or hypergraph, the dimension of the linear code, and the maximum value of the submodular function, respectively. 

The above mentioned works primarily focus on the standard model of computing where the algorithm has unrestricted access to the input. Our interest in this paper is in hypergraph sparsification via linear sketches of small size. Linear sketching algorithms immediately lend themselves to several models of computation including the dynamic streaming model (allowing for insertions \emph{and} deletions of (hyper)edges) and the massively parallel computation (MPC) model \cite{KSV10}, where unrestricted access to the entire input is not readily available.  
On the other hand, the restrictive nature of linear sketching algorithms also makes it more challenging to obtain such sketches with a small space footprint, for complex problems.
In the context of graph algorithms, the power of linear sketching was first illustrated in the work of Ahn, Guha, and McGregor \cite{AGM12, AGM12b} who showed that for any $\eps \in (0,1)$, a linear sketch of size $\widetilde{O}(n / \eps^2)$ suffices to recover with high probability a $(1 \pm \eps)$-(cut-)sparsifier of any graph. This led to a sequence of works studying the capabilities of linear sketching (and, as a consequence, dynamic streaming) for creating sparsifiers of graphs. For instance, the work of Kapralov et. al. \cite{KLMMS14} showed that a linear sketch using $\widetilde{O}(n / \eps^2)$ bits suffices for creating $(1 \pm \eps)$-\emph{spectral} sparsifiers of graphs, and the work of Chen, Khanna, and Li \cite{CKL22} studied cut and spectral sparsification for \emph{weighted} graphs in the \emph{turnstile stream model} using linear sketches. 
Guha, McGregor and Tench \cite{GMT15} initiated the study of \emph{hypergraph} cut-sparsification with linear sketches and showed that in this case a complexity of $\widetilde{O}(nr^2 \log^4(m) / \eps^2)$ bits suffices to recover a $(1 \pm \eps)$ cut-sparsifier.\footnote{The work of \cite{GMT15} focuses on the case when hypergraphs are of \emph{constant} arity, and show that in this case a linear sketch of size $\widetilde{O}(n / \eps^2)$ suffices (i.e. when $r = O(1), m = n^{O(1)}$). If one uses their algorithm for general hypergraphs, the sketch size becomes $\widetilde{O}(nr^2 \log^4(m) / \eps^2)$ bits.} 

\subsection{Our Results}

In this work we present a linear sketching framework for creating hypergraph cut-sparsifiers that achieves nearly-optimal size.

\begin{theorem}\label{thm:main}
 For any $\eps \in (0,1)$, there is a randomized linear sketch of size $\widetilde{O}(nr \log(m) / \eps^2)$ bits that given any $n$-vertex unweighted hypergraph $H$ with at most $m$ edges of arity bounded by $r$, allows recovery of a $(1 \pm \eps)$-sparsifier of $H$ with high probability.\footnote{Note that the $\tilde{O}(\cdot)$ is hiding only logarithmic factors in $\cdot$, i.e., factors of $\log(n)$, $\log(r)$, $\log(1 /\eps)$, and $\log\log(m)$. }
\end{theorem}

Thus while our result maintains the optimal dependence on $n$ and $\epsilon$ as in \cite{GMT15}, we improve the dependence on $r$ and $\log m$ where each of these parameters could be as large as $n$. Indeed for the extremal choice of $r=\Theta(n)$ and $m=2^{\Theta(n)}$, our result improves the space required from $\widetilde{O}(n^7/\epsilon^2)$ to $\widetilde{O}(n^3/\epsilon^2)$. We also show tightness of our bound (up to $\poly\log n$ factors) for all ranges of $n$, $r$ and $m$, as we elaborate later. 

\begin{remark}
In fact, our result is actually slightly stronger than stated above. The sparsifiers we recover are the so-called \emph{$k$-cut sparsifiers}, meaning that for any $k \in [2, \dots n]$, and any partition of the vertex set into $V_1, \dots V_k$, the weight of cut hyperedges (that is, hyperedges that are not completely contained in any single $V_i$) is preserved to within a $(1 \pm \eps)$ factor. See \cref{rmk:thmforkcuts} for an elaboration.
\end{remark}

Our sketching algorithm is obtained by putting together two ingredients. The first is the framework of $k$-cut strengths in hypergraphs developed by Quanrud \cite{Qua23} originally used for \emph{fast} $k$-cut sparsification algorithms for \emph{static} hypergraphs. 

Instead, we adopt, extend and then ultimately implement this framework using a linear sketch. In this framework, we perform a careful analysis of the degradation of error in our sparsification procedure, and subsequently add a pre-processing phase to our linear sketch which identifies \say{extremely} well-connected components, together saving a factor of $\log^3(m)$ over the work of \cite{GMT15}. Our final ingredient is to introduce our technique of \emph{random fingerprinting}, which we use to save an additional factor of $r$ over a naive implementation, leading to the stated theorem.

We now briefly explain our ideas regarding fingerprinting for obtaining an improved dependence on $r$. In order to use the $k$-cut characterizations of hyperedge strengths \cite{Qua23}, an essential step is to be able to recover all of the hyperedges of \emph{low} strength as these must be preserved exactly (the sampling rate needed for a hyperedge is inversely proportional to its strength). The standard approach towards recovering important edges initializes $\ell_0$-samplers with correlated randomness defined for various components of the graph, and then uses these samplers to recover random hyperedges incident to these components. Unfortunately, when using $k$-cut characterizations of strength, there can be many large arity hyperedges with low enough strength that they must all be exactly recovered. One consequence of the large arity is that each hyperedge may be incident on multiple components, and thus when using correlated $\ell_0$-samplers to recover incident hyperedges, multiple components may output the same hyperedge. Thus in a single round which may consume $\Omega(n)$ $\ell_0$-samplers, one might only recover $\widetilde{O}(\frac{n}{r})$ distinct hyperedges. Recovery of all relevant hyperedges may thus require $\tilde{\Omega}(nr)$ $\ell_0$-samplers overall, leading to a quadratic dependence on $r$ in sketch size as in the previous work (the second factor of $r$ comes from the space to store each $\ell_0$-sampler for hyperedges of arity $r$). To overcome this, we introduce a new technique of \emph{random fingerprinting} of hyperedges. For each hyperedge, we independently sample a random subset of its vertices to induce a \say{fingerprint} of the hyperedge, and now run the recovery procedure on this fingerprinted hypergraph. Because the arities of hyperedges are smaller in this fingerprinted hypergraph, we show that we have largely \emph{broken} the correlation between samplers, yet surprisingly, these fingerprinted hypergraphs still maintain sufficient information to recover all low-strength hyperedges of the original hypergraph using only $O(\polylog(n))$ $\ell_0$-samplers per vertex. In other words, using only $\widetilde{O}(n)$ $\ell_0$-samplers total, we can recover all low strength hyperedges, just as in the graph case. 

Our techniques for removing the super-linear dependence on $\log(m)$ are similarly involved, though we defer their discussion to the detailed technical overview \cref{sec:detailedTech}.

As an aside, note that by using the standard geometric grouping idea, our linear sketch can also be extended to weighted hypergraphs with integer hyperedge weights between $1$ and $W$ using space of $\widetilde{O}(nr \log(m) \log W/ \eps^3)$ bits (see, for instance, \cite{KLMMS14} on how this is done in prior work). We focus here on the unweighted case.

In general, even for static instances of hypergraphs of arity $r$, the bit complexity of a sparsifier is $\widetilde{\Omega}(nr)$ \cite{KKTY21b}. So, ignoring the $\log(m)$ term, our linear sketch has an essentially optimal dependence on $n, r$.
One might then conjecture that, in fact, there should be \emph{no} dependence on $\log(m)$ in the sketch size, particularly as our result already shaves off a factor of $\log^3(m)$ from previously known bounds. However, we complement the preceding theorem with a general lower bound, showing that our size bound (\emph{including the dependence on $\log(m)$}) is in fact nearly-tight.

\begin{theorem}\label{thm:mainLB}
    For any $\eps \in (0,1)$, any randomized linear sketch 
    that can be used to recover a $(1 \pm \eps)$-sparsifier with probability at least $1 - 1 / \text{poly}(n)$ on $n$-vertex unweighted hypergraphs with at most $m$ hyperedges of arity bounded by $r$ requires $\Omega(nr \log(m/n) / \log(n))$ bits of space. 
\end{theorem}

This lower bound follows from a reduction from a variant of the universal-relation problem in the one-way communication setting between Alice and Bob. To do this, we first show that for our variant of universal-relation, Alice must send at least $\Omega(nr \log(m/n))$ bits to solve the problem. Then, we show that for any instance of this problem, there exists an encoding into a family of \say{bipartite} hypergraphs, such that if Alice sends only $O(\log(n))$ independent hypergraph sparsification sketches, Bob can with high probability solve the original problem. Thus, we can immediately conclude the above bound. 

Next, we highlight some easy corollaries of our linear sketching result. As mentioned above, we can use this linear sketching algorithm to create a general \emph{dynamic} streaming algorithm for hypergraphs that tolerates both insertions and deletions of hyperedges.

\begin{corollary}
\label{thm:mainStreaming}
For any $\eps \in (0,1)$, there is a (randomized) dynamic streaming algorithm using $\widetilde{O}(nr \log(m) / \eps^2)$ bits of space that, for any sequence of insertions / deletions of hyperedges in an $n$-vertex unweighted hypergraph $H$ with at most $m$ edges of arity bounded by $r$, allows recovery of a $(1 \pm \eps)$-sparsifier of $H$ with high probability. 
\end{corollary}

The improves upon the best previous space bound of 
$\widetilde{O}(nr^2 \log^4(m) / \eps^2)$ for
hypergraph sparsification in dynamic streams \cite{GMT15}. It also improves upon the best previous space bound of $\widetilde{O}(n r \log^4(m) / \eps^2)$ for the 
simpler insertion-only model \cite{CKN20}. Note that although their algorithm achieves an optimal space dependence on $n$ and $r$, it is strictly tailored for \emph{insertion-only} streams and cannot be extended to handle deletions. In particular, in the setting of dense hypergraphs of large arity, namely, when $m = 2^{\Omega(n)}$, and $r = \Omega(n)$, our sketch requires  $\widetilde{O}(n^3 / \eps^2)$ bits, while the sketches in \cite{GMT15} and \cite{CKN20} guarantee only $\widetilde{O}(n^7 / \eps^2)$ and $\widetilde{O}(n^6 / \eps^2)$ bits, respectively.

Likewise, our linear sketching scheme can also be used to obtain efficient algorithms for computing hypergraph sparsifiers in the \emph{massively parallel computation} (MPC) model \cite{KSV10}. Roughly speaking, in this model, the input data (in our case the hyperedges of a hypergraph) are split across, say $k$, machines. Each machine has bounded memory (in our case bounded by $\widetilde{O}(nr\log(m) / \eps^2)$) and the computation is split into rounds, where between rounds machines are allowed to send their local data to other machines, and within rounds, are allowed to perform an arbitrary amount of computation on their data, with the goal of eventually outputting a sparsifier for the hypergraph. 
The total communication that any machine is allowed in a single round is bounded by the size of the machine's memory. 
Our linear sketch for hypergraph sparsification lends itself to a natural MPC algorithm for hypergraph sparsification, with significant improvements over the canonical algorithm.

\begin{corollary}
 For any $\eps \in (0,1)$, there is a randomized MPC algorithm using machines with memory $\widetilde{O}(nr \log(m) / \eps^2)$ bits that given any $n$-vertex unweighted hypergraph $H$ with at most $m$ edges of arity bounded by $r$ arbitrarily partitioned across the machines, allows recovery of a $(1 \pm \eps)$-sparsifier of $H$ with high probability in $\max(2, \lceil \log_n(m) \rceil)$ rounds. 
\end{corollary}

For comparison, the canonical approach to building MPC algorithms for sparsifying hypergraphs \emph{without} linear sketches involves each machine $m_i$ sparsifying its own induced hypergraph, and then recursively combining these hypergraphs in a tree-like manner, in each iteration pairing up two active machines, merging their hypergraphs, and then sparsifying this merged hypergraph. Thus, in each iteration, the number of active machines decreases by a factor of $2$. This approach (which is also used to create sparsifiers for \emph{insertion-only} streams \cite{CKN20}) unfortunately loses in two key parameter regimes. First, the number of rounds required by such a procedure will be $\Omega(\log(m/n))$, as the number of active machines decreases by a factor of $2$ in each round. Further, the memory required by each machine will be $\Omega(nr\log(m)\log^2(m/n) / \eps^2)$, as the deterioration of the error parameter scales with the depth of the recursive process, which will be $\log(m/n)$, and setting $\eps'= \eps / \log(m/n)$ requires more memory. 

As an example, when $m = \poly(n)$, our MPC protocol runs in a \emph{constant} number of rounds (independent of the number of vertices), while the canonical MPC algorithm for sparsification will require $\Omega(\log(n))$ rounds. Further, we will be getting this reduction in rounds \emph{in conjunction} with a smaller memory footprint. 

\subsection{Conclusion}

Extending near-linear size graph sparsifiers to near-linear size hypergraph sparsifiers has proved to be a challenging task. The work of \cite{KK15} shows that if one is willing to pay a factor of $r$ in the sparsifier size, then simple extensions of ordinary graph sparsification suffice but this leads to quadratic-size sparsifiers when $r$ is large. Eventually, linear-size hypergraph sparsifiers were obtained but these constructions utilize unrestricted access to the input hypergraph to implement more complex non-uniform sampling schemes than used in the case of graph sparsification (for instance, sampling based on balanced weight assignments of \cite{CKN20, KKTY21b}, and sampling based on $k$-cut strengths in \cite{Qua23}). We thus view it as somewhat surprising that despite the complexity of these approaches, linear measurements of space complexity almost matching that of optimal hypergraph sparsifiers still suffice to recover a hypergraph sparsifier. In other words, our results show that there is effectively no space overhead incurred in going from the classical setting of creating a near-linear size sparsifier of a static hypergraph to the linear sketching setting that entertains dynamic insertion/deletion updates to the underlying hypergraph.

\subsection{Organization}

In \cref{sec:detailedTech}, we provide a more in-depth discussion of our results on linear sketching. In \cref{sec:preliminaries}, we provide background on $k$-cuts in hypergraphs, recap results from \cite{Qua23}, derive new properties of $k$-cut strengths, and summarize known constructions in linear sketching. In \cref{sec:sparsifiers}, we introduce a linear sketch for creating sparsifiers conditioned on the existence of a \say{recovery} sketch, which is then constructed in \cref{sec:fingerprinting} via our fingerprinting techniques. \cref{sec:lowerBound} proves the reduction from the universal relation problem to lower-bound the size of any linear sketch for hypergraph sparsification. Finally, \cref{sec:streaming} and \cref{sec:MPC} prove our results in the streaming setting and MPC setting, respectively.

\section{Detailed Technical Overview}\label{sec:detailedTech}

\subsection{Graph Sparsification and Hypergraph Sparsification via $k$-cuts}\label{sec:detailedTechBackground}

A key ingredient underlying the seminal graph sparsification works of Karger~\cite{Kar93} and Bencz\'ur and Karger \cite{BK96}, which most other sparsification algorithms have an analog for, is the following \say{cut-counting} bound for graphs:

\begin{theorem}\cite{Kar93}
    For any $t \in \Z^+$, any graph $G$ on $n$ vertices with minimum cut $c$, has at most $n^{2t}$ cuts of size at most $t \cdot c$. 
\end{theorem}

An easy consequence of the cut counting bound above is that in any graph with minimum cut size $c$, if one samples edges at rate $p=O(\log(n) / (\eps^2 c))$ (and re-scales the weight of each sampled edge to be $1/p$), then the weight of every cut is preserved to within a factor of $(1 \pm \eps)$ with high probability. To establish this assertion for cuts of size roughly $t \cdot c$, we can simply use a union bound over all of them since there are at most $n^{2t}$ such cuts. While this uniform sampling scheme suffices to effectively sparsify graphs with a large minimum cut size, additional ideas are needed to sparsify graphs with small cuts. To this end, Bencz\'ur and Karger \cite{BK96} introduced the notion of \say{strength} of an edge that determines its importance in preserving cut sizes. This yields non-uniform edge sampling rates and they used this to show that every graph admits a $(1 \pm \eps)$-sparsifiers with $\tilde{O}(n/\eps^2)$ edges. The proof of this result once again relies on a more careful application of the cut counting bound above. Subsequently, Ahn, Guha, and McGregor \cite{AGM12} showed that a variant of Bencz\'ur-Karger graph sparsification can in fact be implemented using a {\em linear sketch} of size $\widetilde{O}(n / \eps^2)$ that contains enough information to recover a $(1 \pm \eps)$-sparsifier with high probability. 

Early works generalizing graph cut sparsifiers to hypergraph cut-sparsifiers quickly discovered that the cut counting bound that serves as the foundation of graph sparsification algorithms is far from being true in the case of hypergraphs. Indeed, the work of Kogan and Krauthgamer \cite{KK15} observes that in hypergraphs of arity $r$, there can be as many as $2^{\Omega(r)}$ cuts of size within a constant factor of the minimum cut, and more generally, they show that the number of cuts of size $\leq t \cdot c$ can be as large as $n^{\Omega(t)} \cdot 2^{\Omega(r)}$ (for $c$ the minimum cut). 
This blow-up in the number of small cuts in turn implies that hyperedges need to be sampled at a rate that is $\Omega(r)$ times higher if one wishes to directly apply the Bencz\'ur and Karger \cite{BK96} graph sparsification approach to hypergraphs.  
As a consequence, creating sparsifiers with this approach requires $\Omega(nr)$ hyperedges and therefore $\Omega(nr^2)$ bits of space (as each hyperedge can have $\Omega(r)$ description complexity). 
Thus, this adaptation of graph linear sketches to hypergraphs (as in \cite{GMT15}) inherently requires a quadratic dependence on $r$.

To overcome the obstacle posed by the exponentially larger cut counting bound, we instead build on a new approach to hypergraph sparsification developed by Quanrud \cite{Qua23}. Instead of focusing just on the $2$-cuts in hypergraphs, \cite{Qua23} generalizes this notion to $k$-cuts in hypergraphs where $2 \le k \le n$, with the benefit of now getting a direct analog of the cut-counting bound in graphs. 

\begin{definition}\label{def:minkcut}
   For any $k \in [2..n]$, a $k$-cut in a hypergraph is defined by a $k$-partition of the vertices, say, $V_1, \dots V_k$. The \emph{un-normalized size of a $k$-cut} in an unweighted hypergraph is the number of hyperedges that are not completely contained in any single $V_i$ (we refer to these as the crossing hyperedges), denoted by $E[V_1, \dots V_k]$.

    The \emph{normalized size of a $k$-cut} in a hypergraph is its un-normalized size divided by $k-1$. We will often use $\minkcut(H)$ to denote the minimum normalized $k$-cut, defined formally as follows:
    \[
    \minkcut(H) = \min_{k \in [2..n]} \min_{V_1, \cup  \dots \cup V_k = V} \frac{|E[V_1, \dots V_k]|}{k-1}.
    \]
\end{definition}

Note that when we generically refer to a $k$-cut, this refers any choice of $k \in [2..n]$. That is, we are not restricting ourselves to a single choice of $k$, but instead allowing ourselves to range over any partition of the vertex set into any number of parts. 

The work of \cite{Qua23} established the following result regarding normalized and un-normalized $k$-cuts:

\begin{theorem}\cite{Qua23}
    Let $H$ be a hypergraph, with associated minimum normalized $k$-cut size $\minkcut(H)$. Then for any $t \in \Z^{+}$, and $k \in [2..n]$, there are at most $n^{O(t)}$ un-normalized $k$-cuts of size $\leq t \cdot \minkcut(H)$.
\end{theorem}

A direct consequence of the above is that in order to preserve all $k$-cuts (again, simultaneously for every $k \in [2, \dots n]$) in a hypergraph $H$ to a factor $(1 \pm \eps)$, it suffices to sample each hyperedge at rate $p \geq \frac{C \log(n)}{\eps^2 \minkcut(H)}$, and re-weight each sampled hyperedge by $1/p$. 

Similar to Bencz\'ur and Karger's \cite{BK96} approach for creating $\widetilde{O}(n / \eps^2)$ size graph sparsifiers, Quanrud \cite{Qua23} next uses this notion to define \emph{$k$-cut strengths} for each hyperedge. To do this, fix a minimum normalized $k$-cut, and let $V_1, V_2, ..., V_k$ be the corresponding partition of the vertices. For any hyperedge crossing this minimum normalized $k$-cut, we define its strength to be $\minkcut(H)$. Then, the strengths for hyperedges completely contained within the components $V_1, \dots V_k$ are determined recursively (within their respective induced subgraphs) using the same scheme. This allows Quanrud \cite{Qua23} to calculate sampling rates of hyperedges, which when sampled, approximately preserve the size of every $k$-cut (for all $k \in [2, n]$). Unfortunately, Quanrud's \cite{Qua23} algorithm relies on simultanesouly sampling all hyperedges, which is often unachievable with linear sketches. As such, we present a natural alternative using an iterative algorithm for sparsification (building off the frameworks of \cite{BK96, AGM12, GMT15}), which we present below:

\begin{algorithm}[H]
\caption{SimpleSparsification($H, \eps$)}\label{alg:introSimple}
Let $H_0 = H$, let $C$ be a sufficiently large constant.  \\
\For{$i = 0, 1, \dots \log(m)$}{
Let $F_i$ be all hyperedges in $H_i$ of strength $\leq 2C \log(n) / \eps^2$. \\
Store $F_i$. \\
Let $H_{i+1}$ be hyperedges in $(H_i - F_i)$ sampled at rate $1/2$.
}
\Return{$\cup_i 2^i \cdot F_i$.}

\end{algorithm}

The key observation underlying the above algorithm is that after removing all hyperedges of strength $\leq 2 C \log(n) / \eps^2$ from $H_i$, it must be the case that the minimum normalized $k$-cut in the hypergraph $H_i - F_i$ is at least $2 C \log(n) / \eps^2$ (see \cref{clm:removeLowStrength}). Thus, we can afford to sample $H_{i} - F_i$ at rate $1/2$ while still being guaranteed to preserve all cuts to a factor $(1 \pm \eps)$ with all but polynomially small probability. Note that the only guarantee from this procedure is that $H_{i+1}$ is a $(1 \pm \eps)$-sparsifier to $H_i - F_i$, and in turn that $H_i$ is a $(1 \pm \eps)$-sparsifier to $H_{i-1} - F_{i-1}$. Thus, the final returned result is naively a $(1 \pm O(\eps \log(m)))$-sparsifier to $H$. 

However, it remains to show how we can implement this using a linear sketch. In particular, while downsampling can be done simply with hash functions (as done in prior work with linear sketching \cite{AGM12}), the primary difficulty is in finding (and recovering) the hyperedges of low strength under the new definition of $k$-cut strength. One of our key contributions is presenting a linear sketching algorithm using only $\ell_0$-samplers that allows one to recover exactly such a decomposition; we explain the intuition for the algorithm below.

\subsection{Barriers to Finding Low Strength Hyperedges with Linear Sketches}

First, we recap $\ell_0$-samplers. Roughly speaking, an $\ell_0$-sampler is a linear sketch that takes as input a vector $x \in \R^u$, and returns a uniformly random index in the non-zero support of the vector. For any vertex $v$, if we define an $\ell_0$-sampler on the hyperedges incident on $v$, we can recover a random hyperedge incident on $v$. Furthermore, by adding together $\ell_0$-samplers for different vertices (when using the same random seed), they allow us to sample hyperedges that are leaving the component defined by the union of these vertices. These $\ell_0$-samplers are also amenable to linear updates, meaning that if we know an edge is in the support of the $\ell_0$-sampler, we can update the support of the $\ell_0$-sampler to remove this edge from the support.

\begin{definition}
    Consider a turnstile stream $S = s_1, \dots s_t$, where each $s_i = (u_i, \Delta_i)$ ($u_i \in [n], \Delta_i \in \Z$), and the aggregate vector $x \in \R^u$ where $x_i = \sum_{j:u_j = i} \Delta_i$.
    
    Given a target failure probability $\delta$, an $\ell_0$-sampler for a non-zero vector $x$ returns $\perp$ with probability $\leq \delta$, and otherwise returns an element $i \in [n]$ with probability $\frac{|x_i|_0}{|x|_0}$.
\end{definition}

\begin{fact}
    We will use the fact that (for any universe of size $u$, and support of size $\leq m$) there exists a \emph{linear} sketch-based $\delta$-$\ell_0$-sampler using space $O(\log(m)\log(u) \log (1 / \delta))$. Note that $u$ is the length of the aggregate vector $x$ from the previous definition. $m$ is an upper bound on $|x|_0$. 
\end{fact}

For dynamic streams, it is possible that after insertions, the support of the vector $x$ becomes larger than $m$, and then subsequently becomes $\leq m$ (after some deletions). In this case, the space used by the $\ell_0$-sampler is still $O(\log(m)\log(u) \log (1 / \delta))$, with the only difference being that the correctness of the sampler is only promised when the support is not too large. With this, we now explain the family of vectors for which we will create $\ell_0$-samplers.

\begin{definition}\cite{AGM12, GMT15}
    Given an unweighted hypergraph $G = (V, E)$, define the $n \times 2^{[n]}$ matrix $A_G$ with entries $(i, e)$, where $i \in [n]$ and $e \subseteq [n]$ is a hyperedge. We say that 
    \[
    A_{i,e} = \begin{cases}
        1 & \text{ if } i \in e, i \neq \max_{j \in e} j, \\
        -(|e|-1) & \text{ if } i \in e, i = \max_{j \in e} j, \\
        0 & \text{ else.}
    \end{cases}
    \]

    Let $a_1, \dots a_n$ be the rows of the matrix $A$. The support of $a_i$ corresponds with the neighborhood of the $i$th vertex.
\end{definition}
\begin{lemma}\cite{AGM12, GMT15}
    Suppose we have $\ell_0$-samplers for the neighborhoods of all vertices in a connected component $V_i$, denoted by $\calS(v, R): v \in V_i$, and $R$ the random seed. Then, $\sum_{v \in V_i} \calS(v, R)$ is an $\ell_0$-sampler for the hyperedges leaving $V_i$.
\end{lemma}

\begin{remark}
    Suppose we have a linear sketch for the $\ell_0$-sampler of the edges leaving some connected component $V_i$, denoted by $\calS(V_i, R)$. Suppose further that we know there is some edge $e$ leaving $V_i$ (that was found independently of randomness $R$ used for our $\ell_0$-sampler) that we wish to remove from the support of $\calS(V_i, R)$. Then, we can simply add a linear vector update to $\calS(V_i, R)$ that cancels out the coordinate corresponding to this edge $e$, without changing the failure probability of $\calS(V_i, R)$.
\end{remark}

At the most basic level, prior approaches like \cite{AGM12, GMT15} stored roughly $r\cdot\polylog(n,m)/\eps^2$ $\ell_0$-samplers for each vertex, where across all vertices, the $i$th $\ell_0$-sampler uses the same randomness. With this, it is then straightforward to implement an algorithm for finding disjoint spanning forests of a graph (or hypergraph) $H$. In the first iteration, each vertex opens its first $\ell_0$-sampler. The (hyper)edges recovered from these $\ell_0$-samplers induce some connected components $V_1, \dots V_k$ in $H$. Now, in the second round, for each connected component $V_i$, we add together the $\ell_0$-samplers using the second random seed for the corresponding vertices in $V_i$, yielding an $\ell_0$-sampler for the (hyper)edges leaving $V_i$. Because the randomness used for the $\ell_0$-samplers in the second round is independent of the hyperedges sampled in the first round, one can show that the failure probability of the $\ell_0$-samplers does not change. Further, in each iteration, one can maintain the invariant that a constant fraction of the connected components are merged, and thus after $O(\log(n))$ iterations, a spanning forest of the hypergraph is recovered. After running this for $r\cdot\text{polylog}(n,m) / \eps^2$ rounds (removing each recovered spanning forest between rounds), one can recover $r\cdot\text{polylog}(n,m) / \eps^2$ spanning forests, and one can show that recovering these hyperedges suffices for sampling in accordance with the $2$-cuts of a graph or a hypergraph, as the case may be. Unfortunately, storing so many $\ell_0$ samplers immediately yields a space complexity of $\Omega(nr^2 / \eps^2)$ bits (ignoring the $\log(m)$'s), as for each of $n$ vertices, we store $r / \eps^2$ $\ell_0$-samplers, each requiring $\Omega(r \log(m))$ bits of space.

In our case, where the goal is to have only a linear dependence on $r$, we must avoid sampling in accordance with the \say{$2$-cut-strengths} of the hypergraph (recall that even the static sparsifiers created with $2$-cut-strength sampling schemes require $\Omega(nr^2)$ bits to represent), and instead recover edges in accordance with the \emph{$k$-cut strengths} of the hypergraph. 
One might hope that as in graphs and constant arity hypergraphs, naively storing $\polylog(n)$ $\ell_0$-samplers per vertex of the hypergraph suffices for recovering low $k$-cut strength hyperedges, as this would then yield a linear dependence on $r$ in the sketch size. Unfortunately, as we shall see, this is \emph{not} the case, and more complicated techniques are required to ultimately achieve a linear dependence on $r$.

For instance, let us consider a hypergraph $H$ on $n$ vertices with $\sqrt{n}$ cliques $V_1, \dots V_{\sqrt{n}}$, along with $\sqrt{n}$ hyperedges that are crossing between $V_1, \dots V_{\sqrt{n}}$ (i.e. every such hyperedge is of arity $\sqrt{n}$, and has exactly one vertex in each $V_i$). In this example, the low strength (with strength $O(1)$) hyperedges will be exactly those crossing between $V_1, \dots V_{\sqrt{n}}$, and our goal (and indeed requirement) is to recover these $\sqrt{n}$ hyperedges exactly so that we can afford to sample the remaining hypergraph. 

Now, if as before, we attempt to use correlated $\ell_0$-samplers to recover these crossing hyperedges, we very quickly run into issues. In this case, for each component $V_i$, we add together the corresponding $\ell_0$-samplers for the vertices in $V_i$, yielding a sampler for the hyperedges leaving $V_i$. But, because all the $\ell_0$-samplers across the vertices use the same randomness, this means that the $\ell_0$-samplers for the hyperedges leaving the $V_i$'s \emph{are also correlated}. So, when we recover hyperedges from one round of $\ell_0$-samplers all using the same randomness, it will be the case that the $\ell_0$-samplers \emph{all return the same hyperedge} because they have an identical support. 
This is a fundamental issue, as if we wish to recover all $\sqrt{n}$ crossing hyperedges, this will require us to store extra factor of $\sqrt{n}$ $\ell_0$-samplers (and in general, an extra factor of $r$).

Further, using $\ell_0$-samplers with independent randomness \emph{will not solve} this issue. Indeed, if the $\ell_0$-samplers use independent randomness, we cannot add the samplers together to sample from the support of a component $V_i$. Instead, we would be restricted to sampling from the hyperedges leaving each singular vertex, and thus the \emph{vast} majority of hyperedges sampled will be the clique hyperedges within each $V_i$, not the hyperedges crossing between $V_i$'s. Because we do not know the components $V_1, \dots V_{\sqrt{n}}$ beforehand, this is a fundamental shortcoming, and we cannot use uncorrelated random seeds to perform the recovery.

Solving this recovery task with only a linear dependence on $r$ (as we will require to get \cref{thm:main}) therefore requires a new technique, which we introduce in the next section. 

\subsection{Efficient Recovery using Random Fingerprinting}

This leads to one of our key contributions, namely the technique of \emph{random fingerprinting}. Roughly speaking, for each hyperedge in the hypergraph $H$, we independently, randomly subsample the vertices in this hyperedge to create a new hypergraph $H'$, where each hyperedge has smaller arity. Now, on this hypergraph with edges of smaller arity, we can store correlated $\ell_0$-samplers, and use them to recover the crossing hyperedges. For instance, in the above example of $\sqrt{n}$ cliques with $\sqrt{n}$ hyperedges of arity $\sqrt{n}$ intersecting each of these cliques, suppose we \say{fingerprint} each hyperedge randomly at rate $\frac{\log(n)}{\sqrt{n}}$. By this, we mean for every hyperedge $e$ and each vertex $v \in e$, we independently, randomly keep $v$ in the hyperedge $e$ with probability $\frac{\log(n)}{\sqrt{n}}$ (thus after fingerprinting, the expected new size of $e$ is $|e| \frac{\log(n)}{\sqrt{n}}$). Under this operation, any hyperedge crossing between $V_1, \dots V_{\sqrt{n}}$ is now only crossing between a random subset of $\Theta(\log(n))$ of these components with high probability. 

Thus, in this fingerprinted hypergraph we have effectively \emph{broken the correlation} between $\ell_0$-samplers for different components, even when the samplers are initialized with the same random seed. Specifically, for this fingerprinted version of the hypergraph, let us store correlated $\ell_0$-samplers across all the vertices. Then, we can add these samplers together for each component $V_i$, to recover $\ell_0$-samplers for the fingerprinted hyperedges leaving each component $V_i$.
Because it will be very unlikely for the same hyperedge to be crossing between more than $\Theta(\log(n))$ of the components $V_1, \dots V_{\sqrt{n}}$, at most $O(\log(n))$ samplers can return the same fingerprinted hyperedge. One can then verify that in the first round of opening samplers, we expect to recovery $\Omega(\sqrt{n} / \log(n))$ of the crossing hyperedges in this example, which is a significant improvement.

As stated however, the hypergraph we are dealing with has been heavily idealized. In general hypergraphs, the crossing hyperedges may be of different arities (i.e. not all of the same arity $\sqrt{n}$), and further the hyperedges may be non-uniform with respect to the number of vertices they have in each of the components they touch (i.e., in this example each of the crossing hyperedges had exactly $1$ vertex in each component $V_i$). As a consequence, for any crossing hyperedge, it is not immediately clear what the fingerprinting rate should be in order to recover such a hyperedge with high probability. 

Intuitively, we address this by fingerprinting at $\log(n)$ different rates, and show that with high probability, one of these sampling rates will suffice for recovering the crossing hyperedges. The rest of the analysis is rather subtle, so we leave the complete description to \cref{sec:fingerprinting}.

To argue that this procedure indeed recovers sufficiently many distinct hyperedges, we introduce the notion of a \say{unique representative} for any recovered hyperedge. Simply put, for any hyperedge we recover when opening $\ell_0$-samplers, we assign it to a specific component that it is incident upon. This ensures that even if a hyperedge is of large arity and therefore incident on many components, we only count it as a single recovered hyperedge. 
With our fingerprinting technique, and this notion of a unique representative, we are able to prove the following claim, which turns out to be a key building block towards recovering low-strength hyperedges: 
\begin{claim}[Recovery Procedure]\label{clm:recoveryIntro}
    For a parameter $\phi$ of our choosing, with only $\widetilde{O}(\phi \polylog(n))$ $\ell_0$-samplers per vertex (initialized at varying levels of fingerprinting), for any disjoint partition of components $V_1, \dots V_k$, one can recover with high probability for each component $V_i$ either
    \begin{enumerate}
        \item All of the hyperedges leaving $V_i$.
        \item $\phi \log(n)$ hyperedges leaving $V_i$ for which $V_i$ is the \emph{unique representative}.
    \end{enumerate}
\end{claim}

However, this claim on its own is not enough to recover all low strength hyperedges. In particular, if we knew which components $V_1, \dots V_k$ were \say{high-strength} components, we would be able to use the above procedure to recover the low-strength hyperedges crossing between these components. However, for an arbitrary hypergraph, these components will not be known a priori.
With this, in the next subsection we show how to use this procedure to actually compute a strength decomposition and thus complete our sparsification procedure. 

\subsection{Strength Decomposition with Linear Sketches}

Recall that in our idealized sparsification algorithm, our goal will be to recover all hyperedges of strength $\leq 2C \log(n) / \eps^2$ in a hypergraph $H$, using only a linear sketch. Going forward, we will let $\phi = 2C \log(n) / \eps^2$. Thus $\phi$ denotes the cut-off such that we wish to recover any hyperedge of strength $\leq \phi$ in $H$. 

In the previous subsection, we showed how to implement the \say{recovery} procedure. Given a hypergraph $H$ and a disjoint partition of components $V_1, \dots V_k$, we showed that there is a linear sketch which recovers for each component $V_i$ either (1) all of the hyperedges leaving the components or (2) recovers $\phi \log(n)$ distinct hyperedges leaving $V_i$ (here, we use distinct to mean that no hyperedge appears twice even with respect to different components). Immediately, this implies that either case (1) happens for half of the components $V_1, \dots V_k$, or case (2) happens for half of the components. Our goal now will be to show that this procedure can be used to recover all of the hyperedges of low strength.

Because we do not know the strong components a priori, we create the following natural iterative algorithm: we initially start with $n$ components, with each vertex in $V$ constituting its own component. In each iteration, we \say{open} a linear sketch for the recovery problem defined above. 
Naturally, each time we open this sketch, it yields many hyperedges, either exhausting (i.e., recovering all of) the incident hyperedges on some components, or yielding many distinct hyperedges. In this second case, we will be forced to \emph{merge} some components together since they may be connected by high strength hyperedges. As such, the set of vertices slowly contracts to give us a set of components. To analyze this more precisely, let us suppose now then that in the current iteration, we are analyzing a set of components $V_1, \dots V_k$.

Intuitively, if we suppose the hyperedges crossing between these components are of low strength, this should necessarily mean there are not too many crossing hyperedges. We will then argue that the recovery procedure is able to recover all of these hyperedges because we fall into the first case of \cref{clm:recoveryIntro}. However, it is possible that some of these components may be \emph{much} more strongly connected than others. Thus, for some components, we should not expect to recover all of their crossing hyperedges, leading to the second case in \cref{clm:recoveryIntro}. When this occurs, we will show that this necessarily means some components should be \emph{merged} together to create a new component of higher strength. We will be guaranteed that any hyperedge contained in this component has strength much larger than $\phi$, and thus we can be sure that we have not missed out on any low strength hyperedges.

Formally, let us consider the components $V_1, \dots V_k$, for which we wish to recover the crossing hyperedges of low strength (initially these components will simply be each individual vertex). When we run our recovery procedure with these components $V_1, \dots V_k$, either the majority of components have all their crossing hyperedges recovered, or the majority of components recover $\phi \log(n)$ distinct crossing hyperedges.

Intuitively, in the first case it is easy to see that we are making progress. If we recover the entire neighborhood of a majority of the components, then we should be able to simply repeat the algorithm $O(\log(n))$ times (correspondingly, store $O(\log(n))$ independent copies of the sketch) before we have recovered the entire hypergraph. At this point, we can perform any computation we want on the hypergraph, including calculating the strengths explicitly. 

The second case is more nuanced and is where we use key properties of the strength of hyperedges. Indeed, if for a majority of the components, we recover $\phi \log(n)$ distinct crossing hyperedges incident on this component, this means that we have recovered at least $\frac{k}{2}\phi \log(n)$ distinct hyperedges total. We show that for any $k$ components in a hypergraph, the number of hyperedges of \emph{small} strength ($< \phi$) crossing between them is likewise small (bounded by $k \phi$), and thus in particular, at least $1/2$ of the edges we recover must have \say{high} strength. High strength here can be chosen to mean strength at least $2 \phi$, as we require in our decomposition. 
By the pigeonhole principle, this means that at least $1/4$ fraction of the components will have an incident hyperedge of high strength in the recovered hypergraph. Because the strengths of hyperedges are only monotonically increasing as one adds hyperedges, the actual strengths of these hyperedges in $H$ can be only larger than they are in the recovered hypergraph. Now, if a hyperedge of high strength is connecting components, this intuitively means that this group of components should be combined together into a single component of high strength. 
Because at least $k/4$ components have a high strength incident hyperedge, when we merge along these hyperedges, we will decrease the number of connected components by at least $k/8$. Essentially, recovering too many hyperedges (as in the second case) gives us a certificate of the fact that some of components we were considering were actually connected together by high strength edges and can therefore be merged together.

Thus, in both cases we are making progress: either we recover the incident hyperedges on many of the components, and thus reduce the problem to recovering the incident hyperedges on a much smaller graph, or we recover many distinct hyperedges which provides proof that certain components in the graph need to be merged together as they have much higher strength. Because in either case the number of connected components under consideration goes down by a constant fraction, we can repeat this a logarithmic number of times after which the algorithm will return a set of high strength connected components, as well as all hyperedges crossing between these components. We are guaranteed that every component is of high strength, and as a result, it must also be the case that all low strength edges are crossing, and thus recovered. 

In summary, starting with a hypergraph $H$, we can simply run the recovery procedure $O(\log(n))$ times, and be ensured that we recover the strong components, as well as all the low-strength hyperedges crossing between them. Because we perform this only $O(\log(n))$ times, the total space usage is just that of $\widetilde{O}(n \phi)$ $\ell_0$-samplers, which immediately yields our desired dependence on $r$ (as each $\ell_0$-sampler has a linear dependence on $r$ due to the universe size).

\subsection{Simple Sparsification Using Strength Decomposition}

Recall the algorithm presented earlier (\cref{alg:introSimple}). Using the linear sketch discussed above for recovering low-strength hyperedges, we can now implement the algorithm as a linear sketch. Indeed, for each of the $\log(m)$ levels of sampling, we store a linear sketch for recovering low-strength hyperedges of the sampled hypergraph (and this yields the $\log(m)$ factor in our sketch size which is unavoidable). In practice, this involves storing $\log(m)$ independent hash functions mapping $E \rightarrow \zo$. A hyperedge $e$ is present in $H_i$ if and only if it has not already been recovered in some $F_j$ for $j < i$ and if the hyperedge $e$ satisfies $\prod_{j = 1}^i h_j(e) = 1$. It is worth highlighting that we use independent randomness for the linear sketches at each level of sampling the hypergraph. This ensures that the randomness used in the $i$th level is independent of the recovered hyperedges in $F_1, \dots F_{i-1}$, and thus we can afford to simply \emph{remove} the hyperedges in $F_1, \dots F_{i-1}$ from the linear sketch stored for $H_i$.

As discussed in \cref{sec:detailedTechBackground}, a naive analysis of the sparsifier returned by the algorithm guarantees only a sparsifier with accuracy $(1 \pm O(\eps \log(m)))$. Thus, it will be necessary to operate with error parameter $(\eps / \log(m))$ to ultimately get a $(1 \pm \eps)$-sparsifier. This contributes an extra factor of $\log^2(m)$ to the size of the linear sketch that we store. Second, in (most) levels of downsampling, the size of the hypergraph we are dealing with could potentially still be $m^{\Omega(1)}$. This requires us to use $\ell_0$-samplers defined for support sizes as large as $m^{\Omega(1)}$, which costs us an additional $\log(m)$ factor. 

We discuss our approach to removing these $\log(m)$ factors in the next subsection.

\subsection{Improving the Error Accumulation}

First, we show how we can choose our error parameter to be $(\eps / \polylog(n/\eps))$ without changing our algorithm. This will immediately improve our space complexity by a factor of $\log^2(m)$. To see why this holds, let us focus our attention on a single cut in the original hypergraph $H$. We will denote this cut by a set of edges $Q\subseteq E$, and we understand this to be the set of crossing hyperedges for some partition. Now, if we look at the hyperedges inside $Q$, we can calculate the strengths of these hyperedges with respect to the hypergraph $H$. We will denote by $\lambda(Q)$ the maximum strength of any hyperedge in $Q$, i.e., 
\[
\lambda(Q) = \max_{e \in Q}{\lambda_e}.
\]

Note that if a cut $Q$ contains a single hyperedge $e$ of strength $\lambda(Q)$, then in fact it must contain many such hyperedges. This is because any hyperedge $e$ of strength $\lambda(Q)$ is part of a component $C \subseteq V$ in the hypergraph of strength $\geq \lambda(Q)$. Because the cut $Q$ is \say{cutting} the hyperedge $e$, it is necessarily the case that $Q$ is also cutting the component $C$ into two or more pieces. Now, by definition, any cut in a component of strength $\geq \lambda(Q)$ \emph{must} be of size at least $\lambda(Q)$. At the same time, we know that the number of hyperedges in $H$ of strength (say) $\leq \lambda(Q) /  n^{10}$ is at most $\lambda(Q) / n^9$ (this fact has been used before with respect to $2$-cut strength, and we show it holds here with respect to $k$-cut definitions of strength). Thus, a $\geq 1 - 1 / n^{9}$ fraction of the cut hyperedges have strength between $\lambda(Q) /  n^{10}$ and $\lambda(Q)$. Intuitively, this means that we should be able to focus only on preserving the weight of these cut hyperedges of high strength, effectively ignoring those of lower strength. When we adopt this perspective, we can then argue that the degradation in error is much better than the naive inductive analysis may have suggested. 

Indeed, for the first $\log(\lambda(Q) /  n^{11})$ levels of downsampling (i.e., up until the point where we are sampling at rate $\frac{n^{11}}{\lambda(Q)}$), it will still be the case that with \emph{extremely} high probability $1 - 2^{-\text{poly}(n)}$, the total degradation in error will still be bounded by $(1 \pm \eps)$. This is because if we look at the induced subgraph of hyperedges with strength $\geq \lambda(Q) /  n^{10}$, we know this contains most of the mass of the cut $Q$. Further, because the strength of this hypergraph is at least $\lambda(Q) /  n^{10}$, by the cut-counting bound we can afford to sample at any rate $\geq \frac{\log(n) n^{10}}{\lambda(Q) \eps^2}$, while still preserving cuts with high probability.

Beyond this level of downsampling, we lose our guarantee on the rate at which our approximation deteriorates beyond simply the naive factor $(1 \pm \eps)$ per level of downsampling. However, we can now take advantage of the fact that (with high probability), there are only $O(\log(n))$ more levels of downsampling before the cut $Q$ has been entirely removed (i.e., as the low-strength hyperedges removed in each iteration). That is to say, by the time we are sampling at rate $\frac{1}{\lambda(Q) n^{10}}$, all the hyperedges from $Q$ will already have been removed. Thus, there is only a window of size $\text{poly}(n)$ (and thus $O(\log(n)$ levels of downsampling) where we must pay for the degradation in our approximation parameter. Performing this analysis carefully then allows us to remove the superfluous dependence on $\log(m)$ and replace it with only a dependence on $\log(n)$, as we desire. We present this argument more precisely in \cref{sec:cutPerspective}.

\subsection{Preprocessing to Bound Hypergraph Sizes}

Our final improvement in the space for our hypergraph linear sketch will be in optimizing the space each $\ell_0$-sampler requires. Recall that there are three contributing factors to the size of an $\ell_0$-sampler: the universe size (essentially $n^r$, where $r$ is the maximum arity and which we can't hope to optimize), the support size (i.e., the number of hyperedges in the support of each sampler), and the error parameter (which yields only a multiplicative $\log(n)$). Because the universe size cannot be optimized, and the error parameter is already sufficiently small, naturally our goal will be to decrease the support size of the samplers. First, let us recall specifically where the $\log(m)$ is coming from: at each level of downsampling, we will be storing $\ell_0$-samplers for the neighborhoods of vertices. In these downsampled hypergraphs, there may still be as many as $m^{\Omega(1)}$ hyperedges, and thus there may exist some vertices whose degree is also $m^{\Omega(1)}$. Even to recover a single hyperedge then, our $\ell_0$-samplers must be initialized to work on a support size up to $m^{\Omega(1)}$. 

The key insight is that if there are too many hyperedges in the hypergraph, then intuitively this means that there must be some (very) strongly connected components. If we could somehow find these (very) strongly connected components \emph{before} starting to look for our low-strength hyperedges, then we could show that in this meta-graph (where we merge each strongly connected component into a single meta-vertex), the number of crossing hyperedges is bounded by $\poly(n)$. This would then allow us to use $\ell_0$-samplers defined for a smaller support size and thus use only a factor of $\log(n)$ instead of $\log(m)$. Further, if we could guarantee that these components that we merge together are sufficiently strongly connected, then we can also guarantee that there are no low-strength hyperedges which have been lost throughout this procedure, and therefore the recovery procedure on this meta-hypergraph recovers \emph{exactly} the same hyperedges as in the original hypergraph.

Our final contribution is to show that indeed, we can store a separate linear sketch of the hypergraph which we can analyze \emph{before} our sparsification algorithm (a preprocessing phase), and will reveal to us the (exceedingly) strongly connected components in our hypergraph in each iteration. We show that with some careful scheming, the preprocessing linear sketch can be made to use only space $\widetilde{O}(n r \log(m))$, and thus (after saving the final $\log(m)$ term in the sparsifier) our entire linear sketch also only requires space $\widetilde{O}(n r \log(m) / \eps^2)$. This argument is presented in its entirety in \cref{sec:preprocessing}.

\subsection{Lower Bound}

In addition to our upper bound of $\widetilde{O}(n r \log(m) / \eps^2)$ bits for our linear sketch, we also present a lower bound for the size of any sketch which returns a $(1 \pm \eps)$-cut sparsifier, even in the regime where $\eps = \Omega(1)$. To do this, we build a parameterized version of the universal relation problem, that we refer to as the $k$-$\UR_r^{\leq m}$ problem:

\begin{enumerate}
    \item Alice is given a string $x_A \in \zo^{2^r}$. Bob is given a string $x_B \in \zo^{2^r}$ such that $m \geq |\Supp(x_A) - \Supp(x_B)| \geq k$. Alice sends only a message $\calS(x_A)$ to Bob (using public randomness).
    \item Bob has his own string $x_B$ (satisfying the above promises), and receives Alice's message $\calS(x_A)$. Using this (and access to public randomness), he must return $k$ indices $i: (x_A)_i \neq (x_B)_i$ with probability $1 - 1 / r^{5}$.
\end{enumerate}

We show that using $r / (\log(m / k))$ instances of $k$-$\UR_r^{\leq m}$, one can solve the more general problem known as $k$-$\UR_r$, which has a known one-way communication complexity of $\Omega(kr^2)$ (established by the work of \cite{KNPWWY17}). This immediately implies that the one-way communication complexity of $k$-$\UR_r^{\leq m}$ is $\Omega(kr\log(m/k))$.

Finally, to conclude our lower bound, we show given an instance of $n/2$-$\UR_{r/2}^{\leq m}$, Alice can construct a specific type of \say{bipartite} hypergraph with $\leq m$ hyperedges each of arity $\leq r$, such that sending $\log(n)$ independent hypergraph $(1 \pm \eps)$-sparsifier linear sketches (for $\eps < 1$), Bob can recover a solution to the same $n/2$-$\UR_{r/2}^{\leq m}$ instance with all but polynomially small probability. Because of our lower bound on the communication complexity of $n/2$-$\UR_{r/2}^{\leq m}$, this immediately implies an $\Omega(nr \log(m/n) / \log(n))$ lower bound on the size of any linear sketch for hypergraph sparsification. We present this proof in \cref{sec:lowerBound}.

\section{Preliminaries}\label{sec:preliminaries}

\subsection{$\ell_0$-samplers and Vertex Incidence Sketches}

First, we introduce the notion of an $\ell_0$-sampler.

\begin{definition}
    Consider a turnstile stream $S = s_1, \dots s_t$, where each $s_i = (u_i, \Delta_i)$, and the aggregate vector $x \in \R^u$ where $x_i = \sum_{j:u_j = i} \Delta_i$.
    
    Given a target failure probability $\delta$, an $\ell_0$-sampler for a non-zero vector $x$ returns $\perp$ with probability $\leq \delta$, and otherwise returns an element $i \in [n]$ with probability $\frac{|x_i|_0}{|x|_0}$.
\end{definition}

\begin{fact}\cite{CF14}
    We will use the fact that (for any universe of size $u$, and support of size $m$) there exists a \emph{linear} sketch-based $\delta$-$\ell_0$-sampler using space $O(\log(m)\log(u) \log (1 / \delta))$. Note that $u$ is the length of the aggregate vector $x$ from the previous definition. $m$ is an upper bound on $|x|_0$.
\end{fact}

We present a self-contained proof of the existence of such $\ell_0$-samplers in \cref{sec:appendixSampler}.

Going forward, for a vector $x$ and (public) randomness $R$, we will let $\calS(x, R)$ denote an $\ell_0$-sampler for $x$ using the randomness $R$.

\begin{definition}\cite{AGM12, GMT15}\label{def:hyperedgeEncoding}
    Given an unweighted hypergraph $G = (V, E)$, define the $n \times 2^{[n]}$ matrix $A_G$ with entries $(i, e)$, where $i \in [n]$ and $e \subseteq [n]$. We say that 
    \[
    A_{i,e} = \begin{cases}
        1 & \text{ if } i \in e, i \neq \max_{j \in e} j, \\
        -(|e|-1) & \text{ if } i \in e, i = \max_{j \in e} j, \\
        0 & \text{ else.}
    \end{cases}
    \]

    Let $a_1, \dots a_n$ be the rows of the matrix $A$. The support of $a_i$ corresponds with the neighborhood of the $i$th vertex.
\end{definition}

Next, we will use the following result regarding adding together $\ell_0$-samplers that use shared randomness. This property of $\ell_0$-samplers has appeared in many different papers \cite{AGM12, GMT15, CKL22}.
\begin{lemma}\label{lem:sampleCC}
    Suppose we have $\ell_0$-samplers for the neighborhoods of all vertices in a connected component $V_i$, denoted by $\calS(a_v, R): v \in V_i$, and that these samplers share their randomness. Then, $\sum_{v \in V_i} \calS(a_v, R)$ is an $\ell_0$-sampler for the hyperedges leaving $V_i$.
\end{lemma}

\begin{remark}\label{rmk:removeEdge}
    Suppose we have a linear sketch for the $\ell_0$-sampler of the edges leaving some connected component $V_i$, denoted by $\calS(V_i)$. Suppose further that we know there is some edge $e$ leaving $V_i$ that we wish to remove from the support of $\calS(V_i)$. Then, we can simply add a linear vector update to $\calS(V_i)$ that cancels out the coordinate corresponding to this edge $e$. 

    Given a linear sketch of a hypergraph $H$, and some set of hyperedges $S$ in $H$, we will often use $H - S$ to denote the result of updating the linear sketch to remove these hyperedges.
\end{remark}

Finally, we use the following probabilistic bound which underlies many sparsification algorithms:

\begin{claim}\label{clm:concentrationBound}{\rm (\cite{FHH11})}
    Let $X_1, \dots X_{\ell}$ be random variables such that $X_i$ takes on value $1 / p_i$ with probability $p_i$, and is $0$ otherwise. Also, suppose that $\min_i p_i \geq p$. Then, with probability at least $1 - 2e^{-0.38 \eps^2 \ell p}$,
    \[
    \sum_i X_i \in (1 \pm \eps) \ell.
    \]
\end{claim}

\subsection{Strength in Hypergraphs}

In this section, we introduce some definitions of $k$-cut strength, and show that it behaves intuitively, with many convenient closure properties. These properties will be used frequently in the rest of the paper as we create sketches for recovering low-strength hyperedges.

First, we recall the definition of strength that we use \cite{Qua23}.
\begin{definition}
    For a hypergraph $H = (V, E)$, the \emph{minimum normalized $k$-cut} is defined to be 
    \[
    \min_{k \in [n]} \min_{V_1 \cup V_2 \cup \dots \cup V_k = V} \frac{|E[V_1, \dots V_k]|}{k-1}.
    \]
    $|E[V_1, \dots V_k]|$ refers to the number of edges which cross between (any subset) of $V_1, \dots V_k$. This is a generalization of the notion of a $2$-cut in a graph, which is traditionally used to create cut sparsifiers in ordinary graphs. Further, note that $V_1, \dots V_k$ form a partition of $V$. As mentioned in the introduction, we will often use the following to denote the minimum normalized $k$-cut:
    \[
    \minkcut(H) = \min_{k \in [n]} \min_{V_1, \cup  \dots \cup V_k = V} \frac{|E[V_1, \dots V_k]|}{k-1}.
    \]
    We also refer later to \emph{un-normalized $k$-cuts}, which is simply $|E[V_1, \dots V_k]|$, for some partition $V_1, \dots V_k$ of $V$.
\end{definition}

Now, to define strength, we iteratively use the notion of the minimum $k$-cut.
\begin{definition}
    Given a hypergraph $H = (V, E)$, let $\minkcut(H)$ be the value of the minimum normalized $k$-cut, and let $V_1, \dots V_k$ be the components achieving this minimum. For every edge $e \in E[V_1, \dots V_k]$, we say that $\lambda_e = \minkcut(H)$. Now, note that every remaining edge is contained entirely in one of $V_1, \dots V_k$. For these remaining edges, we define their strength to be the strength inside of their respective component.
\end{definition}

\begin{remark}\label{rmk:monotone}
    Note that the strengths assigned via the preceding definition are non-decreasing. Indeed if the minimum normalized $k$-cut has value $\phi$ and splits a graph into components $V_1, \dots V_k$, it must be the case that the minimum normalized $k$-cuts in each $H[V_i]$ are $\geq \phi$, as otherwise one could create an even smaller original normalized $k$-cut by further splitting the component $V_i$.
\end{remark}

We will also refer to the strength of a component.

\begin{definition}
    For a subset of vertices $ S \subseteq V$, we say that the \emph{strength of $S$ in $H$} is $\lambda_S = \min_{e \in H[S]} \lambda_e$. That is, when we look at the induced subgraph from looking at $S$, $\lambda_S$ is the minimum strength of any edge in this induced subgraph. 
\end{definition}

\begin{definition}\label{def:contractedHypergraph}
    For a hypergraph $H$ and partition $V_1, \dots V_k$ of the vertex set, let $H / (V_1, \dots V_k)$ denote the hypergraph obtained by contracting all vertices in each $V_i$ to a single vertex. For a hyperedge $e \in H$, we say that the corresponding version of $e \in H / (V_1, \dots V_k)$ (denoted by $e / (V_1, \dots V_k)$) is incident on a super-vertex corresponding to $V_i$ if there exists $v \in V_i$ such that $v \in e$.
\end{definition}

We will take advantage of the following fact when working with these \say{contracted} versions of hypergraphs:

\begin{claim}\label{clm:minkcutcontracted}
    Let $H$ be a hypergraph, and let $V_1, \dots V_k$ be a set of connected components of strength $> \kappa$. Then, the hyperedges of strength $\leq \kappa$ in $H$ are exactly those hyperedges of strength $\leq \kappa$ in $H/(V_1, \dots V_k)$.
\end{claim}

\begin{proof}
It is clear to see that if a hyperedge $e \in H$ is completely contained in some component $V_i$, then $e$ will correspond to a self-loop in the graph $H / (V_1, \dots V_k)$. Thus, the crossing edges in $E_H[V_1, \dots V_k]$ will make up the entirety of $H / (V_1, \dots V_k)$ up to self-loops.

Now, we claim that for any edge $e \in E_H[V_1, \dots V_k]$, the strength of $e \in H$ is $\leq \kappa$ if and only if the strength of $e/(V_1, \dots V_k) \in H / (V_1, \dots V_k)$ is $\leq \kappa$. Further, if the strength of $e/(V_1, \dots V_k) \in H / (V_1, \dots V_k)$ is $\leq \kappa$, then the strength is \emph{exactly} equal to the strength of $e \in H$. It follows then that this yields an algorithm for finding all of the edge of strength $\leq \kappa$ in $H$. We simply look at the contracted graph $H / (V_1, \dots V_k)$, find all edges $e/(V_1, \dots V_k)$ of strength $\leq \kappa$, and we will know the corresponding strength in $H$. Note that by definition, any self-loop edge in $H$ has strength $> \kappa$ because the components $V_i$ have strength $> \kappa$.

We will first show that the minimum normalized $k$-cut in $H$ obtains the same value as the minimum normalized $k$-cut in $H / (V_1, \dots V_k)$. Indeed, consider the minimum normalized $k$-cut in $H$ and suppose it has value $\phi < \kappa$ and components $V'_1, \dots V'_{k'}$. Because the components $V_1, \dots V_k$ each have strength $\kappa$, it must be the case that the partition $V'_1, \dots V'_{k'}$ does not split any component $V_i$, as otherwise this would mean some edge $e \in V_i$ is assigned strength $\phi < \kappa$ which is a contradiction. Thus, the partition $V'_1, \dots V'_{k'}$ also forms a valid partition of $V_1, \dots, V_k$ in the sense that each $V_i$ is contained in exactly one $V'_j$. Thus, we can interpret $V'_1, \dots V'_{k'}$ to be a partition of the contracted super vertices in the canonical manner. We write this as $V'_1 / (V_1, \dots V_k), \dots V'_{k'} / (V_1, \dots V_k)$. Next, the crossing edges $E_H[V'_1, \dots V'_{k'}]$ will be in exact correspondence with the crossing edges $E_{H / (V_1, \dots V_k)}[V'_1 / (V_1, \dots V_k), \dots V'_{k'} / (V_1, \dots V_k)]$ because an edge which is crossing from $V'_i, V'_j$ is only crossing if it also crosses between $V'_{i} / (V_1, \dots V_k), V'_{j} / (V_1, \dots V_k)$. Thus, the cut corresponding to $E_{H / (V_1, \dots V_k)}[V'_1 / (V_1, \dots V_k), \dots V'_{k'} / (V_1, \dots V_k)]$ will have normalized value $\phi$ in $H / (V_1, \dots V_k)$. Further, for any minimum normalized cut in $H / (V_1, \dots V_k)$, the corresponding partition of $V_1, \dots V_k$ that it makes will also be a valid $k$-partition of $H$. Thus, we have shown that the minimum normalized $k$-cut in $H / (V_1, \dots V_k)$ is both $\geq$ and $\leq$ the minimum normalized $k$-cut of $H$.

As pointed out in the above paragraph, as long as the value of the minimum normalized $k$-cut is $\leq \kappa$, the edges involved in any minimum normalized $k$-cut in $H / (V_1, \dots V_k)$ are in an exact bijection with $H$. Thus, the strength for edges in $E_{H / (V_1, \dots V_k)}[V'_1 / (V_1, \dots V_k), \dots V'_{k'} / (V_1, \dots V_k)]$ will be exactly the same as $E_H[V_1, \dots V_k]$, and can be calculated directly from $H / (V_1, \dots V_k)$. We can then inductively apply this to the components $V'_1, \dots V'_{k'}$ that result from removing the crossing edges. This means that as long as the strength of the hypergraph we are operating on is $\leq \kappa$, we will correctly assign strength values to the hyperedges involved in the minimum normalized $k$-cut. This means that all hyperedges with strength $\leq \kappa$ will have their strengths correctly calculated, as we desire. 

Note that the claim follows because calculating strengths in $H / (V_1, \dots V_k)$ can be done exactly. This uses the fact that self-loops do not play a role in cut-sizes, so our lack of knowledge of the edges in each $V_i$ does not impact our calculations. 
\end{proof}

There is also the following equivalence between the strengths of hyperedges and induced subgraphs:

\begin{claim}
    For a hypergraph $H = (V, E)$ and a hyperedge $e \in E$, 
    \[
    \lambda_e \leq \max_{e \subseteq S \subseteq V} \minkcut(H[S]).
    \]
\end{claim}

\begin{proof}
    This follows because when we calculate the strength decomposition, we iteratively find the minimum $k$-cut of induced subgraphs. The first time that $e$ is a \say{crossing edge}, i.e., not completely contained in one component is when $e$ has its strength assigned. This means that the strength of $e$ is ultimately assigned to be the value of the minimum normalized $k$-cut of some induced subgraph that contains $e$. In the above proposition, we consider the \emph{maximum} over such induced subgraphs.
\end{proof}

\begin{claim}
    For a hypergraph $H = (V, E)$ and a hyperedge $e \in E$, 
    \[
    \lambda_e \geq \max_{e \subseteq S \subseteq V} \minkcut(H[S]).
    \]
\end{claim}

\begin{proof}
    We will show that $\lambda_e \geq \max_{e \subseteq S \subseteq V} \minkcut(H[S])$. To do this, let $\hat{S}$ denote the optimizing subset for the above expression. Let us suppose for the sake of contradiction that $\lambda_e < \minkcut(H[\hat{S}]) = \phi$. There are three cases: 
    \begin{enumerate}
        \item One case is that in the strength calculation, when $\lambda_e$ was assigned, $e$ was a crossing edge for some partition of an induced subgraph $H[S]$, for $\hat{S} \subset S$. If this is the case, we want to argue that there is in fact a smaller normalized $k$-cut that one can create in $H[S]$ for which $e$ is not a crossing edge. Indeed, let the optimal min $k$-cut be given by the partition $V_1, \dots V_{k}$. Note that by assumption, $e \in E[V_1, \dots V_{k}]$ and $\gamma = \frac{|E[V_1, \dots V_{k}]|}{k'-1} < \phi$. Now, because $e$ is a crossing edge, it must be the case that the partition $V_1, \dots V_{k}$ splits $\hat{S}$ (as $e \subseteq \hat{S}$ would otherwise not be a crossing edge). Now, we claim that this means that $V_1, \dots V_{k}$ is actually not the minimal normalized $k$-cut. Indeed, consider $W_1 = \{i: V_i \cap \hat{S} \neq \emptyset \}$ which is the set of connected components which intersect $\hat{S}$. WLOG, let us assume there are $\ell$ such components and that they are the first $\ell$ in our list (note that $\ell < k$ as otherwise there would be more than $\gamma(k-1)$ edges being cut). Now, consider the new partition defined with the connected components $W = \bigcup_{i \in [\ell] }V_i, V_{\ell+1}, \dots V_{k}$. In words, we are simply merging all the connected components which split $\hat{S}$, and leaving the other connected components un-touched. Let us calculate the new value of this cut: we will have $k - \ell + 1$ connected components, and the number of crossing edges will be $|E[W, V_{\ell+1}, \dots V_{k}]| \leq |E[V_1, \dots V_{k}]| - \phi(\ell - 1)$ because we have removed all the edges in $\hat{S}$ that were cut in this partition. Thus, the value of this normalized $k$-cut will be 
        \[
        \leq \frac{|E[V_1, \dots V_{k}]| - \phi(\ell - 1)}{k - \ell} = \frac{\gamma(k-1) - \phi(\ell - 1)}{k-\ell} < \frac{\gamma(k - \ell)}{k - \ell} < \gamma,
        \]
        which is thus smaller than the original $k$-cut defined by $V_1, \dots V_{k}$ and yields a contradiction.
        \item Another case is that in the strength calculation, when $\lambda_e$ was assigned, $e$ was a crossing edge for some partition of an induced subgraph $H[S]$ for $e \subseteq S \subset \hat{S}$. This means that at some point in the strength calculation, there was a partition into components $V_1, \dots V_{k}$ such that $\hat{S}$ was split into different parts. Further, note that by \cref{rmk:monotone} it must be the case that the minimum normalized $k$-cut defined by $V_1, \dots V_{k}$ must be $\leq \lambda_e < \phi$ because $e$ has not yet had its strength assigned. However, now we can again invoke the logic from the previous point. This means that $\hat{S}$ was split into different parts by the partition $V_1, \dots V_{k}$, which achieves value $< \phi$, despite the fact that every $k$-cut of $\hat{S}$ is of size $\geq \phi$. Thus, we can merge all the parts of the partition that separate $\hat{S}$ to get a $k$-cut of smaller size. This will contradict the fact that $V_1, \dots V_{k}$ was the minimum $k$-cut. 
        \item The final case is that $\lambda_e$ is assigned when $e$ is a crossing edge of the induced subgraph $H[\hat{S}]$. Then, the strength will be exactly the minimum $k$-cut of $H[\hat{S}]$, as we desire.
    \end{enumerate}
    Thus, we have shown that in every case, it must be that $\lambda_e \geq \minkcut = \phi$.
\end{proof}

\begin{corollary}\label{cor:strengthCharacterization}
For a hypergraph $H = (V, E)$ and a hyperedge $e \in E$, 
    \[
    \lambda_e = \max_{e \subseteq S \subseteq V} \minkcut(H[S]).
    \]
\end{corollary}

A simple consequence of the above is that adding more hyperedges to a graph can only increase (or keep the same) the strengths of existing hyperedges, a fact that we will use throughout the paper. 

We now prove some basic facts about this strength decomposition. 

\begin{claim}\label{clm:strengthNumber}
    In an unweighted hypergraph with $n$ vertices, the number of hyperedges with $\lambda_e \leq w$ is at most $(n-1)\cdot w$.
\end{claim}

\begin{proof}
    Suppose the claim is true by induction for hypergraphs with $n' < n$ vertices. We will show it is true for hypergraphs on $n$ vertices. The base case follows trivially when $n =1$. Indeed, consider a hypergraph $H$ with $n$ vertices, and consider the minimum $k$-cut in $H$ with value $\phi'$ that splits $H$ into $k'$ components. If $\phi' \leq w$, this means that we will get $(k'-1)\cdot \phi' \leq (k'-1)\cdot w$ hyperedges assigned strength $\lambda_e \leq w$, before splitting $H$ into $k'$ connected components. Now, by induction, the maximum number of hyperedges with strength $\leq w$ contained in these $k'$ connected components is $\leq \sum_{V_i \in \{ V_1, \dots V_{k'} \}} (|V_i|-1) \cdot w \leq (n - k') \cdot w$. Adding together the hyperedges crossing the cuts, we get that the total number of potential hyperedges with strength $\leq w$ is at most $(n - k') \cdot w + (k'-1) \cdot w = (n-1) w$, as we desire.
\end{proof}

\begin{claim}\label{clm:removeLowStrength}
    Let $H$ be an unweighted hypergraph on $n$ vertices, and let $\lambda \in \R$. Let $E_{< \lambda} = \{e \in E: \lambda_e < \lambda\}$ be all hyperedges of strength $< \lambda$ in $H$. Then, in the hypergraph $H - E_{< \lambda}$, every hyperedge has strength $\geq \lambda$.
\end{claim}

\begin{proof}
    It suffices to show that if a hyperedge $e$ has strength $\geq \lambda$ in $H$, then the same hyperedge has strength $\geq \lambda$ in $H - E_{< \lambda}$.
    
    So, consider any such hyperedge $e \in H$. Recall from \cref{cor:strengthCharacterization} that we can characterize its strength in $H$ with
    \[
    \lambda_e = \max_{e \subseteq S \subseteq V} \minkcut(H[S]).
    \]
    In particular, since $e$ has strength $\geq \lambda$ in $H$, there must exist an $S \subseteq V$ for which $e \subseteq S$ and $\minkcut(H[S]) \geq \lambda$. However, this means that for every other hyperedge $e' \in H$ such that $e' \subseteq S$, it must be the case that 
    \[
    \lambda_{e'} = \max_{e' \subseteq S' \subseteq V} \minkcut(H[S']) \geq \minkcut(H[S]) \geq \lambda.
    \]

    So, every hyperedge contained in $H[S]$ has strength $\geq \lambda$ and therefore every hyperedge in $H[S]$ remains in the graph $H - E_{< \lambda}$, as none of them are in the set $E_{< \lambda}$. So, the induced sub-hypergraphs $H[S]$ and $(H - E_{< \lambda})[S]$ are the same. This means that the strength of the hyperedge $e$ is still at least $\lambda$ in $H - E_{< \lambda}$ because the strength of $e$ in $H - E_{< \lambda}$ (denoted by $\gamma_e$) satisfies
    \[
    \gamma_e = \max_{e \subseteq S'' \subseteq V} \minkcut((H - E_{< \lambda})[S''])\geq \minkcut((H - E_{< \lambda})[S]) = \minkcut(H[S]) \geq \lambda.
    \]
    
\end{proof}

\begin{remark}
    An immediate consequence of \cref{clm:removeLowStrength} is that if one removes all hyperedges of strength $< \lambda$, the resulting hypergraph has a minimum normalized $k$-cut of size $\geq \lambda$. If this were not the case, then there would exist hyperedges of strength $< \lambda$, which contradicts the above.
\end{remark}

\begin{claim}\label{clm:unionStrength}
    If a set of connected components $V_1, \dots V_r$ in $H = (V, E)$ all have strength $\geq \lambda$ and are connected by a hyperedge whose strength in the overall graph is $\geq \lambda$, it follows that the connected component $\bigcup_{i \in [r]} V_i$ has strength $\geq \lambda$ as well. 
\end{claim}

\begin{proof}
    Suppose for the sake of contradiction that $S = \bigcup_{i \in [r]} V_i$ has strength $< \lambda$. This implies that there is a hyperedge $e$ in $H[S]$ whose strength is $< \lambda$ in $H$. Now, let us consider the procedure by which strength is assigned. We start by finding the minimum normalized $k$-cut value $\minkcut$ in $H$, assign all edges participating in the $k$-cut strength $\minkcut$, and recurse on the connected components left once we remove all these edges that crossed the cut. This procedure thus yields strengths of increasing amounts (see \cref{rmk:monotone}). Thus, in order for an edge in $H[S]$ to be assigned strength $< \lambda$, it must have been the case that $e$ was a crossing edge in some minimum $k$-cut of an induced subgraph and that the value of this $k$-cut was $< \lambda$. Note that because the assigned strengths \emph{increase}, this means that the component $S$ is split apart in some $k$-cut of value $< \lambda$; the cut which splits $e$ also splits $S$, but is certainly possible that $S$ is split apart earlier too, but this again means the $k$-cut splitting $S$ must have had value $< \lambda$ by \cref{rmk:monotone}.

    To summarize, this means that there was some set $S \subseteq A \subseteq V$ such that the minimum $k$-cut in $H[A]$ attained value $< \lambda$, and that the components in this minimum $k$-cut split $S$ apart. Let us denote the components in this minimum $k$-cut by $V'_1, \dots V'_{k'}$. In particular, it must have been the case that $S \cap V'_i \neq S$, i.e., that $S$ must have been split into separate components, as otherwise $S$ would not have separated by this cut. Now, however, we run into a contradiction. Note that since $S$ is split into separate non-empty components, it must either be the case that some $V_i$ is split by $V'_1, \dots V'_{k'}$, or that some of the components $V_1, \dots, V_r$ are separated from one another by $V'_1, \dots V'_{k'}$. We make this more formal below:
    \begin{enumerate}
        \item Suppose that for some $i \in [r]$, it is the case that $ \forall j \in [k'], V_i \cap V'_j \neq V_i$. This means that the partition $V'_1, \dots V'_{k'}$ splits one of our original connected components into at least 2 separate non-empty pieces. We denote these pieces by $V_i \cap V'_j$ for $j \in [k']$. Now, because $V_i$ is connected, this implies that there is an edge in $V_i$ which crosses between at least two of these pieces (as they form a partition). This however is a contradiction, as this would imply that this edge is in $E[V'_1, \dots V'_{k'}]$, and therefore would have been assigned strength $< \lambda$. But, we are told all edges in $H[V_i]$ have strength $\geq \lambda$.
        \item Suppose that it is not the case that $\exists i \in [r]: \forall j \in [k'], V_i \cap V'_j \neq V_i$. This implies that it is not the individual $V_i$ which are split by the partition $V'_1, \dots V'_{k'}$, but rather that the split happens between some of the $V_i$. However, by our hypothesis, we assume that $V_1, \dots V_r$ are connected by an edge $\hat{e}$ of strength $\geq \lambda$. Thus, in this case $\hat{e} \in E[V'_1, \dots V'_{k'}]$, which again yields a contradiction, as this would imply that $\hat{e}$ would have been assigned a strength $< \lambda$.
    \end{enumerate}
    Thus, in either case we reach a contradiction. So, it must be the case that the component $S$ has strength $\geq \lambda$.
\end{proof}

\begin{corollary}
    Suppose two connected components $V_1, V_2$ both have strength $>\lambda$ and share a common vertex. Then, $V_1 \cup V_2$ has strength $> \lambda$ as well. 
\end{corollary}

\begin{proof}
    Let the common vertex be $u$, and consider an edge in $V_1$ which neighbors on $u$ (guaranteed to exist because $V_1$ is connected). This edge will have strength $> \lambda$, because $V_1$ has strength $> \lambda$. But, because $u \in V_2$, this means that $V_1$ and $V_2$ are connected by an edge of strength $> \lambda$, so we can invoke the preceding claim. 
\end{proof}

Next, we mention some facts that have been previously proved about these values $\lambda_e$.

\begin{claim}\cite{Qua23}\label{clm:subsamplingGood}
    If one samples each edge $e$ of a hypergraph $H = (V, E)$ at rate $p_e \geq C \frac{\log(n)}{\lambda_e \eps^2}$ for $n \geq |V|$, and with corresponding weight $1/p_e$, then the size of all $k$-cuts in $H$ are preserved to a $(1 \pm \eps)$-factor with probability $\geq 1 - (|V|-1) n^{-100}$.
\end{claim}

This result relies on the following counting bound from \cite{Qua23} along with a Chernoff bound. 

\begin{theorem}\cite{Qua23}
    Let $H$ be a hypergraph, then, the number of \emph{un-normalized} $k$-cuts of size $\leq t \cdot \minkcut(H)$ is at most $n^{2t}$.
\end{theorem}

\begin{remark}
    A consequence of the above theorem is that if one samples the hyperedges of a hypergraph at rate $\frac{n^c}{ \eps^2 \phi}$, then all cuts are preserved to factor $(1 \pm \eps)$ with probability $\geq 1 - 2^{-n^{c- o(1)}}$, simply by taking a union bound over each cut. 
\end{remark}

\section{Linear Sketching Sparsifiers}\label{sec:sparsifiers}

\subsection{Linear Sketching a Strength Decomposition Algorithm and Analysis}\label{sec:strengthDecomp}

In this section, we will present an algorithm which stores only linear sketches of the neighborhoods of vertices, yet allows us to decompose a graph $H$ into connected components of high strength and return all the edges crossing between these connected components of high strength. 

We will make use of the following notion:

\begin{definition}\label{def:uniqueRep}
    For a hypergraph $H = (V,E)$, a set of components $V_1, \dots V_k$, and a hyperedge $e$ crossing between components $\{V_i: i \in T\}$ ($T \subseteq [k]$), we can arbitrarily assign a component $V_j: j \in T$ to be the \textbf{unique representative} component for $e$, so long as $e$ is a crossing hyperedge incident upon $V_j$, and $V_j$ is the only component assigned to $e$.
\end{definition}

Throughout this section, we will make use of the following theorem, which we prove in \cref{sec:fingerprinting}, and is one of our main technical contributions:

\begin{theorem}\label{thm:recoveryProblem}[Recovery Algorithm]
    For a hypergraph $H$ on $n$ vertices and a parameter $\phi$, there exists a linear sketch (parameterized by the edge set $H$, parameter $\phi$) storing only $\widetilde{O}(\phi \polylog(n))$ $\ell_0$-samplers for suitably restricted neighborhoods of each vertex, such that given any disjoint components $V_1, \dots V_k$, with probability $1 - 2^{-\Omega(\log^2(n))}$ returns a set of edges $S$ such that for each $V_i$ either:
    \begin{enumerate}
        \item All of the hyperedges incident on $V_i$.
        \item At least $\phi \log(n)$ incident hyperedges to $V_i$ for which $V_i$ is assigned as the unique representative.
    \end{enumerate}
Additionally, for each component $V_i$, the algorithm indicates whether the component is in case 1 or case 2. 
\end{theorem}

The unique representative assignment in the second case of the theorem above is to rule out the possibility that multiple components simultaneously recover the same hyperedge among the $\phi \log(n)$ hyperedges recovered by each of them. For instance, a large arity hyperedge may be incident upon \emph{all} of the components $V_i$, yet its recovery is allowed to be claimed by only a \emph{single} unique representative component. Thus, the number of distinct hyperedges recovered must scale with the number of components. Finally, it is not required for the algorithm to say which component is the unique representative for each hyperedge, only to guarantee that among the returned hyperedges, there exists an assignment of unique representatives satisfying the above statement.

We call the above sketch a \say{recovery sketch}, and we will denote it by $\text{Recovery}$.
The proof of such a linear sketch will be provided in the next section, as it is fairly involved. Here, we instead show how such a sketch is powerful, as it yields hypergraph sparsifiers.

\subsubsection{Finding Strong Components and Crossing Hyperedges}

Below, we present an algorithm using the recovery sketch to perform a strength decomposition of a hypergraph. Roughly speaking, the intuition is that using the recovery sketch  \begin{enumerate}
    \item Either, for a large fraction of the connected components the algorithm recovers \emph{all} of the incident hyperedges. If this happens, we have objectively made good progress, as we have recovered a very non-trivial fraction of the entire graph.
    \item Otherwise, a large fraction of the connected components have recovered many unique hyperedges. In particular, just by looking at the recovered hyperedges (which is a subset of the actual hypergraph), we will be able to find many hyperedges of high strength. Because this is only a subset of the original hypergraph, we know that the strengths of these hyperedges can only be larger in the original hypergraph. So, we show that we can in fact merge components connected by strong hyperedges, reducing the number of connected components remaining in the graph.
\end{enumerate}
In either case, the algorithm is making progress by decreasing the number of connected components that we still have to consider. So, we start with a set of components just being each of the individual vertices in the hypergraph, and recovery incident hyperedges via the recovery sketch. Some components may recover many incident hyperedges and thus be merged into other components, while others may simply recover their entire neighborhood, after which we consider them exhausted.

We present an algorithm implementing the above logic:

\begin{algorithm}[H]
\caption{StrengthDecompositionRecovery($G, \phi$).}\label{alg:strengthDecomposition}
Initialize the active connected components to be $V^{(1)}_1, \dots V^{(1)}_n$ to be $1, 2, \dots, n$ (one for each vertex). \\
Let $K_1 = n$ denote the current number of active connected components.\\
$S = \emptyset$ (the set of hyperedges recovered so far), $T = \emptyset$ (the final set of components). \\
\For{$i \in [8\log(n)]$}{
Initialize $V / (V_1^{(i)}, \dots V_{K_i}^{(i)})$ to be the vertex set, (i.e, contract the corresponding components to super-vertices).\\
$S_i \leftarrow $Recovery($G - S, \phi \log(n), (V_1^{(i)}, \dots V_{K_i}^{(i)})$).\\
\If{less than $K_i / 2$ of the components $V_j^{(i)}$ have recovered all incident hyperedges}
{
The recovery has returned $\geq K_i \cdot \phi \log^2(n) / 2$ edges incident on $V^{(i)}_1, \dots V^{(i)}_{K_i}$. \\ 
Calculate the strengths of the recovered hyperedges $S_i$ on the vertex set $V / (V_1^{(i)}, \dots V_{K_i}^{(i)})$. Merge any components that are connected by a hyperedge of strength $> 2 \phi \log(n)$ in this meta-graph to create the components $V_j^{(i+1)}$.
}
\Else{
Set $S \leftarrow S \cup S_i$. \\
For any component $V_j^{(i)}$ which has recovered all incident hyperedges, remove $V_j^{(i)}$ from the remaining active components, and add $V_j^{(i)}$ to the set $T$.
}
Let $K_{i+1}$ denote the number of remaining connected components, and let $V_j^{(i+1)}$ for $j \in [K_{i+1}]$ represent the remaining connected components. 
}
\Return{
$S, T$.
}
\end{algorithm}

\begin{claim}
    After $8\log(n)$ iterations in \cref{alg:strengthDecomposition}, the set of active components, $\{V_j^{(8 \log(n))}\}$ is empty. 
\end{claim}

\begin{proof}
    Consider an iteration $i$ of the algorithm in which we start with $K_i$ connected components $V_1, \dots V_{K_{i}}$ under consideration in the graph. There are two cases:
    \begin{enumerate}
        \item Suppose the sampling procedure has recovered all incident hyperedges on at least $K_i/2$ of the connected components. This means that it has found all incident edges on at least $K_i/2$ of the connected components, so the algorithm removes these connected components from future iterations. In this case, the number of connected components goes down by a factor of $1/2$, i.e. $K_{i+1} \leq K_i / 2$.
        \item Suppose that the sampling procedure has not exhausted the incident hyperedges on at least $K_i / 2$ component. Thus, for at least $K_{i}/2$ of the connected components, the Recovery procedure has recovered $\geq \phi \log^2(n)$ hyperedges for which they are the unique representative. In particular, this means that the sampling returns at least $K_i \phi \log^2(n) / 2$ distinct hyperedges. Now, note that the number of hyperedges of strength $< 2\phi\log(n)$ can be at most $2K_i\phi\log(n)$ by \cref{clm:strengthNumber}. This means there must be at least $K_i \phi \log^2(n) / 2 - 2K_i\phi\log(n) \geq K_i \phi \log^2(n) / 4$ hyperedges of strength at least $2\phi\log(n)$ just in the subhypergraph on the contracted super-vertices $V^{(i)}_1, \dots V^{(i)}_{K_i}$ with these sampled edges. For this subhypergraph, we can exactly compute the strengths of the hyperedges (see \cite{Qua23} for instance) and find those hyperedges with strength at least $2\phi \log(n)$. Now, by \cref{clm:minkcutcontracted}, it follows that hyperedges we find of strength $\geq 2\phi \log(n)$ are exactly those of strength $\geq 2 \phi \log(n)$ in the \emph{unconctracted} hypergraph. Thus, in the original hypergraph (which contains only more hyperedges), their strengths can only be larger, and thus will still be $\geq 2 \phi \log(n)$.
        
        Further, because each of these $\geq K_i/2$ connected components are the unique representative for $\phi\log^2(n)$ hyperedges, this means that by the PHP, at least $\lceil \frac{K_i \phi \log^2(n) / 4}{\phi\log^2(n)} \rceil \geq K_i / 4$ connected components have incident hyperedges with strength at least $\phi \log(n)$ for which they are the unique representative. In particular, we can then merge the connected components that this hyperedge crosses between, as we are guaranteed that they are all contained in a component of strength $2\phi\log(n)$ (this follows from \cref{clm:unionStrength}). Note that each of the $K_i/4$ connected components with a neighboring edge of strength $\phi\log(n)$ participates in a union, so the number of connected components decreases by at least $\frac{K_i/4}{2} = K_i/8$.
    \end{enumerate}
    Note that in either case, the number of remaining connected components decreases by at least a factor of $1/8$. Starting with $n$ connected components and repeating this $8\log(n)$ times then exhausts the entire graph. Hence, every connected component is removed after at most $8 \log(n)$ iterations.
\end{proof}

\begin{claim}\label{clm:removeVertices}
    In \cref{alg:strengthDecomposition}, whenever a connected component $V_j^{(i)}$ is removed from consideration, it is either combined with another component to form a component of strength at least $\phi \log(n)$, or all of its incident hyperedges have been exhausted.
\end{claim}

\begin{proof}
    This follows by definition. Either a connected component is merged into a different connected component, or a connected component has all of its incident edges recovered, and is therefore removed.
\end{proof}

\begin{claim}\label{clm:componentStrength}
    Any connected component considered during \cref{alg:strengthDecomposition} is a singleton vertex or has strength at least $\phi\log(n)$.
\end{claim}

\begin{proof}
    Suppose a connected component is not a singleton vertex. Then, it follows that the connected component is the result of merging other connected components (possibly vertices). Let us suppose by induction that every connected component has strength at least $\phi \log(n)$. Then, to get our new connected component, we merge connected components that share an edge with strength $\geq \phi \log(n)$. It suffices to show then that if a set of connected components shares a hyperedge of strength $\phi \log(n)$, and each connected component also has strength $\phi \log(n)$, then the union of these connected components has strength $\phi \log(n)$. This follows exactly from \cref{clm:unionStrength}.
\end{proof}

\begin{claim}
    \cref{alg:strengthDecomposition} returns a set of connected components, each either a singleton vertex or of strength $\geq \phi \log(n)$, as well as all of the hyperedges crossing between these connected components. 
\end{claim}

\begin{proof}
This follows from \cref{clm:componentStrength} and \cref{clm:removeVertices}. Indeed, the list of components we return includes only those components which were removed during an iteration of \cref{alg:strengthDecomposition}. A component is removed only when all of its incident edges are recovered. The strength follows because every component that appears in the above algorithm has strength $\phi \log(n)$.
\end{proof}

\begin{remark}
    Note that as a consequence of the above algorithm, we are able to find the minimum normalized $k$-cut in the graph if it is of size $\leq \phi \log(n)$. This is because any minimum normalized $k$-cut of size $\leq \phi \log(n)$ will not cut any component of strength $\geq \phi \log(n)$, and thus the cut is entirely defined in the edges crossing between the components returned by the above algorithm. 
\end{remark}

More formally, we have the following:

\begin{claim}\label{clm:minkcutstructure}
    Let $H$ be a hypergraph, and let $V_1, \dots V_k$ be a set of connected components of strength $> \kappa$. Suppose we know all of the hyperedges in $E_H[V_1, \dots V_k]$, then we can correctly identify exactly all hyperedges in $H$ of strength $\leq \kappa$.
\end{claim}

\begin{proof}
This follows from \cref{clm:minkcutcontracted}. If we know all of the crossing hyperedges $E_H[V_1, \dots V_k]$, we can construct the contracted hypergraph $H / (V_1, \dots V_k)$. We know that in this hypergraph, the edges of strength $\leq \kappa$ are exactly those of strength $\leq \kappa$ in the original hypergraph $H$. Thus, we can simply find these corresponding hyperedges in $H / (V_1, \dots V_k)$.
\end{proof}

\subsubsection{More Exact Strength Decomposition}

This suggests the following algorithm, where $\kappa < \phi \log(n)$:

\begin{algorithm}
    \caption{ConditionalEdgeRecovery($G, \phi, \kappa$)}
    Recover $V_1, \dots V_p$ of strength $\geq \phi \log(n)$ as well as all crossing hyperedges between these components by running StrengthDecompositionRecovery($G, \phi$). \\
    Let $S$ denote all hyperedges of strength $\leq \kappa$ in $H / (V_1, \dots V_p)$. \\
    \Return{$S$}
\end{algorithm}

\begin{claim}\label{clm:edgeRecovery}
    If $\kappa < \phi \log(n)$ and there is a normalized $k$-cut of size $\leq \kappa$ in $G$, ConditionalEdgeRecovery returns all hyperedges of strength $\leq \kappa$.
\end{claim}

\begin{proof}
    The correctness follows from \cref{clm:minkcutstructure}. Because the strengths of $V_1, \dots V_p$ are all at least $\phi \log(n)$, one can find the exact edge strengths in $H / (V_1, \dots V_p)$ for any edge of strength $\leq \phi \log(n)$. We are then simply returning these hyperedges.
\end{proof}

\subsubsection{Space Analysis}

\begin{claim}\label{clm:strengthRecoverySpace}
    \cref{alg:strengthDecomposition} can be implemented as a linear sketch using only $\widetilde{O}(n r \phi \log(m) \log(1 / \delta))$ bits, where $\delta$ is the failure probability per $\ell_0$-sampler, $r$ is the maximum arity of $H$, and $m$ is the number of hyperedges in $H$.
\end{claim}

\begin{proof}
    The only space used by the linear sketch is in the $\ell_0$-samplers that are used in the recovery sketch. By assumption, we are storing $\widetilde{O}(\phi \polylog(n))$ $\ell_0$-samplers per vertex (and there are $n$ vertices). We can observe that the universe size of these $\ell_0$-samplers is bounded by $n^{2r} = 2^{2r \log(n)}$, and we can bound the support size of these $\ell_0$-samplers by $m$ (the total number of hyperedges in the hypergraph). It follows then that the total space required to store the $\ell_0$-samplers is 
    \[
    \leq \widetilde{O}(n \phi  \polylog(n) \cdot \log(m) \cdot (2r \log(n)) \cdot \log( 1 / \delta)) = \widetilde{O}(n r \phi \log(m) \log(1 / \delta)).
    \]
\end{proof}

\subsection{Sparsification}\label{sec:sparsification}

Now, we use the strength decomopsition algorithm as a building block in our sparsification algorithm. 

\subsubsection{Idealized Algorithm}

Our algorithm will attempt to implement the following sparsification algorithm in a linear sketch. We present a simple idealized sparsification procedure corresponding to \cite{Kar93, BK96} (and used in many subsequent works, for instance \cite{GMT15}). 

\begin{algorithm}
    \caption{IdealSparsify($H, \eps, m)$)}
    Initialize $H_{-1} = H, F_{-1} = \emptyset$.\\
    \For{$i = 0, 1, \dots \log(m)$}{
    Let $F_i$ contain all edges of strength $\leq 100 C \log(n) / \eps^2$ in $H_{i-1}$. \\
    Store $2^{i} \cdot F_i$. \\
    Let $H_i$ be the result of downsampling $H_{i - 1} - F_i$ at rate $1/2$. 
    }
\end{algorithm}

A simple way (as in \cite{GMT15}) to analyze this algorithm is presented below:

\begin{claim}\label{clm:correctnessIdeal}
   If $H$ has $m$ edges, the above algorithm returns a $(1 \pm O(\eps \log(m)))$-sparsifier for $H$ with probability $1 - 1 / \text{poly}$(n). 
\end{claim}

\begin{proof}
Consider any iteration $i$ and the corresponding hypergraph in that iteration $H_i$. We claim that with high probability, $F_i \cup 2 \cdot H_{i+1}$ is a $(1 \pm \eps)$-sparsifier for $H_i$. To see this, note that $H_i = F_i + (H_i - F_i)$. Now, the strength of every hyperedge in $(H_i - F_i)$ is at least $100 C \log(n) / \eps^2$ (see \cref{clm:removeLowStrength}), so it follows that sampling at rate $1/2$ (and reweighing by a factor $2$) will preserve every cut in $(H_i - F_i)$ to a factor $(1 \pm \eps)$ with probability $1 - n^{-100}$. Thus, since $H_{i+1}$ is this downsampled graph, it follows that $F_i \cup 2 \cdot H_{i+1}$ is a $(1 \pm \eps)$-sparsifier for $H_i$ with probability $1 - n^{-100}$.

Now, we claim inductively that the hypergraph under consideration after $j$ iterations is a $(1 \pm 2\eps j)$-sparsifier for $H$. The base case is easy to see, as the preceding paragraph proves the case when $j = 1$. Let us suppose the claim holds by induction up to the $j$th iteration. Then, it follows that $F_0 \cup 2 F_1 \cup \dots \cup 2^j F_j \cup H_{j+1}$ is a $(1 \pm 2\eps j)$-sparsifier for $H$. Now, by the preceding paragraph, it follows that $2 F_{j+1} \cup H_{j+2}$ is a $(1 \pm \eps)$-sparsifier for $H_{j+1}$ with high probability. Thus, $F_0 \cup 2 F_1 \cup \dots \cup 2^{j+1} F_{j+1} \cup H_{j+2}$ is at least as good as a $(1 \pm \eps)$-sparsifier to $F_0 \cup 2 F_1 \cup \dots \cup 2^j F_j \cup H_{j+1}$, and thus by composition, must be a $(1 \pm 2\eps (j+1))$-sparsifier for $H$.

Next, we must argue that the algorithm itself terminates within $\log(m)$ iterations. This follows because after $\log(m)$ iterations, the original hypergraph is being downsampled at rate $1 / m$, so there will be $O(\log(n))$ surviving hyperedges with probability $1 - 1 / \text{poly(n)}$, and these will be recovered exactly as the edges of low strength. Next, we know that $\log(m) \leq n$, so we can take a union bound over the at most $n$ levels of sparsification. Each level of sparsification returns a $(1 \pm \eps)$-sparsifier with probability $1 - n^{-100}$, so in total, the probability of getting a sparsifier is at least $1 - n^{-99} - 1 / \text{poly(n)}$, as we desire.
\end{proof}

\begin{remark}\label{rmk:simpleforkcuts}
    Although the argument in \cref{clm:correctnessIdeal} is only stated for preserving $2$-cuts, note that $F_i \cup 2 \cdot H_{i+1}$ is actually a \emph{$k$-cut}-sparsifier for $H_i$ by the reasoning from \cref{clm:subsamplingGood}. That is, every hyperedge in $H_i - F_i$ has $k$-cut strength at least $2C \log(n) / \eps^2$, and thus we can afford to sample at rate $1/2$ while preserving the weight of all $k$-cuts (simultaneously for every value of $k \in [n]$) to a factor $(1 \pm \eps)$.
\end{remark}

\subsubsection{Linear Sketch Implementation}

Next, we will show how to implement the above algorithm more carefully in a linear sketching framework. Consider the following algorithm which takes as input a hypergraph $H$, an approximation parameter $\eps$, the number of edges in $H$, denoted by $m$, as well as (uniformly random) filter functions $f_1, \dots f_{\log(m)}$, $f_i: 2^{[n]} \ra \zo$:

\begin{algorithm}
    \caption{LinearSketchSparsify($H, \eps, m, (f_1, f_2, \dots f_{\log(m)})$)}\label{alg:LSS}
    Initialize $H_{-1} = H, F_{-1} = \emptyset$. \\
    \For{$i = 0, 1, \dots \log(m)$}{
    Let $H_i$ contain all edges $e$ from $H_{i-1} - F_{i-1}$ such that $\prod_{j = 1}^i f_j(e) = 1$. \\
    $F_i \leftarrow$ ConditionalEdgeRecovery$(H_i, \phi, \kappa)$, with $\phi = C \log(n) / \eps^2$, and $\kappa = 100\phi$. \\
    Store $2^i \cdot F_i$. \\
    }
\end{algorithm}

There are a few key claims that we will show about the above algorithm. 

\begin{claim}
    The above algorithm returns a $(1 \pm O(\eps \log(m)))$-sparsifier for $H$ with probability $1 - 1 / \text{poly}(n)$.
\end{claim}

\begin{proof}
    This follows by the exact same proof as \cref{clm:correctnessIdeal}. Indeed, consider the execution in the $i$th step of the algorithm. By \cref{clm:edgeRecovery}, it must be the case that all edges of strength $\leq 100 C \log(n) / \eps^2$ are removed from $H_i$ and stored in $F_i$. Then, with probability $1 - 1 / \text{poly}(n)$, downsampling $H_i - F_i$ at rate $1/2$ to get $H_{i+1}$ will yield $H_{i+1}$ which is a $(1 \pm \eps)$-sparsifier for $H_i - F_i$, and thus $F_i \cup 2 \cdot H_{i+1}$ is a $(1 \pm \eps)$-sparsifier for $H_i$ with the same probability. 
    
    It follows then that if we inductively repeat this, we will get a $(1 \pm O(\eps \log(m)))$-sparsifier for $H$ with probability $1 - 1 / \text{poly}(n)$.
\end{proof}

\begin{claim}
    The above algorithm can be implemented with a linear sketch of size $\widetilde{O}(n r \log^4(m) / \eps^2)$ to get a $(1 \pm \eps)$-sparsifier for $H$.
\end{claim}

\begin{proof}
    The only space we use for the linear sketch is in storing independent copies of the sketch required for ConditionalEdgeRecovery. We do this for $O(\log(m))$ different levels (before $H$ is empty), and at each level we invoke ConditionalEdgeRecovery with $\phi = O(\log(n) / (\eps/\log(m))^2)$. By \cref{clm:strengthRecoverySpace}, each sketch will require $\widetilde{O}(nr \log^3(m) \log(1 / \delta) / \eps^2)$ bits, and thus over the $\log(m)$ possible levels, the total space is $\widetilde{O}(nr \log^4(m) \log(1 / \delta) / \eps^2)$.
    
    Because there are $\widetilde{O}(n \log(m) / \eps^2)$ $\ell_0$-samplers, it suffices to choose $\delta = \eps^2 / \text{poly}(n)$. For this choice of $\delta$ then, it follows that the total space requirement is $\widetilde{O}(n r  \log^4(m) / \eps^2)$.
\end{proof}

Note that because our sketch is linear, the operation of removing $F_{i-1}$ from $H_{i-1}$ is allowed, as this simply corresponds with updating the support of $\ell_0$ samplers. In particular, we only ever update \emph{later rounds} of $\ell_0$ sampling which are initialized with independent random seeds.

\subsection{Cut-Perspective for Getting Rid of $O(\log^2(m))$ Terms}\label{sec:cutPerspective}

In this section, we will re-analzye the above algorithm to show that we can get rid of an extra $\log^2(m)$ factor. At the core of this analysis is showing that it suffices to set our error parameter to be $\eps / \text{polylog}(n)$ as opposed to $\eps / \log(m)$. We do this by carefully analyzing the rate at which the accuracy of each cut deteriorates as we continue to downsample the hypergraph. 

We next present some definitions for the above algorithm.

\begin{definition}
    Consider any $k$-cut of the hypergraph $H$. We denote this cut by $Q$ (i.e. denoting the set of edges in the cut). Let $\lambda(Q)$ be the maximum strength (in $H$) of any hyperedge which is in $Q$.
\end{definition}

\begin{definition}
    Let $H$ be a hypergraph and $Q$ be a cut in $H$, and let $H_i$ be a version of $H$ which results from running our linear sketching algorithm for $i$ iterations. We say that the cutoff for $Q$ is $\lambda(Q) / n^{8}$. With this we have some definitions:
    \begin{enumerate}
        \item We say that $Q$ is \textbf{inactive} in $H_i$ if $2^i \leq \eps^5 \lambda(Q) / n^{24}$.
        \item We say that $Q$ is \textbf{active} in $H_i$ if $\lambda(Q) \cdot n^{10} \geq 2^i \geq \eps^5 \lambda(Q) / n^{24}$.
        \item We say that $Q$ is \textbf{exhausted} in $H_i$ if $\lambda(Q) \cdot n^{10} < 2^i $.
    \end{enumerate}
\end{definition}

\begin{definition}
    We let $Q_{\leq \kappa}$ denote the edges in $Q$ that have strength $\leq \kappa$ in $H$, and likewise $Q_{\geq \kappa}$ denotes the edges in $Q$ of strength $\geq \kappa$ in $H$. 
\end{definition}

\begin{definition}
    Let $E_i \subseteq E$ denote the set of edges which survive $i$ rounds of downsampling from filter functions. 
\end{definition}

Intuitively, if $2^i$ is below the cutoff, we are going to argue that the majority of $Q$ has had its weight preserved in the sparsification routine so far. While $2^i$ is slightly above the cutoff, we will show that this is where the sparsification of the majority of $Q$ is happening, and that indeed, most of the cut is preserved to the right size. Finally, when $2^i$ is far too large, we will argue that all of the edges from the cut have already been removed. We will use the following claim in a key way:

\begin{claim}\label{clm:downsampleRemove}
    Let $H$ be a hypergraph. With probability $1 - {n^{8}}$, for all edges $e$, $e$ will not be in $H_i$ for $2^i \geq n^{10} \cdot \lambda_e$ (where $\lambda_e$ denotes the strength of $e$).
\end{claim}

\begin{proof}
    Note that there can only be $n$ different strengths in $H$, as each strength corresponds with some $k$-partition which increases the number of connected components. So, fix one of these $n$ strength values $\lambda$. We then know that there can be at most $(n-1)\lambda$ hyperedges of strength $\leq \lambda$. Thus, the probability that a single one of these hyperedges survives at the given sampling rate is $\leq 1 / (n^{10} \cdot \lambda_e)$. Taking the union bound over all hyperedges, we know that no hyperedges of strength $\lambda_e$ survive with probability $\geq 1 - \frac{(n-1)\lambda_e} {n^{10} \cdot \lambda_e}\geq 1 - {n^{9}}$. Finally, we can take a union bound over all $n$ possible strength values to conclude that with probability $1 - {n^{8}}$ any hyperedge $e$ is not in from $H_i$ when $2^i \geq n^{10} \cdot \lambda_e$.
\end{proof}

\begin{claim}\label{clm:errorAccumulation}
    Let $H$ be a hypergraph, and let $Q$ be some cut of $H$ corresponding to the hyperedges crossing between components $V_1, \dots V_k$. Let $i$ be the first iteration in which $Q$ is active when running \cref{alg:LSS}. Then, 
    \begin{enumerate}
        \item Let $Q_{\geq \eps\lambda(Q) / n^{20}}$ be the hyperedges in $Q$ with strength at least $\eps \lambda(Q) / n^{20}$ in $H$. It follows that by the $i$th iteration, $2^i \cdot |H_i \cap Q_{\geq \eps\lambda(Q) / n^{20}}| \in (1 \pm \eps)|H  \cap Q_{\geq \eps\lambda(Q) / n^{20}}|$.
        \item $|Q| \geq |Q_{\geq \eps\lambda(Q) / n^{20}}| \geq \lambda(Q)$.
        \item In the resulting sparsifier for $H$, the total contribution to $Q$ from hyperedges of strength $\leq \eps \lambda(Q) / n^{20}$ is $\leq \eps \lambda(Q) / n^3$  with probability $1 - 2^{-\Omega(n^3)}$
        \item In the resulting sparsifier for $H$, the weight of edges crossing cut $Q$ is preserved to a factor $(1 \pm \eps / n^3)(1 \pm \eps)^{\log(n^{24 + 10} / \eps^5)}$ with high probability. 
        \item By setting $\eps* = \frac{\eps}{\log^2(n / \eps)}$, and creating a sparsifier for $H$ by calling \cref{alg:LSS} with error parameter $\eps^*$, every cut $Q$ is preserved to a factor $(1 \pm \eps)$ with probability $1 - n^{-8}$.
    \end{enumerate}
\end{claim}

\begin{proof}
\begin{enumerate}
    \item 
    First, consider the hypergraph $H^*$ which contains only those hyperedges of $H$ with strength at least $\eps \lambda(Q) / n^{20}$. 
    It follows that if one subsamples $H^*$ at any rate $p \geq \frac{n^{24}}{2 \eps^5 \lambda(Q)}$ to get $H^{*'}$, all cuts in $H^*$ will be preserved (after reweighting) to a factor $(1 \pm \eps)$ with probability $1 - 2^{-\Omega(n^4)}$ (this follows from \cref{clm:subsamplingGood}). 
    
    In particular, this means that every non-empty normalized k-cut in $H^{*'}$ will be of size at least $(1 - \eps) \cdot n^4/\eps^2$ with probability $1 - 2^{-\Omega(n^4)}$. Taking a union bound over all $n$ possible rates of downsampling to get $H^{*'}$, it follows that in successive iterations leading up to $Q$ becoming active, no hyperedges from $H^*$ will ever be removed in the strength decomposition (since their strength remains above $(1 - \eps) \cdot n^4/\eps^2$), and that at every step, we maintain a $(1 \pm \eps)$-approximation to the size of $Q$ in $H^*$.

    \item Note that by definition, $Q$ cuts a component of strength $\lambda(Q)$. It therefore follows that if we restrict our attention to only the hyperedges of strength $\geq \lambda(Q)$, $Q$ must have at least $\lambda(Q)$ crossing hyperedges among these.
    
    \item First, from the previous item, we know that the only hyperedges which will be stored prior to the $i$th iteration are those that correspond to hyperedges in $H$ of strength $\leq \eps \lambda(Q) / n^{20}$. 
    
    Next, we know that the number of hyperedges with strength $\leq \eps\lambda(Q) / n^{20} = W$ is at most $n \cdot \eps\lambda(Q) / n^{20} \leq \eps\lambda(Q) / n^{19} = nW$. We call these edges the low strength edges. Now, we can upperbound the total contribution from these low strength hyperedges in the sparsifier by considering the filter functions $f_i$. We know that a given hyperedge survives a single downsampling iteration with probability $1/2$, at which point the hyperedge is given weight at most $2$. Thus, after $i$ levels of downsampling, it is still the case that the expected weight of remaining edges is $nW$. We also know that by the time we are sampling edges at rate $1/ (Wn^{10})$, all the edges will have been removed with high probability. Now, because each hyperedge can only be stored once (after which it is removed from future sketches), we can get a crude upper bound for the total weight contributed by these edges by summing the total weight of the surviving edges after each level of downsampling. To summarize,
    \begin{align*}
        \text{total contribution of low strength edges} \\
        \leq \sum_{j = 1}^{\log(W n^{10})} 2^j \cdot |Q_{\leq W} \cap F_j| \leq  \sum_{j = 1}^{\log(W n^{10})} 2^j \cdot |Q_{\leq W} \cap E_j|
    \end{align*}
    Next, note that $|Q_{\leq W} \cap E_j|$ is simply distributed as a Binomial($nW, 2^{-j}$) variable. Thus, we note that 
    \[
    \Pr[2^j \cdot \text{Binomial($nW, 2^{-j}$)} \geq \eps\lambda(Q) / n^4] = \Pr[\text{Binomial($nW, 2^{-j}$)} \geq \frac{\eps\lambda(Q)}{n^4 \cdot 2^j}] 
    \]
    \[
    = \Pr[\text{Binomial($nW, 2^{-j}$)} \geq \frac{W \cdot n^{20}}{n^4 \cdot 2^j}] = \Pr[\text{Binomial($nW, \frac{1}{2^j}$)} \geq \frac{W \cdot n^{16}}{2^j}]
    \]
    \[
    \leq \Pr[\text{Binomial($nW, \frac{1}{Wn^{10}}$)} \geq \frac{W \cdot n^{20}}{n^4 W n^{10}}] = \Pr[\text{Binomial($nW, \frac{1}{Wn^{10}}$)} \geq n^{6}].
    \]
    The inequality follows from the fact that $\Pr[\text{Binomial}(\ell, p_1) \geq K \cdot p_1] \leq \Pr[\text{Binomial}(\ell, p_2) \geq K \cdot p_2]$ whenever $K \geq \ell, p_1 \geq p_2$. To see why this is true, this is equivalent to 
    \[
    \Pr[\text{Binomial}(\ell, p_1) \geq (K / \ell) (\ell p_1)] \leq \Pr[\text{Binomial}(\ell, p_2) \geq (K/\ell)(\ell p_2)].\]

    Then, it follows that 
    \[
    \Pr[\text{Binomial($nW, \frac{1}{Wn^{10}}$)} \geq n^6] \leq \Pr[\text{Binomial($nW, \frac{n^3}{W}$)} \geq n^6],
    \]
    which is bounded by $2^{-\Omega(n^3)}$ by a Chernoff Bound. Thus, it follows that with probability $1 - 2^{-\Omega(n^3)}$, the total contribution from low strength edges is at most $n \cdot \eps\lambda(Q) / n^4 \leq \eps\lambda(Q) / n^3$.
    \item Let us denote the sparsifier we obtain by $\hat{H}$. Further, let us denote the accuracy parameter we obtain for $Q_{\geq \eps \lambda(Q) / n^{20}}$ by $\eps'$.
    
    First we will show that with high probability $|E_{\hat{H}}[V_1, \dots V_k]| \geq (1 - \eps') \cdot (1 - \eps / n^{19}) |E_{H}[V_1, \dots V_k]| = (1 - \eps) \cdot (1 - \eps / n^{19}) |Q|$. Indeed, we know that in $H$, the total contribution to $|Q|$ from $|Q_{\leq \eps\lambda(Q) / n^{20}}|$ was $\leq n \cdot \eps\lambda(Q) / n^{20} = \eps\lambda(Q) / n^{19}$ by the previous part. Because $|Q| \geq \lambda(Q)$, it follows that the edges of low strength contribute at most a $\eps / n^{19}$ fraction of the hyperedges to $|Q|$. Thus, if we get a $(1 \pm \eps')$ approximation to the cut-sizes of $Q_{\geq \lambda(Q) / n^{20}}$, this will be at least a $(1 - \eps') \cdot (1 - \eps / n^{19})$ factor approximation to $Q$.

    Next, we will show that with high probability $|E_{\hat{H}}[V_1, \dots V_k]| \leq (1 + \eps')(1 + \eps / n^3)|E_{H}[V_1, \dots V_k]| = (1 + \eps')(1 + \eps / n^3)|Q|$. This follows because we get a $(1 \pm \eps')$ approximation to $Q_{\geq \lambda(Q) / n^{20}}$, and the remaining low strength edges contribute a factor of at most $\eps |Q| / n^3$ with high probability. Thus, we get an upper bound on our approximation factor of $(1 + \eps')(1 + \eps / n^3)$.

Finally, the exact factor of $\eps'$ that we achieve is the level of approximation that we achieve for $Q_{\geq \lambda(Q) / n^{20}}$. Note that in $H_i$, $Q_{\geq \lambda(Q) / n^{20}}$ has all cuts preserved to a factor $(1 \pm \eps)$. Then for each iteration in which $H_i$ is active, we lose a factor of $(1 \pm \eps)$ in the approximation. Thus, because $H_i$ is active for $\log(n^{24 + 10} / \eps^5)$ iterations, we get an approximation factor of $(1 \pm \eps)^{\log(n^{34} / \eps^5)}$.

\item It follows that if we run the above algorithm with $\eps^* = \frac{\eps}{\log^2(n / \eps)}$, then the approximation factor we achieve for any cut $Q$ is $(1 \pm \eps / n^3) \cdot (1 \pm \frac{\eps}{\log^2(n / \eps)})^{\log(n^{34 } \log^2(n/\eps) / \eps)}$. Because $\log^2(n / \eps) \geq 2 \log(n^{34} \log^2(n / \eps) / \eps)$, we can bound this second term by $(1 \pm \eps / 2)$. Likewise, the first term $(1 \pm \eps^*/n^3)$ will have error bounded by $(1 \pm \eps/2)$, and thus the total error in preserving the cut $Q$ is $\leq (1 \pm \eps)$.

Next, we will analyze the probability with which this will hold. We define some \say{bad} events in the execution of \cref{alg:LSS}. Note that some of these bad events are global in the sense that the bad event is defined without mention of a specific cut. Some of the bad events are local, meaning they depend on a specific cut. In the local case is where we will have to ensure that the probabilities are sufficiently low so as to survive a union bound. First, we define the global bad events:

\begin{enumerate}
    \item $B_1$ is the event that \cref{clm:downsampleRemove} fails to happen.
    \item $B_2$ is the event that in the execution of \cref{alg:LSS}, there is some iteration $j$ in which $F_j \cup 2 \cdot H_{j+1}$ is not a $(1 \pm \eps)$-cut sparsifier for $H_{j}$.
\end{enumerate}

Next, we define the local bad events for a cut $Q$:

\begin{enumerate}
    \item $B_3$ is the event that $2^i \cdot |H_i \cap Q_{\geq \lambda(Q) / n^{20}}| \notin (1 \pm \eps)|H  \cap Q_{\geq \lambda(Q) / n^{20}}|$, where $i$ is the first iteration in which $Q$ is active.
    \item $B_4$ is the event that in the resulting sparsifier for $H$, the total contribution to $Q$ from hyperedges of strength $\leq \lambda(Q) / n^{20}$ stored in the first $\log(Wn^{10})$ iterations is $> \lambda(Q) / n^3$.
\end{enumerate}

Now, by our previous logic, if none of these happen for any cut $Q$, we will have our desired result. So, it suffices to bound the probability that any of these happen. We know that $\Pr[B_1] \leq n^{8}$ from \cref{clm:downsampleRemove}. Next, for $B_2$, we know that in each iteration $F_j \cup 2 \cdot H_{j+1}$ is not a $(1 \pm \eps)$ sparsifier for $H_j$ with probability at most $n^{-10}$. Because there are at most $n$ levels of downsampling, the total failure probability here is at most $\Pr[B_2] \leq n^{-9}$.

Next, for our local events, we know that we must take a union bound over at most $n^n$ choices of $Q$. For any such choice, it is the case that $\Pr[B_3] \leq 2^{-\Omega(n^4)}$, by the first item in this claim. Taking the union bound over all $n^n$ choices of $Q$, we get that $\Pr[B_3 \text{ occurs for any $Q$}] \leq 2^{-\Omega(n^3)}$. Likewise for a given $Q$, $B_4$ occurs with probability $\leq 2^{-\Omega(n^3)}$, so $\Pr[B_4 \text{ occurs for any $Q$}] \leq 2^{-\Omega(n^2)}$. 

Thus, the total probability of any bad event happening is $\leq n^{8} + n^{-9}+ 2^{-\Omega(n^3)} +2^{-\Omega(n^2)} \leq n^{-8}$, so with high probability, our algorithm sparsifies all cuts to factor $(1 \pm \eps)$.
\end{enumerate}
\end{proof}

\begin{lemma}
    There exists a linear sketching algorithm that with high probability returns a $(1 \pm \eps)$ sparsifier for a hypergraph $H$ of maximum arity $r$ using only $\widetilde{O}(nr \log^2(m) / \eps^2)$ bits of space.
\end{lemma}

\begin{proof}
    The correctness follows from \cref{alg:LSS} called with error parameter $\eps / \log^2(n / \eps)$. The only sketch we store is for ConditionalEdgeRecovery at each level of downsampling. We do this for $O(\log(m))$ different levels (before $H$ is empty), and at each level, we use ConditionalEdgeRecovery with $\phi = O(\log(n) / (\eps/\log^2(n/\eps))^2)$. By \cref{clm:strengthRecoverySpace}, each sketch will require $\widetilde{O}(nr \log(m) \log(1 / \delta) / \eps^2)$ bits, and thus over the $\log(m)$ possible levels, the total space is $\widetilde{O}(nr \log^2(m) \log(1 / \delta) / \eps^2)$.
    
    Because there are $\widetilde{O}(n \log(m) / \eps^2)$ $\ell_0$-samplers, it suffices to choose $\delta = \eps^2 / \text{poly}(n)$. For this choice of $\delta$ then, it follows that the total space requirement is $\widetilde{O}(nr \log^2(m) / \eps^2)$.
\end{proof}

\subsection{Getting Rid of the Final $O(\log(m))$ via Preprocessing}\label{sec:preprocessing}

Our goal in this section will be to get rid of an additional $O(\log(m))$ term. Roughly speaking, this extra factor of $\log(m)$ comes from the fact that the $\ell_0$-samplers must be defined for a support size as large as $m$. Here, we will show that with a preprocessing step, we can reduce the number of hyperedges under consideration in every level of downsampling to be bounded by $\text{poly}(n)$. This then allows us to store $\ell_0$-samplers of size $\widetilde{O}(r \log(n))$ instead of potentially as large as $\widetilde{O}(r \log(m))$.

\subsubsection{Idealized Algorithm}

 To get this reduction, we will consider the following idealized algorithm:

\begin{algorithm}[H]
    \caption{SparsifyWithStrongComponents($H, \eps, m, (f_1, f_2, \dots f_{\log(m)}), (V^{(i)}_1, \dots V^{(i)}_{p_i})_{i = 0}^{\log(m)}$)}\label{alg:LSSStrongComp}
    Initialize $H_{-1} = H, F_{-1} = \emptyset$. \\
    \For{$i = 0, 1, \dots \log(m)$}{
    Let $H_i$ contain all edges $e$ from $H_{i-1} - F_{i-1}$ such that $\prod_{j = 1}^i f_j(e) = 1$. \\
    $F_i \leftarrow$ ConditionalEdgeRecovery$(H_i /(V^{(i)}_1, \dots V^{(i)}_{p_i}) , \phi, \kappa)$, with $\phi = C \log(n) / \eps^2$, and $\kappa = 100\phi$. \\
    Store $2^i \cdot F_i$. \\
    }
    \Return{all stored hyperedges.}
\end{algorithm}

\begin{claim}\label{clm:sameBehavior}
    \cref{alg:LSSStrongComp} behaves exactly the same as \cref{alg:LSS} if for each $i \in [\log(m)]$, $V^{(i)}_1, \dots V^{(i)}_{p_i}$ is a partition of $V$ such that each $V^{(i)}_{\ell}$ is of strength $\geq n^{10} / \eps^2$ in $H_i$, and each hyperedge of strength $\geq n^{100} / \eps^2$ in $H_i$ is completely contained in some $V^{(i)}_{\ell}$.\\
\end{claim}

\begin{proof}
    It suffices to prove that the edges $F_i$ that are recovered are the same. This follows exactly from \cref{clm:minkcutstructure}. Indeed, because the components $V_{\ell}^{(i)}$ are of strength $\geq n^{10} / \eps^2$, the hyperedges in $H_i /(V^{(i)}_1, \dots V^{(i)}_{p_i})$ of strength $\leq C \log(n) / \eps^2$ are exactly the same as the hyperedges in $H_i$ of strength $\leq C \log(n) / \eps^2$. Thus, recovering these edges in the contracted version of $H_i$ is the same as recovering these edges in the original version of $H_i$.
\end{proof}

\begin{claim}\label{clm:boundedSupport}
    In \cref{alg:LSSStrongComp}, if for each $i \in [\log(m)]$, $V^{(i)}_1, \dots V^{(i)}_{p_i}$ is a partition of $V$ such that each $V^{(i)}_{\ell}$ is of strength $\geq n^{10} / \eps^2$ in $H_i$, and each hyperedge of strength $\geq n^{100} / \eps^2$ in $H_i$ is completely contained in some $V^{(i)}_{\ell}$, then we can implement each $\ell_0$-sampler for ConditionalEdgeRecovery with support size $\text{poly}(n / \eps)$ instead of $m$. 
\end{claim}

\begin{proof}
    In the $i$th level of downsampling, we run ConditionalEdgeRecovery on the hypergraph $H_i /(V^{(i)}_1, \dots V^{(i)}_{p_i})$. We are told that every hyperedge of strength $\geq n^{100} / \eps^2$ in $H_i$ is completely contained in some $V^{(i)}_{\ell}$, so it follows that in the contracted graph $H_i /(V^{(i)}_1, \dots V^{(i)}_{p_i})$, each such edge has been contracted away (to a self-loop). Thus, the only crossing edges in $H_i /(V^{(i)}_1, \dots V^{(i)}_{p_i})$ will be a subset of those edges of strength $\leq n^{100} / \eps^2$ in $H_i$. Now, by \cref{clm:strengthNumber}, there can be at most $n^{101} / \eps^2$ such edges, so it follows that $H_i /(V^{(i)}_1, \dots V^{(i)}_{p_i})$ has $\leq n^{101} / \eps^2$ edges. 

    So, we know that each $\ell_0$-sampler using correlated randomness in the sketch for ConditionalEdgeRecovery only requires a support of size $\text{poly}(n / \eps)$. This is because these $\ell_0$-samplers will always be added together to create a component that has at most $n^{101}/\eps^2$ crossing hyperedges incident upon it. Further, in \cref{thm:recoveryProblem} each $\ell_0$-sampler is assumed to be defined on a subset of the support, so in particular, the upper-bound of $\poly(n/\eps)$ remains.
\end{proof}

\subsubsection{Strong Component Recovery With Smaller Sketches}

As we showed in the previous section, if we can create a method which identifies these \say{exceedingly strong} components before running our sparsification routine, then we can afford to save a factor of $\log(m)$ in the size of the $\ell_0$ samplers that we use. Unfortunately, we cannot afford to use the \say{Recovery} algorithm we defined before, as this is exactly the algorithm we are trying to optimize.

Instead, we use an algorithm which iteratively samples the hypergraph $H$, and at each level of sampling, only stores enough $\ell_0$-samplers to check the connectivity of the sampled hypergraph. We show that (1) this connectivity sketch suffices for identifying strong components, and (2) \emph{if we open the sketches in reverse} (starting with the version of the hypergraph that has undergone the most levels of sampling), we can actually implement that sketch with only $\widetilde{O}(nr \log(m))$ bits. 

To this end, consider the following algorithm:

\begin{algorithm}
\caption{RecoverStrongComponents$(H)$}\label{alg:recoverStrongComponents}
    Let $\widetilde{H}_i$ be $\widetilde{H}_{i-1}$ downsampled at rate $1/2$, starting with $\widetilde{H}_0 = H$. \\
    Let the initial starting vertex set be $[n]$, so $\widetilde{V}_i^{(\log(m)+1)}= i$, $p_{\log(m)+1} = n$.\\
    \For{$i = \log(m), \dots, 1, 0$}{
    Let $\widetilde{V}^{(i)}_1, \dots \widetilde{V}^{(i)}_{p_i}$ be the connected components in $\widetilde{H}_i / (\widetilde{V}^{(i+1)}_1, \dots \widetilde{V}^{(i+1)}_{p_i})$. \\
    }
    \Return{$(\widetilde{V}^{(i)}_1, \dots \widetilde{V}^{(i)}_{p_i})_{i = 0}^{\log(m) + 20\log(n)}$}
\end{algorithm}

\begin{claim}
    With probability $1 - 2^{-\Omega(n^3)}$, for every $k$-cut $Q$, and for every $i$, it must be the case that if $|Q \cap \widetilde{H}_{i}| \geq n^5$, then $|Q \cap \widetilde{H}_{i+1}| \geq 1$.
\end{claim}

\begin{proof}
    This follows from a Chernoff bound. 
\end{proof}

\begin{claim}\label{clm:boundedinprep}
    With probability $1 - 2^{-\Omega(n^3)}$, for every $i \in [\log(m)]$ the degree of every (super)-vertex in $\widetilde{H}_i / (\widetilde{V}^{(i+1)}_1, \dots \widetilde{V}^{(i+1)}_{p_i})$ is bounded by $\text{poly(n)}$.
\end{claim}

\begin{proof}
    The algorithm works from the bottom up. Clearly, in $\widetilde{H}_{\log(m)}$, there will be fewer than $n^5$ hyperedges surviving total with probability $1 - 2^{-\Omega(n^3)}$ (by Chernoff), and thus with high probability every (super)-vertex will have degree bounded by $n^5$.

    Now, consider the $i$th iteration of the above process. Because in $\widetilde{H}_{i-1}$ we merge together all vertices that are connected, it follows that the number of hyperedges in the contracted graph is $0$. Now, it must be the case that the surviving hyperedges in this contracted graph corresponds with some $k$-cut $Q$ in the original graph. Thus, by the previous claim, we know that (with high probability) because $|Q \cap \widetilde{H}_i| = 0$, it must be that $|Q \cap \widetilde{H}_{i-1}| \leq n^5$. Thus, we get that the number of surviving hyperedges in the up-sampled version of the graph is bounded by $\text{poly(n)}$.
\end{proof}

\begin{claim}\label{clm:estimatingComponents}
    Let $\widetilde{H}_j, H_j$ be independently downsampled hypergraphs where $H_i$ is $H_{i-1}$ downsampled at rate $1/2$ and $H_0 = H$ (and the same respectively for $\widetilde{H}_i$). Let $\widetilde{V}^{(i)}_1, \dots \widetilde{V}^{(i)}_{p_i}$ denote the connected components recovered by \cref{alg:recoverStrongComponents}.
    Then, with probability $1 - 3n^{-8}$, it must be the case that 
    \begin{enumerate}
        \item Any connected component of strength $\geq n^{100} / \eps^2$ in $H_{j}$ will remain connected in $\widetilde{H}_{j + \log(n^{20}/\eps^2)}$.
        \item Any connected component in $\widetilde{H}_{j + \log(n^{20}/\eps^2)}$ will have strength at least $n^{10} / 2\eps^2$ in $H_{j}$. 
    \end{enumerate}
\end{claim}

\begin{proof}

First, let us invoke \cref{clm:downsampleRemove} twice for both the sequences of downsampling defined by $\widetilde{H}_i$ and $H_i$. This states that with probability $1 - n^{-8}$, all edges $e \in H$ will be removed from $H_i$ when $2^i \geq \lambda_e \cdot n^{10}$ (and the same respectively for $\widetilde{H}$).

    Now, let us show the first point, let $C \subseteq V$ denote some component in $H_{j}$ of strength $\geq n^{100}/\eps^2$. This means with probability at least $1 - 2^{-\Omega(n^{10})}$, $C$ will have strength $(1/2) \cdot n^{50}/\eps^2 \cdot 2^j$ in the graph $H_{j + 50 \log(n)}$. In particular, this will mean that the component $C$ will still be connected in $H_{j + \log(n^{50}/\eps^2)}$. Because $2^{j + \log(n^{50}/\eps^2)} = 2^j \cdot n^{50}/\eps^2$, this means that all edges of strength $\leq 2^j \cdot n^{40}/\eps^2$ in $H$ have been removed from $H_{j +\log(n^{50}/\eps^2)}$. Thus, because the component $C$ is still connected in $H_{j +\log(n^{50}/\eps^2)}$, this means that $C$ must be connected by hyperedges of strength $\geq 2^j \cdot n^{40}/\eps^2$ in $H$, and therefore have strength $\geq 2^j \cdot n^{40}/\eps^2$ in $H$. Thus, with probability $1 - 2^{-\Omega(n^{10})}$, when we downsample by $2^j \cdot n^{20}/\eps^2$ in $\widetilde{H_0}$, $C$ will continue to have strength $\geq (1/2)n^{20}/\eps^2$, and therefore $C$ constitute a connected and be merged together in $\widetilde{H}_{j + \log(n^{20}/\eps^2)}$. Therefore, the constituent vertices of $C$ will be combined together into a single component $\widetilde{V}^{j + \log(n^{20}/\eps^2)}_{\ell}$.

    Again because \cref{clm:downsampleRemove} holds, this means that all edges with strength $\leq n^{10} \cdot 2^j/\eps^2$ will be removed in $\widetilde{H}_{j +\log(n^{20}/\eps^2)}$. Thus any connected component $V^{j + \log(n^{20}/\eps^2)}_{\ell}$ that forms in $\widetilde{H}_{j + \log(n^{20}/\eps^2)}$ must have strength $\geq n^{10 }\cdot 2^j/\eps^2$ in the original $H$. 
    Now, when we downsample to get $H_{j}$ (at rate $1 / 2^j$), it follows that with probability $1 - 2^{-\Omega(n^9)}$, $\widetilde{V}^{j + \log(n^{20} / \eps^2)}_{\ell}$ has
    strength $\geq (1/2) 2^j n^{10} / (\eps^2 2^j) = (1/2)n^{10}/\eps^2$ in $H_{j}$.

    Now, to see our probability bound, note that we must only invoke \cref{clm:downsampleRemove} twice globally after which it holds for every edge strength. Then, for each possible component that we can see, we can take a union bound over the probability of any of the above bad events. There are at most $n^n$ components, and $n$ rounds of sampling, for a total of $n^{n+1}$ possible components seen. The probability of failure for any given component is bounded by $2 \cdot 2^{-\Omega(n^{10})} + 2^{-\Omega(n^{9})}$ and thus remains overwhelmingly small after the union bound. In total then, the failure probability can be bounded by $1 - 3n^{-8}$.
\end{proof}

\subsubsection{Complete Algorithm}

We now present the complete algorithm for sparsification:

\begin{algorithm}[H]
\caption{StrengthRecoverySparsification($H, \eps, m, (f_1, f_2, \dots f_{\log(m)})$)}\label{alg:StrengthRecoverySparsification}
    Let $\eps_* = \frac{\eps}{\log^2(n / \eps)}$. \\
    Let $(\widetilde{V}^{(i)}_1, \dots \widetilde{V}^{(i)}_{p_i})_{i = 0}^{\log(m)+\log(n^{20}/(\eps_*)^2)} =$RecoverStrongComponents($H$). \\
    For $i = 0, \dots \log(m)$, let $V^{(i)}_{\ell} = \widetilde{V}^{(i + \log(n^{20}/(\eps_*)^2))}_{\ell}$. \\
    \Return{SparsifyWithStrongComponents($H, (\eps_*), m, (f_1, f_2, \dots f_{\log(m)}), (V^{(i)}_1, \dots V^{(i)}_{p_i})_{i =0 }^{\log(m)}$)}
\end{algorithm}

\begin{claim}\label{clm:finalProb}
    \cref{alg:StrengthRecoverySparsification} returns a $(1 \pm \eps)$-sparsifier for $H$ with probability $1 - 4n^{-8}$.
\end{claim}

\begin{proof}
    This follows from \cref{clm:sameBehavior} and \cref{clm:estimatingComponents}. Indeed, \cref{alg:LSSStrongComp}, behaves the same as \cref{alg:LSS} under the condition that the components under consideration in the $i$th iteration are of strength $\geq n^{10}/\eps^2$, and contain all edges of strength $\geq n^{100}/\eps^2$ in $H_i$. By \cref{clm:estimatingComponents}, we know that this holds with probability $1 - 3n^{-8}$ for the components returned by \cref{alg:recoverStrongComponents}. Thus, with probability $1 - n^{-7}$, the above algorithm returns results from the same distribution as \cref{alg:LSS}, which we know returns a $(1 \pm \eps)$-sparsifier with probability $1 - n^{-8}$. Thus, with probability $\geq 1 - 4n^{-8}$, the above algorithm returns a $(1 \pm \eps)$-sparsifier for $H$.
\end{proof}

\begin{claim}\label{clm:finalSpace}
    \cref{alg:StrengthRecoverySparsification} can be implemented with a linear sketch of size $\widetilde{O}(nr \log(m) / \eps^2)$.
\end{claim}

\begin{proof}
    First, we consider \cref{alg:recoverStrongComponents}. By \cref{clm:boundedinprep}, each $\ell_0$-sampler must only be defined on a support of size $\poly(n)$. In each level of downsampling, we only require $\ell_0$-samplers sufficient for computing the connectivity structure of the hypergraph. From \cite{GMT15}, this can be done by storing $\log(n)$ $\ell_0$-samplers per vertex (with correlated randomness). Combined over the $\log(m)$ levels of downsampling and $n$ vertices, this means we must store $\widetilde{O}(n \log(m))$ $\ell_0$-samplers total, using $\widetilde{O}(nr \log(m))$ bits (where we have used that the support size is bounded by $\poly(n)$ to avoid an extra factor of $\log(m)$ in the representation size of each $\ell_0$-sampler).

    Next, we consider \cref{alg:LSSStrongComp}. By \cref{clm:boundedSupport}, it suffices to use $\ell_0$-samplers defined on a support of size $\text{poly}(n/\eps)$ and so each sketch for ConditionalEdgeRecovery requires only $\widetilde{O}(nr \log(1 / \delta) / \eps^2)$ bits. Taking the union of the $\log(m)$ sketches of ConditionalEdgeRecovery, and setting $\delta = \frac{\eps^2}{\text{poly}(n)}$, we then get our desired bound.
\end{proof}

\begin{theorem}\label{clm:finalMain}
There exists a linear sketch for arbitrary hypergraphs on $n$ vertices and $\leq m$ hyperedges, with arity $\leq r$ which recovers a $(1 \pm \eps)$ hypergraph sparsifier with probability $1 - 4n^{-8}$, using only $\widetilde{O}(nr \log(m) / \eps^2)$ space.
\end{theorem}

\begin{proof}
    This follows from \cref{clm:finalProb} and \cref{clm:finalSpace}. 
\end{proof}

\begin{remark}\label{rmk:thmforkcuts}
    Since our sparsification procedure preserves $k$-cuts (see  \cref{rmk:simpleforkcuts}) and our error accumulation analysis is already done with respect to any $k$-cut of the hypergraph $H$ (see \cref{clm:errorAccumulation}), it follows that the linear sketch from \cref{clm:finalMain} recovers a sparsifier that preserves the weight of every $k$-cut to within a $(1 \pm \eps)$-factor simultaneously for every $k \in [2..n]$.
\end{remark}

\section{Fingerprinting Approach to \cref{thm:recoveryProblem}}\label{sec:fingerprinting}

In this section, we will detail a \say{fingerprinting} approach towards proving \cref{thm:recoveryProblem}. 
As we will see, this fingerprinting allows us to implement the \say{recovery} step before computing our hypergraph decomposition. As mentioned before, the goal of this recovery step is to construct a linear sketch with only a near-linear number of $\ell_0$-samplers such that given a list of connected components $V_1, \dots V_k$, we can recover for each component $V_i$ either:
\begin{enumerate}
    \item All of the crossing hyperedges incident on $V_i$. 
    \item At least $\phi \log(n)$ distinct hyperedges for which $V_i$ is the unique representative (see \cref{def:uniqueRep}).
\end{enumerate}

We call this task the \say{recovery problem}. As we saw in the preceding section, this is \emph{sufficient} for computing a strength decomposition of the hypergraph, and ultimately calculating sampling rates, and thus constructing our sparsifiers. As discussed in the introduction, performing this recovery step is non-trivial, as large-arity hyperedges can correlate the $\ell_0$-samplers for different components, and thus the task of recovering unique representatives for components is not as simple as just opening that number of $\ell_0$-samplers. Thus, one of our key contributions is to introduce the notion of, and then analyze, fingerprinting of hyperedges.

\begin{definition}
    For a hypergraph $H = (V, E)$, and a hyperedge $e \in E$, we say that a random fingerprint of $e$ at rate $p$ is the result of independently keeping each vertex in $e$ with probability $p$. 
    We denote this fingerprinted version of $e$ by $\hat{e}$. We refer to the vertices in $\hat{e}$ as the \say{fingerprinted vertices} of $e$.
\end{definition}

Note that this operation can be implemented in a linear sketch. For each hyperedge, we can randomly sample its representatives and correspondingly update the $\ell_0$-samplers to use only the encoding of fingerprinted hyperedge $\hat{e}$ (using \cref{def:hyperedgeEncoding}).

\subsection{Conditional Algorithm}

In this section, we will present a linear sketching algorithm that solves the general hyperedge recovery problem conditioned on the existence of a specific linear sketch. We then show that such a linear sketch exists in the following subsection. This linear sketch uses two new definitions, which we describe below:

\begin{definition}
    We say that a hyperedge $e$ \textbf{touches} a component $V_i$, if $e \cap V_i \neq \emptyset$, and $e \cap (V - V_i) \neq \emptyset$.
\end{definition}

\begin{definition}
    We say that a hyperedge $e$ \textbf{places q vertices} in a component $V_i$ if $e \cap V_i = q$.
\end{definition}

We call this the \say{RestrictedRecovery} task:

\begin{lemma}[RestrictedRecovery]\label{clm:assumption}
    Consider a hypergraph $H$, and any partition into components $V_1, \dots V_k$. Suppose further that we are guaranteed there is a subset of the components, denoted $\{V_i \}_{i \in T}$, for $T \subseteq [k]$, where we are guaranteed that all the hyperedges incident on $\{V_i \}_{i \in T}$ are either touching $O(\log^2(n\phi))$ of the $\{V_i \}_{i \in T}$, or placing at most $O(\log^2(n\phi))$ vertices in each of the $\{V_i \}_{i \in T}$. Then, there exists an algorithm / linear-sketch RestrictedRecovery using only $\widetilde{O}(\phi \polylog(n))$ $\ell_0$-samplers for suitably restricted neighborhoods of each vertex, which returns a set of hyperedges $S$ such that for each $V_i \in \{V_i \}_{i \in T}$, either 
    \begin{enumerate}
        \item $S$ contains all incident hyperedges on $V_i$.
        \item $S$ contains $\Omega(\phi\log(n))$ hyperedges for which $V_i$ is the unique representative. 
    \end{enumerate}
\end{lemma}

Now, we will show that this RestrictedRecovery sketch lends itself towards a sketch solving the more general recovery problem of \cref{thm:recoveryProblem}. The intuition is that we create a sequence of hypergraphs that are fingerprinted in a \say{nested} manner. I.e., the $\ell$th hypergraph is the result of fingerprinting the $(\ell-1)$st hypergraph at rate $1/2$. Then, we show that if we work from the final hypergraph backwards (i.e., in the direction of less fingerprinting), the hypergraphs will inductively satisfy the necessary conditions for the RestrictedRecovery algorithm. 

\begin{algorithm}
\caption{Recovery$(H, \phi, (V_1, \dots V_k))$}\label{alg:Recovery}
    Initialize the components under consideration to be $\{ V_i \}_{i \in T_{\log(n)}}$, where $T_{\log(n)} = [k]$. \\
    Let $H_0 = H$ (no fingerprinting) and for $\ell = 0, 1, \dots \log(n)$, let $H^{(\ell)}$ be the result of fingerprinting $H^{(\ell-1)}$ at rate $1/2$. \\
    Let $S = \emptyset$ be the set of hyperedges recovered so far. \\
    \For{$\ell = \log(n), \dots 1, 0$}{
    $\hat{S} = \mathrm{RestrictedRecovery}(H^{(\ell)} - S, \phi, (V_1, \dots V_k))$. \\
    $S \leftarrow S \cup \hat{S}$.\\
    (For analysis, let $\{V_i\}_{i \in T_{\ell-1}}$ be the subset of $\{V_i\}_{i \in T_{\ell}}$ for which case 1 of \cref{clm:assumption} occurs.)
    }
    \Return{$S$}
    
\end{algorithm}

Note that we do not assume that the linear sketch from \cref{clm:assumption} needs to know which components are in the set $T$. It is simply given a guarantee that there is some such set $T$ for which the conditions hold, as this is purely a tool we use in analysis.

First, we prove some facts about the fingerprinting procedure. We will let $e^{\ell}$ denote the corresponding fingerprinted version of the hyperedge $e$ in the hypergraph $H^{\ell}$. Note that it may be the case that $e^{\ell}$ is empty or a singleton. 

\begin{claim}\label{clm:boundedVertices}
    Suppose the number of crossing hyperedges $e \in E[V_1, \dots V_k]$ is at most $\poly(n\phi)$. Then, with probability $1 - 2^{-\Omega(\log^2(n\phi))}$, for any such hyperedge $e$, and any component $V_i$ on which $e$ is incident, if $\ell'$ is the level of fingerprinting at which $e^{\ell'} \cap V_i = \emptyset$, at level $\ell' -1$, $|e^{\ell'-1} \cap V_i| \leq \log^2(n\phi)$.
\end{claim}

\begin{proof}
    Suppose for the sake of contradiction that $|e^{\ell'-1} \cap V_i| > \log^2(n\phi)$. Note that for each vertex in $e^{\ell'-1}$, we keep it with probability $1/2$ in the next level of fingerprinting. Thus, the probability that none of them survive for the next iteration is bounded by $2^{-\log^2(n\phi)}$. Taking the union bound over all $\poly(n\phi)$ hyperedges, and $\log(n)$ levels of fingerprinting, we conclude that the probability of ever going from $> \log^2(n\phi)$ vertices of $e^{\ell'-1}$ in $V_i$ to $0$ is bounded by $2^{-\Omega(\log^2(n\phi))}$.
\end{proof}

\begin{claim}\label{clm:boundedComponents}
    Suppose the number of crossing hyperedges $e \in E[V_1, \dots V_k]$ is at most $\poly(n\phi)$. Then, with probability $1 - 2^{-\Omega(\log^2(n\phi))}$, for any hyperedge $e \in E[V_1, \dots V_k]$, if we let $\ell'$ be the level of fingerprinting at which $|\{i: V_i \cap e^{\ell'} \neq \emptyset \} | \leq 1$, then $|\{i: V_i \cap e^{\ell'-1} \neq \emptyset \} | \leq \log^2(n\phi)$.
\end{claim}

\begin{proof}
        Suppose for the sake of contradiction that $|\{i: V_i \cap e^{\ell'-1} \neq \emptyset \} | > \log^2(n\phi)$. Note that for each vertex in $e^{\ell'-1}$, we keep it with probability $1/2$ in the next level of fingerprinting, and therefore each component $V_i$ should (independently) remain incident to $e^{\ell'}$ with probability $\geq 1/2$. Thus, the probability that all but $1$ of these components should no longer be incident after sampling is bounded by $2^{-\Omega(\log^2(n\phi))}$, and after taking a union bound over all $\leq n$ components, and $\poly(n\phi)$ hyperedges, we conclude the bound with probability $2^{-\Omega(\log^2(n\phi))}$.
\end{proof}

\begin{claim}\label{clm:innerLoopRecovery}
    In the inner loop of \cref{alg:Recovery}, the components $\{V_i \}_{i \in T_{\ell-1}}$ in the hypergraph $H^{\ell-1}-S$ satisfy the guarantees of \cref{clm:assumption}, that is, each hyperedge incident on any component in $\{V_i \}_{i \in T_{\ell-1}}$ is either crossing between $O(\log^2(n\phi))$ of the components, or places at most $O(\log^2(n\phi))$ vertices in each such component (with probability $1 - 2^{-\Omega(\log^2(n\phi))}$). 
\end{claim}

\begin{proof}
    First, let us consider the base case, when $\ell = \log(n)$. In this case, we are fingerprinting the hypergraph $H$ at rate $1/n$. Because there are only $n$ vertices in the hypergraph with probability $1 - 2^{- \Omega(\log^2(n\phi))}$, it follows that every hyperedge will have at most $\log^2(n\phi)$ vertices surviving the fingerprinting process. Necessarily then, for every component, hyperedges are both placing $\leq \log^2(n\phi)$ vertices in each component, and crossing between $\leq \log^2(n\phi)$ components. 

    Now, let us suppose that claim holds by induction down to $\ell$, and we will show it necessarily must hold for $\ell-1$. If it holds by induction down to level $\ell$, then for each component $V_i: i \in T_{\ell}$, we either recover $\Omega(\phi\log(n))$ hyperedges for which $V_i$ is the unique representative, or recover \emph{all} of the hyperedges incident on $V_i$ in $H^{\ell}$. If we are in the first case, we remove $V_i$ from $T_{\ell-1}$, and therefore it is not relevant to the inductive hypothesis.
    
    So instead, let us consider components in the second case, i.e. the components $V_i: i \in T_{\ell-1}$. Note that for these components, at the $\ell$th level of sampling, every crossing incident hyperedge to these components was recovered. Now, let us consider the hyperedges which are crossing between $V_i: i \in T_{\ell-1}$ in the \emph{unfingerprinted} hypergraph. There are two ways in which such a hyperedge $e$ can \emph{stop being a crossing} hyperedge after $\ell$ levels of fingerprinting. 
    
    The first way is that all of the hyperedges vertices in $V_i: i \in T_{\ell-1}$ have been removed (i.e. were not sampled) in the hypergraph $H^{(\ell)}$. That is, $\forall i \in T_{\ell-1}, |e^{\ell}\cap V_i| = 0|$. For any such hyperedge, by \cref{clm:boundedVertices}, in the $\ell-1$st level of fingerprinting, every component $V^{\ell-1}_i$ on which it is incident will have at most $\log^2(n\phi)$ vertices. 

    The second way for a hyperedge to no longer be crossing is if \emph{exactly} $1$ component (out of the components $V_i: i \in T_{\ell-1}$) which has a non-zero number of vertices in the hyperedge, and \emph{all} other $V_i: i \in [k]$ have an empty intersection. I.e., there is some component $V_i: i \in T_{\ell-1}$ for which $T \cap e^{\ell} \neq \emptyset$, yet no other component \emph{in the entire hypergraph} has a non-zero number of surviving vertices in the hyperedge (if any other component had a non-zero intersection, then the hyperedge would still be crossing in $H^{(\ell)}$). For any such hyperedge, by \cref{clm:boundedComponents}, it must be the case that in level $\ell-1$ of sampling, the hyperedge crosses between $\leq \log^2(n\phi)$ components.

    Thus, in either case, the components $V_i: i \in T_{\ell-1}$ in the hypergraph $H^{\ell-1}-S$ satisfy the guarantees of \cref{clm:assumption}.
\end{proof}

\begin{lemma}\label{lem:proofOfRecovery}
    For each component $V_i: i \in [k]$, \cref{alg:Recovery} returns either 
    \begin{enumerate}
        \item $\Omega(\phi\log(n))$ hyperedges for which $V_i$ is the unique representative.
        \item All incident hyperedges on $V_i$.
    \end{enumerate}
\end{lemma}

\begin{proof}
By \cref{clm:innerLoopRecovery}, we know that at every iteration of the inner loop, the components $V_i: i \in T_{\ell}$ satisfy the conditions of \cref{alg:Recovery}. Thus, for each such component, we either recover all of the neighboring hyperedges, or sufficiently many hyperedges for which it is the representative.

    Now, consider any of the original components $V_i: i \in [k]$. If for some value of $\ell $ $V_i$ is no longer one of the components $V^i: i \in T_{\ell}$, then this means in some iteration, we recovered $\Omega(\phi\log(n)))$ hyperedges for which $V_i$ is the unique representative, and therefore satisfies the first condition above. Otherwise, if we never recover $\Omega(\phi\log(n)))$ hyperedges for which $V_i$ is the unique representative, this must mean that in every iteration (including when $\ell = 0$), we recovered all incident hyperedges on $V_i$. In particular, when $\ell = 0$, we are doing no fingerprinting at all, so this means we must have recovered each of the original hyperedges incident on $V_i$ in the hypergraph $H$, yielding the above theorem. 
\end{proof}

\begin{proof}[Proof of \cref{thm:recoveryProblem}]
    By \cref{lem:proofOfRecovery}, \cref{alg:Recovery} is an algorithm satisfying the conditions of \cref{thm:recoveryProblem}. Further, the total space required by the sketch is $O(\log(n))$ independent copies of the linear sketch used by \cref{clm:assumption}, which by assumption uses only $O(\phi \polylog(n))$ $\ell_0$-samplers for suitably restricted neighborhoods of each vertex. This yields the claim.
    
    As an aside, note also that \cref{lem:proofOfRecovery} guarantees only $\Omega(\phi \log(n))$ recovered hyperedges, in order to get exactly $\phi \log(n)$, we can simply store a constant number of independent copies of the skech.
\end{proof}

Now, it remains to prove \cref{clm:assumption}. 

\subsection{Proof of \cref{clm:assumption} with Random Fingerprinting}

In this section, we will present a linear sketch / algorithm and analysis that achieves \cref{clm:assumption}. We will assume simply that we are given a hypergraph $H$ and connected components $V_1, \dots V_k$, and that there exists some subset of these components which we are interested in (for analysis). We denote this subset of components that we are interested in by $V_i: i \in T$. Our assumption tells us that for these components of interest, any hyperedge placing mass on these components is either (a) touching at most $\log^2(n\phi)$ of these components, or (b), placing at most $\log^2(n\phi)$ vertices in each component. We call hyperedges in case (a) \textbf{Type I hyperedges}, and hyperedges in case (b) \textbf{Type II hyperedges}.

With this, we will introduce some terminology which will be essential in our analysis. 

\begin{definition}
    For a component $V_i$ in the hypergraph $H$, we say $\mathrm{deg}(V_i) = |\{e \in H: e \cap V_i \neq \emptyset \}|$.
\end{definition}

\begin{definition}\label{def:Ws}
    For a range of degrees $[d, 2d]$, we let $V_i: i \in T^{(d)}$ denote the corresponding subset of $V_i: i \in T$ with degree in that range, and for which we have not yet recovered $\Omega(\log(n))$ hyperedges for which they are the unique representative. Note that these are continuously re-defined with respect to the hypergraph $H$, as when we recover hyperedges and remove them from $H$, the degree will necessarily decrease. 
\end{definition}

\begin{definition}
    For a parameter $j \in \N$, we say that $E_j^{(d)}$ consists of hyperedges crossing between $[j, 2j]$ of the components $V_i: i \in T^{(d)}$.
\end{definition}

\begin{definition}
    We say that $D^{(d)} = \sum_{e \in E[V_1, \dots V_k]} |\{i \in T^{(d)}:V_i \cap e \neq \emptyset \}|$, and likewise, $D^{(d)}_j = \sum_{e \in E_j^{(d)}} |\{i \in T^{(d)}:T_i \cap e \neq \emptyset \}|$.
\end{definition}

By definition, it follows that $D^{(d)} = \sum_{\log(j) = 0}^{\log(n)} D^{(d)}_j$. Additionally, note that $D^{(d)} = \sum_{i \in T^{(d)}} \deg(V_i)$, as we are counting each hyperedge with multiplicity of the number of components $V_i: i \in T^{(d)}$ that it touches. 

\begin{remark}\label{rmk:PHP}
    For any $d$, there exists a value of $j \in \{1, 2, 4, \dots n/2 \}$ for which $D^{(d)}_j \geq D^{(d)} / \log(n) \geq \frac{d |T^{(d)}|}{\log(n)}$. This follows from the PHP and the relation to total degree. As a consequence, for this value of $j$, there must be at least $\frac{|T^{(d)}|}{4\log(n)}$ components $V_i: i \in T^{(d)}$, each of which is incident upon $\frac{d}{4\log(n)}$ hyperedges from $E^{(d)}_j$.
\end{remark}

As stated above, we know there must exist some value of $d$ for which components of degree $[d, 2d]$ constitute an $\Omega( 1 /\log(n))$ fraction of the total degree. For this value of $d$, we also know there must be some value of $j$ for which an $\Omega (1 /\log(n))$ fraction of the hyperedges are crossing between $[j, 2j]$ of these components of degree $[d, 2d]$. Using this, we will show that there is in fact an explicit, good rate for fingerprinting which will ensure that we make progress when opening our $\ell_0$-samplers. Intuitively, by our assumption, we know that hyperedges can only be type I hyperedges or type II hyperedges (before fingerprinting). If a hyperedge is a type I hyperedge, then the analysis is very easy. Any such hyperedge is crossing between $O(\log^2(n\phi))$ components, meaning we can essentially think of such a hyperedge as having arity bounded by $O(\log^2(n\phi))$. In general, such small arity hypergraphs are not too different than graphs, and just by storing an extra factor of $O(\log^2(n\phi))$ $\ell_0$-samplers, we will be able to perform the recovery step. The more nuanced analysis happens for type II hyperedges. Here, we use the fact that if a type II hyperedge is crossing between $[j, 2j]$ components, then the right fingerprinting rate is roughly $\frac{1}{j}$. We show that indeed, if we fingerprint ($\polylog(n)$ times) at this rate, then indeed we will recover sufficiently many such hyperedges with high probability. 

We make this formal below:

\begin{claim}\label{clm:boundedDegreeRecovery}
    Let $H$ be a hypergraph with a decomposition into components $V_1, \dots V_k$. Suppose that $V_i: i \in T$ is a subset of these components satisfying the conditions of \cref{clm:assumption}. For any choice of $d$, let $V_i: i \in T^{(d)}$ be defined as in \cref{def:Ws}.
    If we repeatedly fingerprint $H$ at rate $\frac{\log^2(n\phi)}{j}$ for $j$ as defined in \cref{rmk:PHP}, and open $\ell_0$-samplers for each $V_1, \dots V_k$ $\phi \log^{10}(n\phi)$ times (after each round of opening samplers, removing the hyperedges that were recovered from future samplers), then with probability $1 - 2^{-\Omega(\log^2(n\phi))}$ either 
    \begin{enumerate}
        \item $D^{(d)}$ decreases by a factor of $(1 - 1 / (2^{12} \log^5(n)))$.
        \item At least a $1 / 8\log(n)$ fraction of the components $V_i: i \in T^{(d)}$ will have recovered $\Omega(\phi \log(n))$ hyperedges for which they are the unique representative.
        \item At least a $1 / 8\log(n)$ fraction of the components $V_i: i \in T^{(d)}$ will have recovered a $1 / 8\log(n)$ fraction of \emph{all} their incident hyperedges.
    \end{enumerate}
\end{claim}

\begin{proof}
    First, by \cref{rmk:PHP}, there must exist a value of $j$ for which $D^{(d)}_j \geq D^{(d)} / \log(n)$. 
    
    As a consequence, this condition means that there must exist $\geq \frac{|T^{(d)}|}{4 \log(n)}$ components $V_i: i \in T^{(d)}$ each of which is touching at least $\frac{d}{4 \log(n)}$ hyperedges from $E^{(d)}_j$. We denote this subset of $ T^{(d)}$ by $\widetilde{T}^{(d)}$.
    
    Now, we remark that if $j \leq \log^2(n\phi)$, we do not need to perform any fingerprinting. This is because there would exist $\geq \frac{|T^{(d)}|}{4 \log(n)}$ components, each of which is receiving $\geq 1 / (4 \log(n))$ fraction of its degree from edges in $E^{(d)}_j$. For this value of $j$, each such hyperedge is touching at most $2\log^2(n\phi)$ of the components $V_i: i \in \widetilde{T}^{(d)}$. Thus, after opening $\phi \log^{10}(n\phi)$ (correlated) $\ell_0$-samplers for each component, there are two cases:
    \begin{enumerate}
        \item For a component $V_i, i \in \widetilde{T}^{(d)}$ all the $\ell_0$-samplers returned incident hyperedges (i.e., the fingerprinted hypergraph always has incident hyperedges on $V_i$). Then the process has recovered $\phi \log^{10}(n\phi)$ distinct hyperedges incident on $V_i$. Because each $\ell_0$-sampler is receiving \emph{uniformly random} samples from the neighborhood of $V_i$, this means we receive a random sample of $\phi\log^{10}(n\phi)$ of the incident hyperedges on $V_i$. Further, since we know that a $\geq 1 / (4 \log(n))$ fraction of the incident hyperedges on $V_i$ touch at most $2\log^2(n\phi)$ components, this means that in expectation we recover at least $\phi\log^9(n\phi) / 4$ hyperedges which are incident on at most $2\log^2(n\phi)$ components. With probability $> 1 - 2^{-\phi\log^2(n)}$ then, we recover at least $\phi\log^8(n\phi)$ hyperedges which are touching at most $2\log^2(n\phi)$ components. For each, we simply choose one of the incident components $V_i$ at random to be the unique representative for the hyperedge. Thus, with probability $> 1 - 2^{-\phi\log^2(n\phi)}$, we will recover at least $\phi\log(n)$ hyperedges for which $V_i$ is the unique representative.
        \item For a component $V_i, i \in \widetilde{T}^{(d)}$, not all the $\ell_0$-samplers returned incident hyperedges. This must mean we have recovered the entire neighborhood of $V_i$, as we have not done any fingerprinting. 
    \end{enumerate}

    Thus, we may assume that $j > \log^2(n\phi)$. 
    
    Now, let us fingerprint hyperedges at rate $\frac{\log^2(n\phi)}{j}$. We consider two distinct cases:  
    \begin{enumerate}
        \item The first case is when $\frac{d \log^2(n\phi)}{j} \leq 1/2$. Note that as an immediate consequence, because each component $V_i: i \in \widetilde{T}^{(d)}$ has degree $\leq 2d$, the number of Type II hyperedges (those placing mass $\leq \log^2(n\phi)$ on each component) is bounded by $\log^2(n\phi)$ with probability $1 - 2^{- \Omega(\phi\log^2(n))}$. This is because there can be at most $d \log^2(n\phi)$ vertices from Type II hyperedges in each $V_i$ for $i \in \widetilde{T}^{(d)}$, and thus when fingerprinting at rate $\log^2(n\phi) / j$, the expected number of vertices (and thus an upper bound on the number of hyperedges) in the fingerprinted hyperedges is bounded by $\log^2(n\phi) /2$.

        Next, we break the components into two parts. Let $V_i: i \in \widetilde{T}^{(d, \geq \phi\log^4(n\phi))}$ denote the subset of $V_i:i \in \widetilde{T}^{(d)}$ for which there are more than $\phi\log^4(n\phi)$ Type I hyperedges that remain incident in expectation when sampling at rate $\frac{\log^2(n\phi)}{j}$, and let $V_i: i \in \widetilde{T}^{(d, < \phi\log^4(n\phi))}$ denote the subset of components for which there are less than $\phi\log^4(n\phi)$ Type I hyperedges that remain incident in expectation when sampling at rate $\frac{\log^2(n\phi)}{j}$. There are two cases here:
        \begin{enumerate}
            \item The components $V_i: i \in \widetilde{T}^{(d, \geq\phi \log^4(n\phi))}$ make up at least half of the components $V_i:i \in \widetilde{T}^{(d)}$. If this is the case, note that for each component $V_i: i \in \widetilde{T}^{(d, \geq\phi \log^4(n\phi))}$, in each round of fingerprinting, there are $\geq \phi\log^4(n\phi)/2$ Type I hyperedges that are incident on $V_i$ with probability $\geq 1 - 2^{-\Omega(\phi\log^4(n)}$ (by assumption, the expectation is this large). 
            
            Because there are at most $\log^2(n\phi) /2$ Type II hyperedges with probability $\geq 1 - 2^{-\Omega(\log^2(n\phi)}$, this means in the first $\phi\log^4(n\phi)/2$ rounds of opening $\ell_0$-samplers, with probability $\geq 1 - 2^{-\Omega(\log^2(n\phi))}$, we will see $\Omega(\phi\log^4(n\phi))$ Type I hyperedges that are incident on $V_i$ as long as the number of type I hyperedges incident has not decreased below $\log^2(n\phi)$ (in this case, we simply move component $V_i$ to $\widetilde{T}^{(d, < \phi\log^4(n\phi))}$). Otherwise, by choosing unique representatives for each such hyperedge at random, we will find $\geq \phi\log(n)$ hyperedges for which $V_i$ is the unique representative. If this happens, then for a $\geq 1 / (8 \log(n))$ fraction of our original components, we have recovered $\Omega(\phi\log(n))$ hyperedges for which they are the unique representative, placing us in condition 2. 
            \item The components $V_i: i \in \widetilde{T}^{(d, < \phi\log^4(n\phi))}$ make up at least half of the components $V_i: i \in \widetilde{T}^{(d)}$. Note that because we are assuming more than half of the $V_i: i \in \widetilde{T}^{(d)}$ satisfy this condition, this means there must be $\geq \frac{|T^{(d)}|}{8 \log(n)}$ such components, each of which has $\geq \frac{d}{4 \log(n)}$ hyperedges from $E_j^{(d)}$. In particular, these components capture at least a $\frac{1}{32 \log^2(n)}$ fraction of the degree of $D^{(d)}_j$. Thus, at least $\frac{1}{128\log^2(n)}$ of the edges in $E_j^{(d)}$ must have $\geq \frac{1}{64 \log^2(n)}$ fraction of their degree coming from components $V_i: i \in \widetilde{T}^{(d, < \phi \log^4(n\phi))}$. We denote this subset of $E_j^{(d)}$ by $E_j^{(d, < \phi\log^4(n\phi))}$. 
            
            Now, consider any hyperedge $e \in E_j^{(d, <\phi\log^4(n\phi))}$. We want to analyze the probability that $e$ is recovered in one round of fingerprinting. To do this, first we note that any hyperedge in $E_j^{(d, <\phi\log^4(n\phi))}$ must be a Type II hyperedge (one that places $< \log^2(n\phi)$ vertices in each component $V_i: i \in \widetilde{T}^{(d)}$, as $j > \log^2(n\phi)$). After fingerprinting at rate $\log^2(n\phi) / j$, any such hyperedge is still crossing between at least $2$ components $V_i: i \in \widetilde{T}^{(d, < \phi\log^4(n\phi))}$ with probability $1 - 2^{-\Omega(\log^2(n\phi))}$ by a Chernoff bound. 

            Next, we observe that for each $e \in E_j^{(d, \leq \phi\log^4(n\phi))}$, it places vertices in $\geq \frac{j}{64 \log^2(n)}$ of the components $V_i: i \in \widetilde{T}^{(d, < \phi\log^4(n))}$. Thus, when we fingerprint at rate $\log^2(n\phi) / j$, the probability that some component $V_i: i \in \widetilde{T}^{(d, < \phi\log^4(n\phi))}$ is still incident to $e$ is $\Omega(1)$. But, the total degree of $V_i: i \in \widetilde{T}^{(d, < \phi\log^4(n\phi))}$ is bounded by $\phi\log^4(n\phi) + \log^2(n\phi)$ (the total number of Type I and Type II hyperedges that can be incident) with probability $1 - 2^{- \Omega(\log^2(n\phi))}$. So, in each round of fingerprinting, $e$ has a $\geq \Omega(1 / \phi\log^4(n\phi))$ chance of being recovered. After repeating this $\phi\log^{10}(n\phi)$ times, we are guaranteed that with probability $1 - 2^{- \Omega(\log^2(n\phi))}$, at least $1/2$ of the edges in $E_j^{(d, <\phi\log^4(n\phi))}$ have been recovered. This captures a $\geq \frac{1}{2^{12} \log^5(n)}$ fraction of $D^{(d)}$, and therefore we end up satisfying condition 1 of the claim we are proving.
        \end{enumerate}
        
        \item Next, we consider the case when $\frac{d \log^2(n\phi)}{j} > 1/2$. Note that as a consequence, in each component $V_i: i \in T^{(d)}$, after fingerprinting we expect at least $1 / 8 \log(n)$ Type II hyperedges to be incident, as each component $V_i: i \in \widetilde{T}^{(d)}$ has at least $d / 4 \log(n)$ edges from $E_j^{(d)}$ incident. As before, we again have two cases for each component $V_i: i \in \widetilde{T}^{(d)}$. 

        \begin{enumerate}
                \item A $> 1 / \log^2(n\phi)$ fraction of $\ell_0$-samplers returned incident hyperedges. This means we recovered $> \phi\log^8(n\phi)$ incident hyperedges. Either $\phi\log^8(n\phi) / 2$ of them must be Type I hyperedges (in which case we are able to choose unique representatives at random, yielding $\Omega(\phi\log(n))$ hyperedges for which this component is the unique representative), or $\phi\log^8(n\phi) / 2$ of them must be Type II hyperedges. We know that among type II hyperedges, in expectation a $\geq 1 / 8 \log(n)$ fraction of them are in $E_j^{(d)}$. Thus, we recover $\Omega(\phi\log^7(n\phi))$ edges from $E_j^{(d)}$ with probability $1 - 2^{-\Omega(\phi\log^7(n))}$. After fingerprinting, each such edge is crossing between $O(\log^4(n\phi))$ components, and we can simply choose a unique representative at random among these. Thus, for the component $V_i$, we recover $\Omega(\phi\log(n))$ hyperedges for which it is the unique representative.
                \item A $< 1 / \log^2(n\phi)$ fraction of $\ell_0$-samplers returned incident hyperedges. Initially, just from $E_j^{(d)}$, we would have expected that with probability $> 1 / 8 \log(n)$ fraction the first $\ell_0$-sampler would return an incident hyperedge. We claim that in order for a $< 1 / \log^2(n\phi)$ fraction of $\ell_0$-samplers to return incident hyperedges, by the end of the $\phi\log^{10}(n\phi)$ $\ell_0$-samplers, we must have recovered at least half of the edges in $E_j^{(d)}$ incident on $V_i$. Indeed, suppose not. Then, by the final iteration, we still expect $1 / 16 \log(n\phi)$ hyperedges to be incident after each round of fingerprinting. Because a hyperedge surviving fingerprinting is simply a Bernoulli random variable, this means that with probability $\Omega(1 /  \log(n\phi))$ we expect at least one hyperedge to survive fingerprinting. But, the probability that we would then only see $<\phi\log^8(n\phi)$ hyperedges sampled out of $\phi\log^{10}(n\phi)$ rounds is bounded by $1 - 2^{-\Omega(\phi\log^9(n\phi))}$. Thus, with high probability, it is the case that we have recovered at least half of the edges in $E_j^{(d)}$ incident on $V_i$. Consequently, because $E_j^{(d)}$ contributed an $\Omega(1 / \log(n))$ fraction of the degree for $V_i$, we have recovered at least an $\Omega(1 / \log(n))$ of the incident hyperedges on $V_i$, placing us in case 3.
            \end{enumerate}

            Note now that either half of the $V_i$ fall in case a or in case b. Either way, this constitutes a $1 / 8 \log(n)$ fraction of the original components $V_i: i \in T^{(d)}$ satisfying either Condition 2 or 3 of the stated claim. This concludes the proof.

    \end{enumerate}
\end{proof} 

Unfortunately, we do not know a priori what the best sampling rate is (i.e., the rate calcualted in the previous claim). So, instead we simply range over all choices of sampling rates, and are guaranteed that for some choice of this sampling rate, we will have recovered sufficiently many hyperedges.

With this, we now present a building block of the algorithm we will analyze. 

\begin{algorithm}[H]
    \caption{IterativeRecovery}\label{alg:IterationRecovery}
    Let $V_1, \dots V_k$ be the set of components. \\
    Initialize $S$ to be the set of hyperedges recovered so far. \\
    \For{$i \in [\phi\log^{10}(n\phi)]$}{
    \For{$p \in \{1, 1/2, 1/4, \dots 1/n\}$}{
    Fingerprint each hyperedge at rate $p$. \\
    Remove the hyperedges in $S$ from each of the relevant $\ell_0$-samplers with this fingerprinting scheme.\\
    \For{each $V_i$}{
    Add together the $\ell_0$-samplers (with correlated randomness) for the vertices in $V_i$. \\
    Open the $\ell_0$-sampler and add the corresponding edge to $S$ (if not already there). \\
    }
    }
    }
\end{algorithm}

\begin{remark}\label{rmk:supersetEdges}
    Note that \cref{alg:IterationRecovery} tries fingerprinting at \emph{all} possible rates $p$. In particular, a subset of the fingerprinting it does is at the optimal rate $\log^2(n\phi) / j$ (or within a factor of $2$). Thus, the hyperedges recovered by \cref{alg:IterationRecovery} are a superset of the hyperedges needed to argue the claim in \cref{clm:boundedDegreeRecovery}.
\end{remark}

Now, we repeat this algorithm many times as a sub-routine to get our final algorithm. 

\begin{algorithm}
    \caption{IterativeRecovery}\label{alg:IterativeRecovery}
    Let $V_1, \dots V_k$ be the set of components. \\
    Initialize $S$ to be the set of hyperedges recovered so far. \\
    \For{$i \in [2^{20}\phi\log^{16}(n\phi)]$}{
    \For{$p \in \{1, 1/2, 1/4, \dots 1/n\}$}{
    Fingerprint each hyperedge at rate $p$. \\
    Remove the hyperedges in $S$ from each of the relevant $\ell_0$-samplers with this fingerprinting scheme.
    \For{each $V_i$}{
    Add together the $\ell_0$-samplers (with correlated randomness) for the vertices in $V_i$. \\
    Open the $\ell_0$-sampler and add the corresponding edge to $S$ (if not already there). \\
    }
    }
    }
\end{algorithm}

\begin{corollary}\label{cor:proofofassumption}
    If one runs \cref{alg:IterativeRecovery} on a hypergraph $H$ with components $V_1, \dots V_k$, and some subset of the components $V_i: i \in T$ satisfying the conditions of \cref{clm:assumption}, then with probability $1 - 2^{-\Omega(\log^2(n\phi))}$, for any component $V_i: i \in T$ we either recover
    \begin{enumerate}
        \item $\Omega(\phi\log(n))$ crossing hyperedges for which $V_i$ is the unique representative.
        \item All of the hyperedges incident upon $V_i$.
    \end{enumerate}
\end{corollary}

\begin{proof}
    Let us start by considering the largest remaining value of $d$ as well as the components $V_i: i \in T^{(d)}$. To start, this is bounded by $d = n^{100}\phi$ (the largest value of $d$ that we will ever encounter by \cref{clm:estimatingComponents}). We then run the \cref{alg:IterationRecovery} $2^{12} \log^5(n\phi)$ times. Note that after each time we run \cref{alg:IterationRecovery}, the components $V_i: i \in T^{(d)}$ are re-defined, as some components may now have degree below $d$. We denote the subset of $T^{(d)}$ the remains in the $p$th iteration by $T^{(d,p)}$ for $p \in [2^{14} \log^5(n\phi)]$.
    
    At this point, we will be guaranteed that either case 1, 2, or 3 of \cref{clm:boundedDegreeRecovery} has occurred $2^{12} \log^5(n\phi)$ times. Thus, either $D^{(d)}$ has gone to $0$, or for the remaining components $V_i: i \in T^{(d,2^{12} \log^5(n\phi))}$ either $V_i$ has recovered at least $1/2$ of its incident edges (meaning that now it will be paired into the next group of components with degree $d/2$), or $V_i$ has recovered at least $\Omega(\phi\log(n))$ distinct crossing hyperedges for which it is the unique representative. For any components in the last case, we simply remove these components from consideration, as they have recovered sufficiently many hyperedges. In the first two cases, the degrees of the components must have decreased, and therefore will be lumped in with the remaining lower-degree components. 

    Note then, that after repeating this for $100 \log(n\phi)$ rounds, the degree of the components under consideration must have gone down to 0. Thus, the components under consideration must have had all of their hyperedges recovered, whereas the components removed from consideration must have at least $\log(n)$ hyperedges for which they are the unique representative. 
    
    The probability bound follows from the fact that we run the algorithm \cref{alg:IterationRecovery} $\polylog(n\phi)$ times, and each round has a failure probability of $2^{-\Omega(\log^2(n\phi))}$. Our stated claim thus follows immediately. 
\end{proof}

Therefore, \cref{alg:IterativeRecovery} is a constructive algorithm which achieves the needs of \cref{clm:assumption}.

Further, we can implement \cref{alg:IterativeRecovery} using only $n \phi\polylog(n\phi)$ $\ell_0$-samplers.

\begin{claim}\label{clm:spaceBound}
    \cref{alg:IterativeRecovery} requires storing only $\widetilde{O}( \phi \polylog(n\phi))$ $\ell_0$-samplers per vertex, each initialized for a suitably restricted subset of the neighborhood.
\end{claim}

\begin{proof}
    For $\phi\polylog(n\phi)$ iterations, \cref{alg:IterativeRecovery} samples from the neighborhood of each component $V_i$ (after fingerprinting). For each iteration, this requires only storing correlated $\ell_0$-samplers for each vertex in the fingerprinted version of the hypergraph.  In order to sample according to a specific component $V_i$, we must only add together the corresponding samplers for each vertex in $V_i$. Thus, because there are only $\phi\polylog(n\phi)$ iterations, this can be done using only $\widetilde{O}(\phi\polylog(n\phi))$ $\ell_0$-samplers per vertex. 
\end{proof}

\begin{proof}[Proof of \cref{clm:assumption}]
    \cref{alg:IterativeRecovery} is an algorithm satisfying the conditions of \cref{clm:assumption}. The correctness follows by \cref{cor:proofofassumption}, and the space bound follows from \cref{clm:spaceBound}.
\end{proof}

This concludes the section, as in this subsection we proved \cref{clm:assumption}, and in the previous subsection, showed that \cref{clm:assumption} can be used to prove \cref{thm:recoveryProblem}.

\section{Lower Bounds on Linear Sketches for Hypergraph Sparsifiers}\label{sec:lowerBound}

\subsection{Preliminaries}

In this section we will show that in fact, any linear sketch for an arbitrary hypergraph $H$ on $\leq m$ edges, and arity $\leq r$ which can be used to recover a $(1 \pm \eps)$ sparsifier for $H$ (with high probability) must use $\widetilde{\Omega}(nr\log(m))$ bits. To do this, we will consider a modification of the following well-known one-way communication problem with public randomness known as the universal relation problem:

\begin{enumerate}
    \item Alice is given a vector $x_A \in \zo^{2^r}$, and must send a possibly randomized encoding of $x_A$ to Bob.
    \item Bob is given a vector $x_B \in \zo^{2^r}$ with the promise that $\Supp(x_B) \subset \Supp(x_A)$, and must return an index $i$ such that $(x_A)_i \neq (x_B)_i$ with probability $1 - 1 / \text{poly}(r)$.
\end{enumerate}

The work of \cite{KNPWWY17} defined a variant of the above problem which has strong lower bounds, and will be of interest to us.

We denote this variant by $k$-$\UR_r$, and define it formally below:
\begin{enumerate}
    \item Alice is given a string $x_A \in \zo^{2^r}$. Bob is given a string $x_B \in \zo^{2^r}$ such that $|\Supp(x_A) - \Supp(x_B)| \geq k$. Alice sends only a message $\calS(x_A)$ to Bob (using public randomness).
    \item Bob has his own string $x_B$ with the promise that $\Supp(x_B) \subset \Supp(x_A)$, and receives Alice's message $\calS(x_A)$. Using this (and access to public randomness), he must return $k$ indices $i: (x_A)_i \neq (x_B)_i$ with probability $1 - 1 / r^{5}$.
\end{enumerate}

The following is known from \cite{KNPWWY17}:

\begin{theorem}\cite{KNPWWY17}
    The one-way communication complexity of $k$-$\UR_n$ (with public randomness) is $\Omega(kr^2)$.
\end{theorem}

However, this still does not suffice for us, as ideally we should have a bound on the support size (to mimic the bound on the number of hyperedges in the hypergraph). So, we make use of the following communication problem building on top of $k$-$\UR_r$, which we denote by $k$-$\UR_r^{\leq m}$:

\begin{enumerate}
    \item Alice is given a string $x_A \in \zo^{2^r}$. Bob is given a string $x_B \in \zo^{2^r}$ such that $m \geq |\Supp(x_A) - \Supp(x_B)| \geq k$. Alice sends only a message $\calS(x_A)$ to Bob (using public randomness).
    \item Bob has his own string $x_B$ with the promise that $\Supp(x_B) \subset \Supp(x_A)$, and receives Alice's message $\calS(x_A)$. Using this (and access to public randomness), he must return $k$ indices $i: (x_A)_i \neq (x_B)_i$ with probability $1 - 1 / r^{5}$.
\end{enumerate}

We will show the following:

\begin{theorem}
    The one-way communication complexity of $k$-$\UR_r^{\leq m}$ (when $m \geq \max(2k, \log^5(r))$) with failure probability $1 - 1 / (2r^{6})$ is $\Omega(kr \log(m/k))$.
\end{theorem}

\begin{proof}
    To prove this, we will show that with $O(r / \log(m/k))$ simultaneous instances of $k$-$\UR_r^{\leq m}$ one can solve $k$-$\UR_r$. It follows then that Alice's message for each instance of $\UR_r^{\leq m}$ requires $\Omega(kr \log(m/k))$ bits, as otherwise this would yield a contradiction to the complexity of $k$-$\UR_r$.

    So, let Alice be given an instance of $k$-$\UR_r$, with her vector $x_A$. Then, Alice makes the following set of $\Theta(r / \log(m/k))$ instances of $k$-$\UR_r^{\leq m}$: 
    for $i = 1, \dots 2r / \log(m/k)$, let $h^{(i)}$ be a uniformly random, independent hash function (from the shared randomness) such that $\forall k \in [2^r], h^{(i)}(k) = 1$ with probability $1 / \sqrt{m/k}$ (and is $0$ otherwise). Let $P^{(i)} \subseteq [2^r]$ be defined as $P^{(i)} = \{\ell \in [2^r]: \prod_{j = 1}^{i-1}h^{(j)}(\ell) = 1  \}$. Let $(x_A)|_{P^{(i)}}$ refer to the the vector in $\zo^{2^r}$, which is obtained by setting to $0$ all the corresponding entries of $(x_A)$ that are at indices not in $P^{(i)}$. Now, let $\calS^{\leq m}$ be the encoding function that Alice uses for instances of $k$-$\UR_r^{\leq m}$. Alice sends the encodings $\calS^{\leq m}((x_A)|_{P^{(i)}})$ for each $i$ to Bob as well as $|\Supp(x_A)|$ to Bob.

    Now, we will show how Bob can use this to recover a solution to the original instance of $k$-$\UR_r$ with high probability. Using the shared randomness, Bob makes $(x_B)|_{P^{(i)}}$ in an analogous manner to Alice (using the same hash functions) and also calculates $|\Supp(x_A)| - |\Supp(x_B)|$ (which we are promised is at least $k$ by the hypothesis of the $k$-$\UR_r$ instance). If $|\Supp(x_A)| - |\Supp(x_B)| \leq m$, it follows that Bob can simply use the full vectors $(x_B)|_{P^{(1)}}$ and $\calS^{\leq m}((x_A)|_{P^{(1)}})$ to recover $k$ indices which is in $\Supp(x_A) - \Supp(x_B)$, as this will then satisfy the requirement of being an instance of $k$-$\UR_r^{\leq m}$. Otherwise, let $W$ denote $|\Supp(x_A)| - |\Supp(x_B)|$. At the $i$th level of downsampling, $|\Supp(((x_A)|_{P^{(i)}}))| - |\Supp((x_B)|_{P^{(i)}})|$ is distributed as $\text{Binomial}\left(W, \frac{1}{\sqrt{m/k}^{i}}\right)$. It follows that there must exist an $i$ such that $k \leq k \cdot (m/k)^{1/4} \leq \mathbb{E}[\text{Binomial}\left(W, \frac{1}{\sqrt{m}^{i}}\right)] \leq k \cdot (m/k)^{3/4} \leq m$. Thus, by a Chernoff bound, for this value of $i$, there will be at least $k$, and at most $m$ indices in $\Supp(((x_A)|_{P^{(i)}})) - \Supp((x_B)|_{P^{(i)}})$ with probability $1 - 2^{-m^{1/4}}$, and thus the corresponding instance of $k$-$\UR_r^{\leq m}$ must return $k$ valid indices in $\Supp(((x_A)|_{P^{(i)}})) - \Supp((x_B)|_{P^{(i)}})$ (which are thus also a valid indices in $\Supp(x_A) - \Supp(x_B)$). Under the condition that $m \geq \log^5(r)$, the success probability is then $\geq 1 - 1/(2r^{6}) - 1 / (r^{5}) = 1 - r^{-6}$, as we desire. 

    Note that the entire size of the sketches used is $\Theta(r / \log(m/k))$ messages for $\UR_r^{\leq m}$, and a single message of size $\leq 2r$ for the size of $|\Supp(x_A)|$. In total then, the size of Alice's message is $\leq 2r + \Theta(r / \log(m/k)) \cdot |k$-$\UR_r^{\leq m}|$. It follows that $|k$-$\UR_r^{\leq m}| \geq \Omega(kr \log(m/k))$, as otherwise this leads to a contradiction with the fact that the problem $k$-$\UR_r$ (with failure probability $1 - 1/r^{6}$) requires messages of size $\Omega(kr^2)$.
\end{proof}

\subsection{Lower Bound}

Now, we are ready to relate the above problem to the problem of creating general hypergraph sparsifiers. In particular, we will show that with $O(\log(n))$, linear sketches of hypergraph sparsifiers on a specific family of hypergraphs (and $\eps < 1$), we can solve the above communication problem. Using the lower bound for the communication problem for $k = n/2$, this then gives us a lower bound on the size of valid linear sketches for hypergraph sparsifiers.

\begin{theorem}
    The linear sketching complexity of $(1 \pm \eps)$ hypergraph sparsification (for $\eps$ constant) on $n$ vertices with $\leq m$ hyperedges, maximum arity $r$ and success probability at least $1 - 1 / n^{7}$ is $\Omega(nr \log(m/n) / \log(n))$.
\end{theorem}

\begin{proof}
We prove this by giving a one-way public randomness communication protocol using linear sketches of hypergraph sparsifiers that solves $(n/2)$-$\UR_{r/2}^{\leq m}$. Indeed, consider an instance $I = (x_A, x_B)$ of $(n/2)$-$\UR_{r/2}^{\leq m}$. We claim that with $100 \log(n)$ linear sketches of hypergraph sparsifiers (each hypergraph with $\leq m$ hyperedges), Alice can send a single message consisting of these linear sketches to Bob, after which he can recover $n/2$ indices such that $(x_A)_i \neq (x_B)_i$. We construct the hypergraphs as follows: for $j = 1, \dots 100 \log(n)$, let $P_j = (S_1, \dots S_{n/2})$ be a (random) partition of $[2^{r/2}]$ into $n/2$ equal sized parts. For each integer in $[2^{r/2}]$, let us bijectively associate it with a subset of $[r/2]$. When we refer to a set $T \subseteq [r/2]$, we will both refer to the subset itself, as well as the corresponding integer in $[2^{r/2}]$. Now, Alice creates the hypergraph $H_j$ on the vertex set $L \cup R$, where $|L| = |R| = n/2$. For each left vertex $v \in [n/2]$, and for each index $T \in S_v$ such that $(x_A)_T = 1$, Alice adds the hyperedge $(v, T)$ to the hypergraph (where $v$ is understood to be in $L$, and $T$ is understood to be $\subseteq R$ - this is a hyperedge of arity $\leq r/2+1$). Now, Alice creates a linear hypergraph sparsifier sketch for each hypergraph $H^{j}_A$ (using different randomness for each one). We denote these sketches by $\calS(H^j_A)$, and sends these to Bob.

Bob receives $\calS(H^j_A)$ for $j = 1, \dots 100 \log(n)$, and wants to recover $n/2$ indices solving the original $(n/2)$-$\UR_{r/2}^{\leq m}$ instance. To do this, Bob uses the shared randomness to create the same partitions $P_j$ as Alice. Likewise, he uses the shared randomness as well as his own string $x_B$ to create his own hypergraphs $H^j_B$, as well as the linear hypergraph sparsifier sketches $\calS(H^j_B)$. Now, by linearity (and using the fact these are instantiated with shared randomness), Bob can subtract his sketches from Alice's to get sketches for $\calS(H^j_A - H^j_B)$. 

Now, let us consider the case when $j = 1$. Bob will open the sketch $\calS(H^j_A - H^j_B)$ and recover a sparsifier for $H^j_A - H^j_B$ with probability $1 - 1 / n^{10}$. We will make use of the following claim:

\begin{claim}
    Let $x_A, x_B \in \zo^{2^r}$ such that $\Supp(x_A) \subseteq \Supp(x_B)$, and let $k = |\Supp(x_B) - \Supp(x_A)|$. Then, in a random partition of $[2^r]$ into $n$ buckets, $\geq 0.01 \cdot (\min(k, n))$ buckets will have an index $i: (x_A)_i \neq (x_B)_i$ with probability $1 - n^{-20}$.
\end{claim}

\begin{proof}
Note that if $k > n$, we can simply focus our attention on the first $n$ indices $i$ such that $(x_B)_i \neq (x_A)_i$. Thus, we may assume that $k \leq n$. Now, let us calculate the probability that $\leq 0.01k$ buckets have an index $i: (x_A)_i \neq (x_B)_i$. We will view the random partitioning as a process where in $\ell$th step, the $\ell$th index in $\Supp(x_B) - \Supp(x_A)$ is randomly assigned a bucket in $[n]$. We want to bound the probability that $k$ indices are all assigned to the same $0.1k$ buckets. In order for this to happen, it must be the case that for at least $0.99k$ of the indices, they are assigned to one of the buckets already populated by the previous indices. Because the indices are contained in $\leq 0.01k$ buckets, the probability that this happens for any given index is at most $\frac{0.01k}{n}$. Because this must happen for $0.99k$ indices, we get the bound
\[
\Pr[\leq 0.01k \text{ buckets s.t. contain } i: (x_A)_i \neq (x_B)_i] \leq 2^k \cdot \left ( \frac{0.01k}{n} \right )^{0.99k}.
\]
Note that if $k \leq 100$, the probability of not having a single bucket contain an index $i: (x_A)_i \neq (x_B)_i$ is zero, we can instead focus on the case $k \geq 100$. In this case, we can bound the above probability with 
\[
2^k \cdot \left ( \frac{0.01k}{n} \right )^{0.99k} \leq 2^{-4k} \cdot (k/n)^{0.99k}.
\]
Now, we split the above into two cases: if $k \leq \sqrt{n}$ or if $k \geq \sqrt{n}$.
\begin{enumerate}
    \item If $k \leq \sqrt{n}$, then we can upper bound the probability of error by the second term: $(k/n)^{0.99k} \leq n^{-1/2(0.99k)} $. Because $k \geq 100$, we can bound the probability of our bad event by $n^{-99/2} \leq n^{-20}$.
    \item If $k \geq \sqrt{n}$, then the first term gives us an error bound of $2^{-4 \sqrt{n}} \leq n^{-20}$.
\end{enumerate}

Thus, in either case we get that 
\[
\Pr[\leq 0.01k \text{ buckets s.t. contain } i: (x_A)_i \neq (x_B)_i] \leq n^{-20},
\]
as we desire.
\end{proof}

By the previous claim, it follows that with probability $1 - n^{-20}$, in the first iteration, the partition $P_1$ created at least $k/100$ buckets $S_{\ell}$ such that $\exists i \in S_{\ell}: (x_A)_i \neq (x_B)_i$, where $k = |\Supp(x_B) - \Supp(x_A)|$. By construction of our hypergraph $H^1_A - H^1_B$, it follows that for these choices of $\ell$, the left vertex $\ell \in L$ must have an incident hyperedge. Because opening the sketch recovers a sparsifier for $H^1_A - H^1_B$, the sketch must recover an incident hyperedge to $\ell$, as otherwise the reported cut size for the set $\{ \ell\}$ would be $0$ (and thus not a $(1 \pm \eps)$ approximation to the true, positive size). Now, this means that Bob can recover $k/100$ indices $i$ for which the original $(x_A)_i \neq (x_B)_i$. Because $k \geq n/2$ originally, this means that we have recovered at least $n/200$ such indices.

Now, because Bob recovers linear sketches of $H^j_A - H^j_B$, he can update the sketches for $j \geq 2$ to remove the hyperedges that he recovered in the first round. Thus, Bob must only recover $\leq \frac{n}{2} (1 - 1/100)$ more indices before he has solved the instance. Inductively, we claim that after the first $j$ rounds of recovery Bob must recover $\leq \frac{n}{2} (1 - 1/100)^j$ more indices. We have already proved the base case. The inductive step follows because in the $j$th iteration, we let $k$ denote the $\min(n, $ remaining number of indices such that $(x_A)_i \neq (x_B)_i$ that we have not yet recovered). Note that $k \geq $ the number of indices that Bob must recover before solving the communication problem. This is because if $k = n$, then $n \geq n/2$ and $n/2$ is an upper bound on the number of indices which must be recovered. In the other case, by our promise that the original $x_A, x_B$ disagreed in at least $n/2$ locations, we are always guaranteed that if we have recovered $\ell$ indices, $k \geq (n/2 - \ell)$.

By the same logic as above, Bob is able to recover $\geq k/100$ of these indices in the $j$th round. 
Thus, the remaining number of indices which Bob must recover is $\leq \frac{n}{2} (1 - 1/100)^{j-1} \cdot (1 - 1/100) = \frac{n}{2} (1 - 1/100)^{j}$.

It follows that after $j = 100 \log(n)$ iterations of this, Bob must only recover $\leq \frac{n}{2} e^{-\log(n)} < 1$ more indices, thus meaning he has solved the instance.

Note that the total error probability in this procedure is bounded by the probability that any partition fails to create enough buckets with at least $1$ index, and the probability the linear sketch fails to return a sparsifier. In total, we can bound this probability by $100 \log(n) \cdot (n^{-20} + 4n^{-8}) \leq n^{-7}$.

Thus, we have shown that by sending $100\log(n)$ hypergraph sparsifier linear sketches (on $\leq m$ edges), Alice can send a message solving the $n/2$-$\UR_{r/2}^{\leq m}$ communication problem with probability $1 - 1 / n^7$. We know that any such message must be of length $\geq \Omega(nr \log(m/n))$, so this means that there must exist hypergraph sparsifier linear sketch instances that require length $\Omega(nr \log(m/n) / \log(n))$.

\end{proof}

\section{Streaming Algorithm}\label{sec:streaming}

From the previous sections, we have shown that there is a linear sketch (we'll denote this by $\calS_{\text{Hypergraph}}(H, R)$ (using public randomness $R$)) of size $\widetilde{O}(nr \log(m) / \eps^2)$ which returns a $(1 \pm \eps)$-sparsifier for a hypergraph $H$ with high probability. It remains now to show how we can use this to create a streaming algorithm. Naively, we can arbitrarily choose the public randomness for the linear sketch, and then start with a linear sketch of the empty hypergraph, $\calS_{\text{Hypergraph}}(\emptyset, R)$. Now, as the streaming algorithm is running, we simply update this sketch with the corresponding hyperedge that has just been seen. I.e., if a hyperedge $e$ is being inserted, we update our sketch by adding $\calS_{\text{Hypergraph}}(e, R)$. The algorithm looks like the following:

\begin{algorithm}[H]
\caption{DynamicHypergraphSparsification$(e_i, u_i)$}
Choose random bits $R$. \\
    Initialize $\calS_{\text{Hypergraph}} = \calS_{\text{Hypergraph}}(\emptyset, R)$. \\
    \For{$i = 1, \dots $}{
    $\calS_{\text{Hypergraph}} \leftarrow \calS_{\text{Hypergraph}} + u_i \cdot \calS_{\text{Hypergraph}}(e_i, R)$.
    }
    \Return{$\calS_{\text{Hypergraph}}$}
\end{algorithm}

The one subtlety is that often the convention with streaming algorithms is that any read-many random bits must count towards the space bound. The problem is that in our setting of hypergraphs, we are operating with uniformly random hash functions from $2^{[n]} \ra \zo$, and thus each hash function naively requires $2^n$ random bits. So, while our linear sketch itself is only taking $\widetilde{O}(nr \log(m))$ bits of space, to actually store the random bits leads to a possible exponential blow-up in size. To combat this, we simply use a variant of Newman's Theorem \cite{NS96} which generally allows us to replace any protocol using small space and public randomness, with a private randomness protocol using slightly more space.

We prove this variant with a few key claims below:

\begin{claim}\label{clm:nonuniform}
    For the linear sketching hypergraph sparsifier, there exists a set $S$ of $2^{10n}$ random seeds such that for an arbitrary hypergraph $H$, with probability $1 - 1 / n^6$ over a random choice $R$ of seed from $S$, the linear sketch using $R$ returns a sparsifier for $H$.
\end{claim}

\begin{proof}
    This follows from the probabilistic method. Let $S$ be a random set of $2^{10n}$ random seeds. We know that for a fixed hypergraph $H$, any random seed chosen at random yields a linear sketch that can be recovered to create a sparsifier for $H$ with probability $\geq 1 - n^{-7}$. Equivalently, we may say that any random seed $R$ for our linear sketch is \say{bad} with probability $1 / n^7$. Now, we want to bound the probability that if we sample $2^{10n}$ such random seeds, that more than a $1 / n^6$ fraction of these random seeds are bad for $H$. Let $X_1, \dots X_{2^{10n}}$ be random variables such that $X_i$ is $1$ if the $i$th random seed is bad for $H$. We want to bound $\Pr[(\sum_i X_i) / 2^{10n} \geq 1/n^{6}]$. We do this using a simple Chernoff bound:
    \[
    \Pr[(\sum_{i=1}^{2^{10n}} X_i) / 2^{10n} \geq 1/n^{6}] \leq  \Pr[(\sum_i X_i) / 2^{10n} \geq 2/n^{7}] \leq 2^{-2^{10n} / \text{poly}(n)} < 2^{-2^{2n}}.
    \]

    Now, note that there are only $2^{2^n}$ possible hypergraphs on $n$ vertices. Thus, we can take a union bound over all possible hypergraphs, and conclude that for a random set $S$ of $2^{2^{10n}}$ random seeds, with very high probability, for an arbitrary hypergraph $H$, using a random choice of seed from $S$ for our linear sketch will create a hypergraph sparsifier for $H$ with probability $\geq 1 - 1 / n^6$. Now, because a randomly constructed set $S$ satisfies this property with high probability, it follows that such a set $S$ must exist, and we can conclude our desired claim. 
\end{proof}

So, we can then create our streaming algorithm as follows. The algorithm is non-uniformly provided with such a set $S$ before execution. Now, it suffices to simply store a uniformly random index to a seed $R$ in $S$. Because $|S| = 2^{10n}$, storing such an index requires only $O(n)$ random bits. For an arbitrary hypergraph $H$, with high probability over the random seed chosen from $S$, the algorithm returns a sparsifier for $H$. This algorithm, as well as a formal statement of the Theorem, is provided below:

\begin{algorithm}\label{alg:DynamicStream}
\caption{DynamicHypergraphSparsification$((e_i, u_i))$}
Choose a random seed $R$ from the set $S$, storing only the index of $R$ in $S$. \\
    Initialize $\calS_{\text{Hypergraph}} = \calS_{\text{Hypergraph}}(\emptyset, R)$. \\
    \For{$i = 1, \dots $}{
    $\calS_{\text{Hypergraph}} \leftarrow \calS_{\text{Hypergraph}} + u_i \cdot \calS_{\text{Hypergraph}}(e_i, R)$.
    }
    \Return{$\calS_{\text{Hypergraph}}$}
\end{algorithm}

\begin{theorem}
    For an arbitrary dynamic stream of hyperedges $(e_i, u_i)$ on $n$ vertices, with the final hypergraph having $\leq m$ hyperedges, and an error parameter $\eps$, \cref{alg:DynamicStream} uses space $\widetilde{O}(nr \log(m) / \eps^2)$, and with probability $\geq 1- 1 / n^6$ returns a $(1 \pm \eps)$ hypergraph sparsifier for the hypergraph $H$ resulting from the stream.
\end{theorem}

\begin{proof}
    The space follows from \cref{clm:finalMain}. Between successive hyperedges, the algorithm stores only the index of the random seed in $S$ (using space $O(n)$), as well as the linear sketch of the hypergraph (using space $\widetilde{O}(nr \log(m) / \eps^2)$). The correctness follows by \cref{clm:nonuniform}. Indeed, for any fixed hypergraph $H$, with probability $1 - 1 / n^6$ over choice of random seed from $S$, our linear sketch returns a $(1 \pm \eps)$-sparsifier for $H$. Because our sketch is linear, it does not matter the order in which the hyperedges in the stream arrive, and rather, it only depends on the final resulting hypergraph induced by the insertions and deletions. Thus, we conclude the above theorem. 
\end{proof}

\section{MPC Algorithm}\label{sec:MPC}

In this section, we detail how to use our linear sketches for hypergraph sparsification to create an MPC algorithm for sparsifying hypergraphs. Recall that in the MPC model, the input data is split evenly across machines, each which has a bounded memory. In this section, we will assume each machine is given memory $\widetilde{O}(nr \log(m))$, that hyperedges have arity bounded by $r$, and that the $m$ hyperedges are split evenly across machines, resulting in each machine having $n$ hyperedges, and therefore a total of $k = \frac{m}{n}$ machines. We denote these machines by $m_1, \dots m_k$. 

At a high level, our MPC protocol will take advantage of the fact that the linear sketches for hypergraph sparsification are actually vertex-incidence sketches. That is, each vertex stores a sketch of its immediate neighborhood. In the first round, each machine creates the linear sketches for the hypergraph induced by the subset of hyperedges that were allocated to this machine. Because these sketches are really vertex-incidence sketches, the machines then coordinate to send their sketches for the first vertex (say $v_1$) to a subset of the machines, and likewise for $v_2$, $v_3$, and so on. We then recursively combine these sketches for individual vertices, until finally in the penultimate iteration, we have the complete sketch for each vertex $v_i$ stored in its own machine. In the final iteration, these machines coordinate and send these sketches to a single coordinator, which then has the entire linear sketch of the hypergraph $H$, and is able to compute a sparsifier. This will yield the following result:

\begin{corollary}
    There exists an MPC protocol which for any hypergraph $H$ on $n$ vertices, $m$ hyperedges, and with arity $\leq r$, uses only $\max(2, \lceil \log_n(m) \rceil)$ rounds of computation, with machines whose memory is bounded by $\widetilde{O}(nr \log(m) / \eps^2)$, and returns a $(1 \pm \eps)$ cut-sparsifier to $H$.
\end{corollary}

For comparison, the canonical approach to building MPC algorithms for sparsifying hypergraphs \emph{without} linear sketches involves each machine $m_i$ sparsifying its own induced hypergraph, and then recursively combining these hypergraphs in a tree-like manner, in each iteration pairing up two active machines, merging their hypergraphs, and then sparsifying this merged hypergraph. Thus, in each iteration, the number of active machines decreases by a factor of $2$. This approach (which is also used to create sparsifiers for \emph{insertion-only} streams \cite{CKN20}) unfortunately loses in two key parameter regimes. First, the number of rounds required by such a procedure will be $\Omega(\log(m/n))$, as the number of active machines decreases by a factor of $2$ in each round. Further, the memory required by each machine will be $\Omega(nr\log(m)\log^2(m/n) / \eps^2)$, as the deterioration of the error parameter scales with the depth of the recursive process, which will be $\log(m/n)$, and setting $\eps'= \eps / \log(m/n)$ requires more memory. 

As an example, when $m = \poly(n)$, our MPC protocol is able to run in a \emph{constant} number of rounds (independent of the number of vertices), whereas the canonical MPC algorithm for sparsification will require $\Omega(\log(n))$ rounds. Additionally, we will be getting this in conjunction with a smaller memory footprint. 

Note that the above algorithm is intended for cases when $k \geq n$ (in particular, $m \geq n^2$). Many times, it may be the case that $k < n$, in which case we have a separate procedure. We first present the algorithm for the case when $k \geq n$, which takes in a set of machines $m_1, \dots m_k$, each with some subset $S_j$ of the hyperedges of the hypergraph $H$:

\begin{algorithm}[H]
    \caption{MPC$((m_j, S_j)_{j = 1}^k)$}\label{alg:MPClargek}
    For each machine $m_j$, compute the hypergraph sparsification linear sketch $\calS(S_j, R)$, where $R$ is a random seed shared across machines. Let $\calS_i(S_j, R)$ denote the corresponding part of the sketch for vertex $j$. \\
    \For{$j \in [k]$}{
    \For{$i \in [n]$}{
    $m_j$ sends $\calS_i(S_j, R)$ to $m_{(j \mod (k/n)) + (k/n) \cdot (i-1)}$. \\
    $K^{(1)}_i = \{(k/n)\cdot (i-1) + 1, \dots (k/n)\cdot (i)\}$ (machines containing sketches for vertex $i$).
    }
    $m_j$ sums together the sketches it received (denote this $\calS^{(1,j)}$).
    }
    \For{$\ell \in [2, \lceil \log_n(m) \rceil]$}{
    \For{$i \in [n]$}{
    \For{$j \in [K^{(\ell-1)}_i]$}{
    Send $m_j$'s sketch to $m_{(j \mod (k / n^{\ell})) + (k / n^{\ell}) \cdot (i-1)}$.
    }
    $K^{(\ell)}_i = \{(k/n^{\ell})\cdot (i-1) + 1, \dots (k/n^{\ell})\cdot (i)\}$.
    \For{$j \in [K^{(\ell)}_i]$}{
    $m_j$ sums together the sketches it received (denote this $\calS^{(\ell, j)}$).
    }
    }
    }
    In the final round, $m_1, \dots m_n$ each send their sketch to $m_1$, which now computes the hypergraph sparsifier. \\
    $m_1$ returns the hypergraph sparsifier.
\end{algorithm}

First, we prove that this procedure does not exceed the memory capacity of any machine.

\begin{claim}\label{clm:MPCspace}
    In every round, each machine uses at most $\widetilde{O}(nr \log(m) / \eps^2)$ bits of memory.
\end{claim}

\begin{proof}
    First, observe that in the first round, when the machines compute the hypergraph sparsifier for their subset of the edges, this creates $\widetilde{O}(\polylog(n) / \eps^2)$ $\ell_0$-samplers for each vertex, each of which requires space at most $\widetilde{O}(r \log(m) \polylog(n))$. Now, by induction, in each subsequent round, the protocol creates groups of $n$ machines, each containing $\widetilde{O}(\polylog(n) / \eps^2)$ $\ell_0$-samplers for a single vertex $v_i$, and sends all of these $\ell_0$-samplers to a single machine. The total space required to receive these samplers (sketches) is bounded by $n \cdot \widetilde{O}(r \log(m) \polylog(n) / \eps^2) = \widetilde{O}(nr\log(m) / \eps^2)$. Now, because these are linear sketches, the protocol simply adds together these sketches, yielding a sketch of size $\widetilde{O}(r\log(m)\polylog(n) / \eps^2)$ because this is still simply a set of $\widetilde{O}(\polylog(n) / \eps^2)$ $\ell_0$-samplers. Thus, inductively, the space required never exceeds $\widetilde{O}(nr \log(m) / \eps^2)$ bits.

    In the final round, $n$ machines, each with $\widetilde{O}(r\log(m)\polylog(n) / \eps^2)$ bits, sends their memory to $m_1$, which now has the complete linear sketch required for hypergraph sparsification, and is able to sparsify the hypergraph $H$. 
\end{proof}

\begin{claim}\label{clm:MPCrounds}
    The number of rounds required for the above procedure is $\lceil \log_n(m) \rceil$.
\end{claim}

\begin{proof}
    Note that in each round, $K_i^{(\ell)}$ is bounded in size by $(k / n^{\ell})$ by construction. Thus, after $\log_n(k) = \log_n(m/n) \leq \lceil \log_n(m) \rceil - 1$ rounds, we have that each $K_i^{(\ell)}$ is of size $1$. In the final round, each machine sends their sketches to $m_1$, which is able to compute the sparsifier and return the result. This yields the desired claim. 
\end{proof}

\begin{claim}\label{clm:MPCcorrect}
    In the final round of the MPC protocol, $m_1$ has a valid hypergraph sparsification sketch for $H$. 
\end{claim}

\begin{proof}
    This follows because in each round, the $\ell_0$-samplers for each vertex are added together. In the final round, $m_1$ receives $\ell_0$-samplers for each vertex defined over the entire hypergraph $H$ (because they have been added together using the hyperedges given to each machine).
\end{proof}

\begin{corollary}
    \cref{alg:MPClargek} is a valid MPC protocol for creating hypergraph sparsifiers with each machine using $\widetilde{O}(nr \log(m) / \eps^2)$ bits of memory, and computing for a total of $\lceil \log_n(m) \rceil$ rounds.
\end{corollary}

\begin{proof}
    This follows from \cref{clm:MPCspace}, \cref{clm:MPCrounds}, and \cref{clm:MPCcorrect}.
\end{proof}

Note that the algorithm presented above is intended for instances where $k \geq n$. When $k < n$, instead of creating multiple machines responsible for the sketches for a single vertex, we create a single machine which is responsible for many vertices. Let us suppose that $n / k$ is an integer for simplicity. Then, the first machine $m_1$ is responsible for creating the sketches for vertices $v_1, \dots v_{n/k}$, and more generally, machine $m_j$ is responsible for the sketches for vertices $v_{(n/k) \cdot (j-1) + 1}, \dots v_{(n/k) \cdot j }$. The first round is spent agglomerating these sketches, and in the final round, these machines send their vertex sketches to a single coordinator who then returns a sparsifier. We present this algorithm below:

\begin{algorithm}
    \caption{SmallMPC$((m_j, S_j)_{j = 1}^k)$}\label{alg:smallMPC}
    For each machine $m_j$, compute the hypergraph sparsification linear sketch $\calS(S_j, R)$, where $R$ is a random seed shared across machines. Let $\calS_i(S_j, R)$ denote the corresponding part of the sketch for vertex $j$. \\
    \For{$j \in [k]$}{
    \For{$i \in [n]$}{
    $m_j$ sends $\calS_i(S_j, R)$ to $m_{ \lceil \frac{jk}{n}\rceil}$. \\
    }
    For each vertex $i \in [(n/k)(j-1) +1, (n/k) j]$, $m_j$ sums together the vertex sketch it receives. Denote these sketches by $\calS^{(i,j)}$.
    }
    \For{$j \in [k]$}{
    \For{$i \in [(n/k)(j-1) +1, (n/k) j]$}{
    $m_j$ sends $\calS^{(i,j)}$ to $m_1$.
    }
    }
    $m_1$ computes the hypergraph sparsifier using the received sketches.
\end{algorithm}

Note that the correctness of the above algorithm follows from the same reasoning as for the original MPC algorithm. Further, by construction, there are only two rounds of communication, once for separating the vertex sketches, and once for recombining them in the coordinator's memory. Thus, it remains to bound the memory usage of each machine. 

\begin{claim}
    Each machine in \cref{alg:smallMPC} uses $\widetilde{O}(nr \log(m) / \eps^2)$ bits of memory.
\end{claim}

\begin{proof}
    Suppose for simplicity that $k$ evenly divides $n$. Note that by assumption, we are also assuming that the total number of hyperedges in the hypergraph is bounded by $kn$. In the first round, each machine receives from $k$ different machines, the $\ell_0$-samplers corresponding to $\frac{n}{k}$ different vertices. From each machine, the total size of the $\ell_0$-samplers stored per vertex is bounded by $\widetilde{O}(r \log(m)\polylog(n) / \eps^2)$. Thus, the total memory required to store the communicated bits is
    \[
    \leq k \cdot \frac{n}{k} \cdot \widetilde{O}(r \log(m)\polylog(n) / \eps^2) = \widetilde{O}(nr \log(m) / \eps^2).
    \]
    Next, each machine is able to add together the corresponding $\ell_0$-samplers for each vertex, thus reducing the total space usage to again only $\widetilde{O}(r \log(m)\polylog(n) / \eps^2)$ bits per vertex. Thus, in the second round, when $m_1$ receives from each of the $k$ machines the $\ell_0$-samplers corresponding with $n/k$ vertices, this is again bounded by $\widetilde{O}(nr \log(m) / \eps^2)$ bits, and $m_1$ is able to compute the sparsifier in its local memory.
\end{proof}

\begin{corollary}
    There exists an MPC protocol which for any hypergraph $H$ on $n$ vertices, $m$ hyperedges, and with arity $\leq r$, uses only $\max(2, \lceil \log_n(m) \rceil)$ rounds of computation, with machines whose memory is bounded by $\widetilde{O}(nr \log(m) / \eps^2)$, and returns a $(1 \pm \eps)$ cut-sparsifier to $H$.
\end{corollary}

\bibliographystyle{alpha}
\bibliography{ref}

\appendix

\section{Reproof of $\ell_0$-samplers}\label{sec:appendixSampler}

We adopt the construction presented in Cormode and Firmani \cite{CF14}. To do this, we re-present their method for perfect 1-sparse recovery. In this setting, we are given a vector $x \in \Z^u$ (and let us suppose that $|x_i| \leq \text{poly}(u)$), and our goal is to either 
\begin{enumerate}
    \item Return $x$ exactly if there is at most one non-zero index in $x$. 
    \item Return $\perp$ with probability $1 - 1 / u^c$, if there is more than $1$ non-zero index in $x$.
\end{enumerate}

To do this, we first choose a prime $p$ which is sufficiently large. For now, we choose $p$ to be in the interval $[u^{c+1}, 2u^{c+1}]$. Next, we choose a random integer $z \in \Z_p$, and store the following quantities:

\begin{enumerate}
    \item $\alpha = \sum_i x_i \cdot i$. 
    \item $\phi = \sum_i x_i$. 
    \item $\tau = \sum_i x_i \cdot z^i \mod p$.
\end{enumerate}

\begin{claim}\label{clm:oneSparseRecovery}\cite{CF14}
    If $x$ is $1$-sparse, then $\tau = \phi \cdot z^{\alpha / \phi } \mod p$. If $x$ is not $1$-sparse, then with probability $\geq 1 - u / p \geq 1 - 1 / u^c$ over the random choice of $z$, $\tau \neq \phi \cdot z^{\alpha / \phi } \mod p$.
\end{claim}

\begin{corollary}\label{cor:oneSparseRecovery}
    There exists a linear sketch of a vector $x$ using $O(c\log(u))$ bits of space, which can recover $x$ exactly if $x$ is $1$-sparse, and otherwise reports that $x$ is not $1$-sparse with probability $\geq 1 - 1 / u^c$.
\end{corollary}

\begin{proof}
    First, we can see that the information we store $(\alpha, \phi, \tau)$ are linear in the vector $x$, thus the sketch itself is linear. 
    
    Second, by \cref{clm:oneSparseRecovery}, we can test if $\tau = \phi \cdot z^{\alpha / \phi } \mod p$ to see whether or not our vector $x$ is truly $1$-sparse (with high probability). If indeed $x$ is one-sparse, then one can find the index $i$ for which $x_i \neq 0$ by dividing $\alpha$ by $\phi$. One can then also recover the value at the index which will be exactly $\phi$.

    The space required for the linear sketch follows from the fact that the prime $p$ requires $O(c \log(u))$ bits to represent. Storing $\alpha$ requires at most $u \cdot \poly(u)$ bits of space (because we are assuming each entry $x_i$ is bounded in magnitude by $\poly(u)$). Likewise $\phi$ is bounded by $\poly(u)$, and $\tau$ is bounded by $p$. Thus storing each of these quantities requires at most $O(c \log(u))$ bits of space. 
\end{proof}

Going forward, we will denote this linear sketch for $1$-sparse recovery by $\calS_{\text{1S}}$.

Now, we can use this method for $1$-sparse recovery to create a linear sketch for $\ell_0$ sampling of a vector $x \in \Z^u$, where each entry is bounded by $\text{poly}(u)$. We will also parameterize this vector $x$ by an upper bound in terms of its support $m$.

\begin{theorem}
    For a vector $x \in \Z^u$, with each entry bounded in magnitude by $\poly(u)$, there exists a linear sketch of size $O(\log(m) \log(1 / \delta) \log(u) \cdot \max(1, \log_u(1 / \delta)))$ which:
    \begin{enumerate}
        \item If the size of the support of $x$ is $\leq m$, returns a uniformly random index $i$, and the corresponding value $x_i$, such that $x_i \neq 0$ with probability $1 - \delta$.
        \item If the size of the support of $x$ is $> m$, either 
        \begin{enumerate}
            \item Returns a uniformly random index $i: x_i \neq 0$, as well as $x_i$,
            \item Outputs $\perp$,
        \end{enumerate}
        with probability $1 - \delta$.
    \end{enumerate}
\end{theorem}

\begin{proof}
First, we create $\log(m)$ uniformly random hash functions from $[u] \ra \zo$. We denote these hash functions by $h_1, \dots h_{\log(m)}$. Now, we create $\log(m)+1$ versions of the vector $x$, where $x^{(j)}$ contains only the indices $i: \prod_{p \leq j} h_p(i) = 1$, and sets all other entries to be $0$. For each of these vectors $x^{(j)}$, $0 \leq j \leq \log(m)$, we store a sketch $\calS_{\text{1S}}(x^{(j)})$.

Now, it follows that if $x$ has $\leq m$ non-zero entries (i.e. support size bounded by $m$), then with constant probability, there will exist a $j \in [\log(m)]$ such that $x^{(j)}$ has only one non-zero entry. 

Thus, if we store $O(\log(1 / \delta))$ (independent) versions of this sketch, we will be ensured that with probability $1 - \delta$, there will exist one version of this sketch with a downsampled vector $x^{(j)}$ such that $x^{(j)}$ has only one non-zero entry. For this vector, by \cref{cor:oneSparseRecovery}, we will exactly recover both the index $i$, and the value $x_i$.

Now, we must also show that we do not recover any incorrect indices in this case. This again follows from \cref{cor:oneSparseRecovery}. There are at most $\log(m) \log(1 / \delta)$ copies of $\calS_{\text{1S}}$ that are stored, and for each, the error probability is bounded by $1 / u^c$. By a union bound, it follows that the total error probability is bounded by $\frac{\log(m) \log(1 / \delta)}{u^c} \leq \frac{\log(u) \log(1 / \delta)}{u^c} \leq \frac{\log(1 / \delta)}{u^{c-1}}$. Setting $c = O(\max(1, \log_u(\delta)))$, we can then also bound this failure probability by $\delta$. Note that this also takes care of the second case, as here we are again only bounding the probability that we fail to correctly identify the vector as being $1$-sparse. 

The space required by this sketch is thus $O(\log(m) \log(1 / \delta) \log(u) \cdot \max(1, \log_u(1 / \delta)))$, as we store $\log(m) \log(1/\delta)$ copies of $\calS_{\text{1S}}$, where we set $c = O(\max(1, \log_u(\delta))$.
\end{proof}

\subsection{Sparse Recovery with Overflow Detection}

As one of the building blocks of $\ell_0$-samplers, we want a linear sketch that satisfies the following conditions: Find a linear map $L: \{ -U, -U+1, \dots U-1, U \}^n \ra \R^k$ such that if $x$ is $s$-sparse, it can be recovered from $L(x)$, and if $x$ is not $s$-sparse, then we say \say{DENSE} with probability at least $1 - \delta$. Here, we are using $n$ to represent the universe size (as opposed to our convention of $u$ so far) in accordance with the coding theory standards.

To do this, we create the following sketch: take any prime $p > \max(2U + 1, n / \delta)$, as well as codes $C_1 \in [n, m, 2s+1]_p$ and $C_2 \in [N, n, (1 - \delta)N]_p$. Note that $[n, k, d]_q$ codes linearly map messages in $\F_q^k$ to codeword in $\F_q^n$ and have distance $d$ between codewords. Given $C_1$, there is a parity check matrix defining a linear function $H: \F_q^n \ra \F_q^{n-m}$ such that every $s$-sparse vector $x$ can be recovered from $H(x)$. Our random function $L$ is obtained by taking $i \in [N]$ uniformly, and letting $L(x) = (H(x), C_2(x)_i)$. We denote this by $\calS_{SR}$ (sparse-recovery).

\begin{claim}
    For a vector $x \in \{ -U, -U+1, \dots U-1, U \}^n$, if $x$ is $s$-sparse, one can recover $x$ exactly from $\calS_{SR}(x)$. If $x$ is not $s$-sparse, one can identify this with probability $1 - \delta$.
\end{claim}

\begin{proof}
    We implement the following recovery-with-detection paradigm: let us use $H(x)$ to recover a candidate $y \in \F_p^n$. If $y$ is $s$-sparse, in $\{ -U, -U+1, \dots U-1, U \}^n$, and satisfies $C_2(y)_i = C_2(x)_i$, then we output $y$. Otherwise, if any of these do not happen, we output \say{DENSE}.

    To see why the above procedure works, note that if $x$ is $s$-sparse, then $y$ will equal $x$ (by virtue of the syndrome decoding via $H$), and satisfy all of the tests. Thus, the interesting case is when $x$ is not $s$-sparse, yet the recovered vector $y$ is $s$-sparse. Then, $x \neq y$, so $C_2(x)$ and $C_2(y)$ will differ in at least $1 - \delta$ coordinates. So, with probability $1 - \delta$ over the choice of $i$, we will output \say{DENSE}.
\end{proof}

\begin{claim}
    For $s$-sparse vectors, we can implement the above with a sketch of size $O(s \log(\max(U, n / \delta)))$.
\end{claim}

\begin{proof}
    We implement the above with two Reed-Solomon codes. We set $C_1$ to be a Reed-Solomon code with $m = n - (2s+1)$, and let $C_2$ have $N = n / \delta$. The size of our message is then $(2s+1) \log(p)$ bits for $H(x)$, and $\log(p)$ bits for $C_2(x)_i$. The amount of randomness used is simply $\log(N) = \log(n / \delta)$.
\end{proof}

Typically, we set $U = \poly(n)$, leading to a linear sketch that uses $O(s \log(n / \delta))$ bits of space. 

\end{document}